\documentclass[fleqn,10pt]{article}
\usepackage{latexsym, graphicx, epsfig, amsmath, amssymb,amsfonts}
\usepackage{natbib,amsthm,version}
\usepackage{amsbsy,bm,multirow,enumerate}
\usepackage[titletoc,page]{appendix}
\usepackage[mathscr]{eucal}
\usepackage{mathtools}
\usepackage{color}
\usepackage[utf8]{inputenc}
\usepackage[english]{babel}
\usepackage{amsthm}
\usepackage{subfigure}

\newcommand{\Order}{{\cal{O}}}             

\newtheorem{theorem}{Theorem}[section]

\newtheorem{remark}{Remark}

\theoremstyle{definition}

\theoremstyle{proposition}

\theoremstyle{remark}

\newcommand{\bs}[1]{\boldsymbol{#1}}

\title{
  A Family of Second-Order Energy-Stable
  Schemes for Cahn-Hilliard Type Equations
} 
\author{
  Suchuan Dong$^1$\thanks{Author of correspondence. Email: sdong@purdue.edu},\; Zhiguo Yang$^1$,\; Lianlei Lin$^{1,2}$ \\
  $^1$Center for Computational and Applied Mathematics \\
  Department of Mathematics \\
  Purdue University, USA \\
  $^2$School of Electrical Engineering and Automation \\
  Harbin Institute of Technology, China
 }

\graphicspath{{./Figures/EvoDrop/} {./Figures/Spin/} {./Figures/Airbubble/} }

\date{(\today)}
\begin{document}
\maketitle




\begin{abstract}

  We focus on the numerical approximation of
  the Cahn-Hilliard type equations, and
  present a family of second-order
  unconditionally energy-stable schemes.
  By reformulating the equation into
  an equivalent system employing a  scalar auxiliary
  variable, we approximate the system at
  the time step $(n+\theta)$ ($n$ denoting the time step index
  and $\theta$ is a real-valued parameter), and
  devise a family of corresponding approximations that are second-order
  accurate and unconditionally energy stable.
  This family of approximations contains the often-used Crank-Nicolson
  scheme and the second-order backward differentiation formula
  as particular cases.
  We further develop an efficient solution algorithm
  for the resultant discrete system of equations to overcome
  the difficulty caused by the unknown scalar auxiliary variable.
  The final algorithm requires only the solution of
  four de-coupled individual Helmholtz type equations 
  within each time step, which involve only constant
  and time-independent coefficient matrices that can be
  pre-computed. A number of numerical examples are
  presented to demonstrate the performance of the family
  of schemes developed herein.
  We note that this family of second-order 
  approximations  can be readily applied to devise
  energy-stable schemes for other types of gradient flows
  when combined with the auxiliary variable approaches.

\end{abstract}


\vspace{0.05cm}

\section{Introduction}
\label{sec:intro}

%
%


Diffuse interface or phase field
approach~\cite{AndersonMW1998,LowengrubT1998,LiuS2003,YueFLS2004}
has become one of the main techniques for modeling and
simulating two-phase and multiphase problems involving
fluid interfaces and the effect of surface tensions.
Cahn-Hilliard type equations~\cite{CahnH1958} are
one of the commonly encountered equations
in such models for describing the evolution of
the phase field functions. Indeed, with an appropriate
free energy density form, the mass balance equations for
the individual fluid components in a multicomponent
mixture will reduce to the Cahn-Hilliard type
equations~\cite{HohenbergH1977,GurtinPV1996,AbelsGG2012,Dong2014,Dong2018}.
Devising efficient numerical schemes for Cahn-Hilliard
 equations therefore has crucial implications to
two-phase and multiphase problems, and this has attracted
a sustained interest from the community~\cite{TierraG2015}.


Nonlinearity and high spatial order (fourth order)
are the main issues encountered when numerically solving
the Cahn-Hilliard equations. The interfacial thickness
scale parameter, when small, also exacts high mesh resolutions
in numerical simulations.
The energy stability property of a numerical scheme, when the computational
cost is manageable, is a desirable
feature for solving the Cahn-Hilliard equations.
While other types (e.g.~conditionally stable, semi-implicit) of schemes
for the Cahn-Hilliard equation also exist in
the literature (see e.g.~\cite{BadalassiCB2003,YueFLS2004,DongS2012,Dong2014obc,GonzalezT2013,Dong2017},
among others),
in what follows we will focus on the energy-stable schemes
in the review of literature.

The nonlinearity of Cahn-Hilliard equation is
induced by the potential free energy density function.
Ensuring discrete energy stability in a numerical scheme
hinges on the treatment of the nonlinear term.
Based on the strategies for treating the nonlinear term,
existing energy stable schemes for the Cahn-Hilliard equations
can be broadly classified into two categories: nonlinear
schemes and linear schemes.
Nonlinear schemes (see e.g.~\cite{ElliotFM1989,Furihata2001,FengP2004,KimKL2004,MelloF2005,Feng2006,Wise2010,HuaLLW2011})
entail the solution of a system of nonlinear algebraic equations
within a time step after discretization.
Convex splitting of the potential energy term
and its variants are a popular approach to treat the nonlinear term
in this category~\cite{ElliotS1993,Eyre1998}.
Other treatments
include the midpoint approximation~\cite{ElliotFM1989,LinLZ2007},
specially designed quadrature formulas~\cite{GomezH2011}, and
Taylor expansion approximations~\cite{KimKL2004} of
the potential term, among others.
%
On the other hand, linear
schemes (see e.g.~\cite{ShenY2010,GonzalezT2013,Yang2016,ShenXY2018})
involve only the solution of a system of linear algebraic equations
after discretization, thanks to certain special treatment
of the nonlinear term or the introduction of certain
auxiliary variables. Adding a stabilization term that is
equivalent to zero while using a potential energy with bounded second derivative
and treating the nonlinear term
explicitly~\cite{ShenY2010,GonzalezT2013}
is a widely used technique in this category.
Using a Lagrange multiplier~\cite{BadiaGG2011,GonzalezT2013}
is another technique to derive unconditionally energy-stable
schemes for the Cahn-Hilliard equation.
The invariant energy quadratization (IEQ)~\cite{Yang2016}
is a general technique that generalizes the Lagrange multiplier approach
and can be applied to a large class of free energy forms.
IEQ introduces an auxiliary field function related to
the square root of the potential free energy density function together with
a dynamic equation for this auxiliary variable, and allows
one to devise schemes to ensure the energy stability
relatively easily. The IEQ method gives rise to a system of
linear algebraic equations involving time-dependent
coefficient matrices after discretization.
%
A further development of the auxiliary variable strategy is
introduced in \cite{ShenXY2018} very recently,
in which an auxiliary variable, which is a scalar value rather than
a field function,
related to the square root of the total potential energy integral
has been employed. 
The scalar auxiliary variable (SAV) method retains the main advantage
of IEQ, and further can lead to a constant coefficient matrix
for the resultant linear algebraic system of equations
after discretization.


In the current work we focus on the numerical approximation of the
Cahn-Hilliard equation, reformulated using the scalar auxiliary variable
appproach. We present a family of second-order accurate 
linear schemes for the system, and show that this family of schemes
is unconditionally energy-stable. This family of approximations contains
the Crank-Nicolson scheme (or trapezoidal rule) and the
second-order backward differentiation formula (BDF2) as
particular cases. The key idea of the schemes lies in
enforcing the system of equations at the time step
$(n+\theta)$, where $n$ is the time step index and $\theta$
is a real-valued parameter ($\frac{1}{2}\leqslant \theta\leqslant\frac{3}{2}$),
and then devising appropriate corresponding
approximations at $(n+\theta)$ with second-order accuracy
that guarantee the
energy stability of the system.
We further present an efficient solution algorithm for
the discretized system of equations to overcome
the numerical difficulty induced by the unknown scalar
auxiliary variable. Our overall algorithm only requires
the computation of four de-coupled individual Helmholtz type equations
within a time step, which involve constant coefficient
matrices that can be pre-computed.


The novelties of this paper lie in two aspects:
(i) the family of second-order accurate energy-stable
schemes, and (ii) the efficient solution algorithm
for the discrete system resulting
from the Cahn-Hilliard type equations.


While we only consider the numerical approximation of
the Cahn-Hilliard equation in this work,
we would like to point out that the family of second-order
energy-stable approximations herein are general,
and are readily applicable to
other types of equations resulting from gradient flows
for designing energy-stable schemes 
when combined with the IEQ or SAV strategy.


The rest of this paper is structured as follows.
In Section \ref{sec:method} we discuss the SAV reformulation
of the Cahn-Hilliard equation, and present the family of
second-order energy-stable approximations of the reformulated
system of equations. We will also present an efficient
solution algorithm for the discretized equations and
an implementation of the algorithm based on the
$C^0$-continuous spectral element method for spatial discretizations.
In Section \ref{sec:tests} we test the performance of
the algorithms using several representative numerical examples,
and demonstrate numerically the stability of computations
with large time step sizes.
Section \ref{sec:summary} concludes the presentation
with some closing remarks. The Appendices A and B
provide proofs for the energy stability and
another property of the presented family of schemes.

\section{A Family of Second-Order Energy-Stable Schemes}
\label{sec:method}

\subsection{Cahn-Hilliard Equation, Boundary Conditions, and Transformed System}

Consider a domain $\Omega$ in two or three dimensions, whose
boundary is denoted by $\partial\Omega$, and
the Cahn-Hilliard equation~\cite{CahnH1958} with a source term
within this domain:
\begin{equation}
  \frac{\partial\phi}{\partial t} = m\nabla^2\left[
    -\lambda\nabla^2\phi + h(\phi)
    \right] + g(\mathbf{x},t)
  \label{equ:CH}
\end{equation}
where $\phi(\mathbf{x},t)$ ($\phi\in [-1, 1]$) is the phase field function, $m>0$ is
the mobility and assumed to be a constant in this work,
and $\mathbf{x}$ and $t$ denote the spatial coordinates and time.
$g(\mathbf{x},t)$ is a prescribed source term for
the purpose of numerical testing (for convergence rates) only, and will be set
to $g(\mathbf{x},t)=0$ in actual simulations.
$\lambda$ is the mixing energy density coefficient and is related to
other physical parameters, e.g.~for two-phase flow problems it is given by
(see~\cite{YueFLS2004})
$
\lambda = \frac{3}{2\sqrt{2}}\sigma\eta,
$
where $\sigma$ is the surface tension and $\eta$ is the characteristic
interfacial thickness.
The nonlinear term $h(\phi)$ is given by $h(\phi) = \frac{dF(\phi)}{d\phi}$,
where $F(\phi)$ is the potential 
free energy density function
of the system.
A double-well potential is often used for $F(\phi)$,
in the form
$
F(\phi) = \frac{\lambda}{4\eta^2}(\phi^2-1)^2,
$
and this form will be used for the numerical tests in
Section \ref{sec:tests}.

We consider the wall-type boundary with
a neutral wettability (i.e.~$90^0$ contact angle)
for $\partial\Omega$,
characterized by
the following boundary conditions~\cite{Jacqmin1999,YueZF2010,Dong2012}
\begin{subequations}
  \begin{equation}
    \mathbf{n}\cdot\nabla\left[-\lambda\nabla^2\phi + h(\phi) \right]
    = g_a(\mathbf{x},t),
    \quad \text{on} \ \partial\Omega
    \label{equ:wbc_1}
  \end{equation}
  \begin{equation}
    \mathbf{n}\cdot\nabla\phi  = g_b(\mathbf{x},t),
    \quad \text{on} \ \partial\Omega
    \label{equ:wbc_2}
  \end{equation}
\end{subequations}
where $\mathbf{n}$ is the outward-pointing unit vector normal to
$\partial\Omega$. $g_a(\mathbf{x},t)$ and
$g_b(\mathbf{x},t)$ are prescribed source terms for
the purpose of numerical testing only, and
will be set to $g_a=0$ and $g_b=0$ in
actual simulations.

The system consisting of the equation \eqref{equ:CH}
and the boundary conditions \eqref{equ:wbc_1} and \eqref{equ:wbc_2},
with $g=0$, $g_a=0$ and $g_b=0$,
satisfies the following energy balance equation:
\begin{equation}
  \frac{\partial}{\partial t}
    \int_{\Omega}\left(\frac{\lambda}{2}\nabla\phi\cdot\nabla\phi
    + F(\phi) \right)
  = -m\int_{\Omega} \left|\nabla\left[
    -\lambda \nabla^2\phi + h(\phi)
    \right]  \right|^2.
  \label{equ:energy_balance}
\end{equation}
This system is to be supplemented by the following initial condition
\begin{equation}
  \phi(\mathbf{x},0) = \phi_{in}(\mathbf{x})
  \label{equ:ic}
\end{equation}
where $\phi_{in}(\mathbf{x})$ denotes the initial phase field
distribution.

We next reformulate this system of equations and boundary conditions
into a modified equivalent system, by introducing
an auxiliary variable associated with the
total potential energy in a way similar to~\cite{ShenXY2018}.
Define the total potential free energy $E(t)$ by
\begin{equation}
E(t) = E[\phi] = C_0 + \int_{\Omega} F(\phi)
\label{equ:def_E1_E2}
\end{equation}
where $C_0$ is a chosen constant such that 
$E(t) > 0$ for all $0 \leqslant t\leqslant T$
($T$ denoting the period of time to find the solution for).
For all the numerical experiments presented in the current work
we employ  $C_0=0$ in the simulations. 
It is important to note that $E[\phi]$ as defined
above is a scalar value,
not a field function.
We define an auxiliary variable $r(t)$ by
\begin{equation}
r(t) = \sqrt{E(t)}.
\label{equ:def_r_q}
\end{equation}
Then $r(t)$ satisfies the dynamic equation
\begin{equation}
\frac{dr}{dt} = \frac{1}{2\sqrt{E[\phi]}}\int_{\Omega} h(\phi) \frac{\partial\phi}{\partial t},
\label{equ:r_equ}
\end{equation}
We re-write the Cahn-Hilliard equation \eqref{equ:CH} into an equivalent form
\begin{subequations}
\begin{equation}
  \frac{\partial\phi}{\partial t} = m\nabla^2 \mathscr{H} \label{equ:CH_trans_1}
  + g,
\end{equation}
\begin{equation}
\mathscr{H} = -\lambda\nabla^2\phi + \frac{r(t)}{\sqrt{E[\phi]}} h(\phi), 
\label{equ:CH_trans_2}
\end{equation}
\end{subequations}
where the definition \eqref{equ:def_r_q} has been used.
The boundary conditions \eqref{equ:wbc_1} 
is also re-written into an equivalent form
\begin{subequations}
\begin{equation}
\mathbf{n}\cdot\nabla\left[
-\lambda\nabla^2\phi + \frac{r(t)}{\sqrt{E[\phi]}} h(\phi)
\right] = g_a, \quad \text{on} \ \partial\Omega,
\label{equ:wbc_trans_1}
\end{equation}
\end{subequations}

The system consisting of equations \eqref{equ:CH_trans_1}--\eqref{equ:CH_trans_2}
and \eqref{equ:r_equ}, the boundary
conditions \eqref{equ:wbc_trans_1} and \eqref{equ:wbc_2}, and 
the initial conditions of \eqref{equ:ic} and the following
\begin{equation}
\begin{split}
&
r(0) = \sqrt{E[\phi_{in}]} = \left[\int_{\Omega} F(\phi_{in}) + C_0 \right]^{1/2}, 
\end{split}
\label{equ:ic_rq}
\end{equation}
is equivalent to the original system consisting of equations
\eqref{equ:CH}, \eqref{equ:wbc_1}--\eqref{equ:wbc_2} and
\eqref{equ:ic}.

\subsection{A Family of Second-Order Energy-Stable Approximations}

We focus on the numerical approximation of the reformulated equivalent system
consisting of equations \eqref{equ:CH_trans_1}--\eqref{equ:CH_trans_2}
and \eqref{equ:r_equ}, the boundary
conditions \eqref{equ:wbc_trans_1} and \eqref{equ:wbc_2}, and 
the initial conditions \eqref{equ:ic} and \eqref{equ:ic_rq}.
We present a family of second-order
energy-stable schemes for this system that allows for
a very efficient implementation.

Let $n\geqslant 0$ denote the time step index, and $(\cdot)^n$
represent the variable $(\cdot)$ at time step $n$,
corresponding to the time $t=n\Delta t$, where $\Delta t$
is the time step size.

Let $\theta$ ($\frac{1}{2}\leqslant \theta\leqslant \frac{3}{2}$)
denote a real-valued parameter. We approximate the 
variables at  time step $(n+\theta)$ (corresponding to
time $(n+\theta)\Delta t$)
as follows with a scheme of second-order accuracy
in time ($\chi$ denoting a generic
variable below),
\begin{subequations}
  \begin{equation}
    \left.\frac{\partial\chi}{\partial t}\right|^{n+\theta}
    = \frac{1}{\Delta t}\left[
      \left(\theta + \frac{1}{2}\right)\chi^{n+1} - 2\theta\chi^n
      + \left(\theta-\frac{1}{2}\right)\chi^{n-1}
      \right],
    \label{equ:def_deriv_n_theta}
  \end{equation}
  \begin{equation}
    \chi^{n+\theta} =
    \left[\theta\left(\frac{5}{2}-\theta\right)-\frac{1}{2}  \right]\chi^{n+1}
    + 2(1-\theta)^2\chi^n
    + \left(\theta-\frac{1}{2}\right)(1-\theta)\chi^{n-1},
    \label{equ:def_var_implicit_n_theta}
  \end{equation}
  \begin{equation}
    \bar{\chi}^{n+\theta} = (1+\theta)\chi^n - \theta\chi^{n-1}.
    \label{equ:def_var_explicit_n_theta}
  \end{equation}
\end{subequations}
In the above $\chi^{n+\theta}$ and $\bar{\chi}^{n+\theta}$ are respectively
an implicit and an explicit approximation of $\chi$
at time step $(n+\theta)$.
The 2nd-order temporal accuracy of these approximations can be verified
by Taylor expansions in a straightforward fashion.
The implicit approximation $\chi^{n+\theta}$ given in
\eqref{equ:def_var_implicit_n_theta} is critical to
the energy stability of 
the current family of schemes, due to the
following crucial property:
\begin{equation}
\begin{split}
&
\chi^{n+\theta}
\left[
\left(\theta +\frac{1}{2} \right)\chi^{n+1}
-2\theta\chi^{n}
+\left(\theta -\frac{1}{2} \right)\chi^{n-1}
\right] \\
& \quad
= \frac{1}{2}\theta\left(\theta-\frac{1}{2} \right)(3-2\theta)\left|
  \chi^{n+1} - 2\chi^n + \chi^{n-1}
\right|^2
+ \frac{1}{2}\left(\frac{3}{2}-\theta \right)\left(
  \left|\chi^{n+1}  \right|^2 - \left|\chi^n \right|^2
\right) \\
& \qquad
+ \frac{1}{2}\left(\theta-\frac{1}{2} \right)\left(
  \left|2\chi^{n+1}-\chi^n \right|^2
  -\left|2\chi^n - \chi^{n-1}  \right|^2
\right).
\end{split}
\label{equ:relation_3}
\end{equation}
This property can be verified by elementary manipulations.
Note that 
the usual 2nd-order implicit approximation,
$
\chi^{n+\theta} = \theta\chi^{n+1} + (1-\theta)\chi^n,
$
cannot guarantee 
the energy stability 
with $\frac{1}{2}<\theta<1$.
The family of approximations given by
\eqref{equ:def_deriv_n_theta}--\eqref{equ:def_var_explicit_n_theta}
contains the often-used
2nd-order backward differentiation formula (or BDF2, corresponding
to $\theta=1$) and the Crank-Nicolson approximation (corresponding
to $\theta=1/2$).

Given $(\phi^n,r^n)$, we compute $(\phi^{n+1},r^{n+1})$
by enforcing the system consisting of \eqref{equ:r_equ}--\eqref{equ:wbc_trans_1}
and \eqref{equ:wbc_2}
at time step $(n+\theta)$ and using
the above approximations
\eqref{equ:def_deriv_n_theta}--\eqref{equ:def_var_explicit_n_theta},
as follows,
\begin{subequations}
  \begin{equation}
    \frac{
      \left(\theta + \frac{1}{2}\right)\phi^{n+1} - 2\theta\phi^n
      + \left(\theta-\frac{1}{2}\right)\phi^{n-1}
    }{\Delta t}
    = m \nabla^2 \mathscr{H}^{n+\theta} + g^{n+\theta},
    \label{equ:alg_disc_1}
  \end{equation}
  \begin{equation}
    \mathscr{H}^{n+\theta}=-\lambda\nabla^2\phi^{n+\theta}
    + S(\phi^{n+1}-2\phi^n + \phi^{n-1})
    + \frac{r^{n+\theta}}{\sqrt{E[\bar{\phi}^{n+\theta}]}}h(\bar{\phi}^{n+\theta}),
    \label{equ:alg_disc_2}
  \end{equation}
  \begin{equation}
    \frac{
      \left(\theta + \frac{1}{2}\right)r^{n+1} - 2\theta r^n
      + \left(\theta-\frac{1}{2}\right) r^{n-1}
    }{\Delta t}
    = \int_{\Omega} \frac{h(\bar{\phi}^{n+\theta})}{2\sqrt{E[\bar{\phi}^{n+\theta}]}}
    \frac{
      \left(\theta + \frac{1}{2}\right)\phi^{n+1} - 2\theta\phi^n
      + \left(\theta-\frac{1}{2}\right)\phi^{n-1}
    }{\Delta t},
    \label{equ:alg_disc_3}
  \end{equation}
  \begin{equation}
    \mathbf{n}\cdot\nabla\left[
      -\lambda\nabla^2\phi^{n+\theta}
      + S(\phi^{n+1}-2\phi^n + \phi^{n-1})
      + \frac{r^{n+\theta}}{\sqrt{E[\bar{\phi}^{n+\theta}]}}h(\bar{\phi}^{n+\theta})
      \right] = g_a^{n+\theta},
    \quad \text{on} \ \partial\Omega,
    \label{equ:alg_disc_4}
  \end{equation}
  \begin{equation}
    \mathbf{n}\cdot\nabla\phi^{n+\theta}
    =g_b^{n+\theta}, \quad \text{on} \ \partial\Omega.
    \label{equ:alg_disc_5}
  \end{equation}
\end{subequations}
%
In the above equations \eqref{equ:alg_disc_2} and \eqref{equ:alg_disc_4},
$S\geqslant 0$ is a chosen constant that satisfies a condition to be
specified later. Note that because $\phi^{n+1}-2\phi^n+\phi^{n-1}=\Order{(\Delta t)^2}$
the extra term $S(\phi^{n+1}-2\phi^n+\phi^{n-1})$ in \eqref{equ:alg_disc_2}
and \eqref{equ:alg_disc_4} does not spoil the second-order accuracy of
the overall scheme.
In the above equations the variables $\phi^{n+\theta}$,
$\bar{\phi}^{n+\theta}$, and $r^{n+\theta}$  are
given by the equations \eqref{equ:def_var_implicit_n_theta}
and \eqref{equ:def_var_explicit_n_theta}.
$g^{n+\theta}$, $g_a^{n+\theta}$ and $g_b^{n+\theta}$
are the prescribed functions $g(\mathbf{x},t)$,
$g_a(\mathbf{x},t)$ and $g_b(\mathbf{x},t)$
evaluated at time $t=(n+\theta)\Delta t$, respectively.

The above scheme has the following  property:
\begin{theorem}
  \label{thm:thm_1}
  
  The scheme consisting of equations
  \eqref{equ:alg_disc_1}--\eqref{equ:alg_disc_5}, in the absence of
  the source terms (i.e.~$g=0$, $g_a=0$ and $g_b=0$),
  satisfies the discrete energy balance equation
  for $\frac{1}{2}\leqslant \theta \leqslant\frac{3}{2}$
  and $S\geqslant 0$,
  \begin{equation}
    \begin{split}
    &
    W^{n+1} - W^n 
    + \theta\left(\theta-\frac{1}{2}\right)(3-2\theta)
    \left(|r^{n+1}-2r^n+r^{n-1}|^2  \right. \\
    & \qquad
    \left.
    + \int_{\Omega}\frac{\lambda}{2}\left|\nabla\phi^{n+1}-2\nabla\phi^n+\nabla\phi^{n-1} \right|^2
    \right) 
    + \theta S\int_{\Omega} \left|\phi^{n+1}-2\phi^n + \phi^{n-1}  \right|^2 \\
    &
    = -m\Delta t\int_{\Omega} \left|\nabla \mathscr{H}^{n+\theta}  \right|^2
    \end{split}
    \label{equ:discrete_energy_balance}
  \end{equation}
  where
  \begin{equation}
    \begin{split}
    W^{n} =& \left(\frac{3}{2}-\theta  \right)\left(
    |r^n|^2 +  \int_{\Omega} \frac{\lambda}{2}\left|\nabla\phi^{n} \right|^2
    \right) \\
    &+ \left(\theta-\frac{1}{2}\right)\left(
    \left|2r^n - r^{n-1}\right|^2 
    + \int_{\Omega} \frac{\lambda}{2}\left|2\nabla\phi^n - \nabla\phi^{n-1}  \right|^2
    \right) \\
    &+ \frac{S}{2}\int_{\Omega} \left|\phi^n - \phi^{n-1} \right|^2.
    \end{split}
    \label{equ:discrete_energy}
  \end{equation}
  
\end{theorem}
\noindent A proof of this theorem is provided in the Appendix A.

%

\begin{remark}

  Based on this theorem the scheme given by equations
  \eqref{equ:alg_disc_1}--\eqref{equ:alg_disc_5} constitutes a family
  of unconditionally stable algorithms for
  $\frac{1}{2}\leqslant \theta\leqslant \frac{3}{2}$.
  Note that this energy stability property holds regardless of the
  specific form of the potential free energy density  $F(\phi)$,
  as long as it is such that an appropriate constant
  $C_0$ in \eqref{equ:def_E1_E2} can be chosen to ensure
  $E(t)>0$ for $0\leqslant t\leqslant T$.
  
\end{remark}
\begin{remark}

  The first term on the right hand side of \eqref{equ:relation_3}
  determines the dissipativeness of the
  approximations \eqref{equ:def_deriv_n_theta}--\eqref{equ:def_var_implicit_n_theta}.
  The algorithm with $\theta = \frac{2}{3}+\frac{\sqrt{7}}{6}\approx 1.11$
  is the most dissipative among this family of approximations,
  while both $\theta=\frac{1}{2}$ (Crank-Nicolson) and $\theta=\frac{3}{2}$
  are non-dissipative.
  In terms of numerical dissipation,
  BDF2 ($\theta=1$) is close to, but not as dissipative as,
  the scheme with $\theta = \frac{2}{3}+\frac{\sqrt{7}}{6}$.

\end{remark}
\begin{remark}

  While we consider only the Cahn-Hilliard type equations in this work,
  the application of the family of 2nd-order approximations
  \eqref{equ:def_deriv_n_theta}--\eqref{equ:def_var_explicit_n_theta}
  is not limited to this class of equations.
  They can be readily applied to other types of equations describing
  gradient flows. For example, by combining these approximations
  and the auxiliary variable approaches of~\cite{Yang2016,ShenXY2018},
  one can derive a family of energy-stable schemes for a large class
  of gradient flows.
  
\end{remark}


\subsection{Efficient Solution Algorithm}

The scheme represented by equations \eqref{equ:alg_disc_1}--\eqref{equ:alg_disc_5}
involves integrals of the unknown field variable $\phi^{n+1}$
over the domain $\Omega$.
Despite this apparent complication, the formulation allows for a simple and
very efficient solution algorithm. We present such an
algorithm below.

To facilitate subsequent discussions and make the
representation more compact, we first introduce several notations
($\chi$ denoting a generic variable):
\begin{subequations}
  \begin{equation}
    \gamma_0 = \gamma_0(\theta) = \theta + \frac{1}{2}, \qquad
    \omega_0 = \omega_0(\theta) = \theta\left(\frac{5}{2}-\theta \right) - \frac{1}{2},
    \label{equ:notation_1}
  \end{equation}
  \begin{equation}
    \hat{\chi} = 2\theta\chi^n - \left(\theta - \frac{1}{2} \right)\chi^{n-1},
    \label{equ:notation_2}
  \end{equation}
  \begin{equation}
    \tilde{\chi} = 2(1-\theta)^2\chi^n
    + \left(\theta-\frac{1}{2} \right)(1-\theta)\chi^{n-1}.
    \label{equ:notation_3}
  \end{equation}
\end{subequations}
Then the approximations in \eqref{equ:def_deriv_n_theta}
and \eqref{equ:def_var_implicit_n_theta} can be
written as
\begin{subequations}
\begin{equation}
  \left.\frac{\partial\chi}{\partial t} \right|^{n+\theta}
  = \frac{\gamma_0\chi^{n+1}-\hat{\chi}}{\Delta t},
  \label{equ:notation_4}
\end{equation}
\begin{equation}
  \chi^{n+\theta} = \omega_0\chi^{n+1} + \tilde{\chi}.
  \label{equ:notation_5}
\end{equation}
\end{subequations}

Combining equations \eqref{equ:alg_disc_1} and \eqref{equ:alg_disc_2}
leads to
\begin{equation}
  \frac{\gamma_0\phi^{n+1}-\hat{\phi}}{\Delta t}
  = m\nabla^2\left[
    -\lambda\nabla^2\left(\omega_0\phi^{n+1}+\tilde{\phi} \right)
    + S(\phi^{n+1} - \bar{\phi}^{n+1})
    + \left(\omega_0r^{n+1}+\tilde{r} \right)
    \frac{h(\bar{\phi}^{n+\theta})}{\sqrt{E[\bar{\phi}^{n+\theta}]}}
    \right]
  + g^{n+\theta}
  \label{equ:CH_disc_1}
\end{equation}
where $\gamma_0$ and $\omega_0$ are given by
\eqref{equ:notation_1}, $\hat{\phi}$ is defined by \eqref{equ:notation_2},
$\tilde{\phi}$ and $\tilde{r}$ are defined by \eqref{equ:notation_3},
$\bar{\phi}^{n+1}=2\phi^n - \phi^{n-1}$ based on
equation \eqref{equ:def_var_explicit_n_theta},
and we have used equations \eqref{equ:notation_4}
and \eqref{equ:notation_5}.
It is important to note that $\phi^{n+1}$, $\hat{\phi}$, $\tilde{\phi}$,
$\bar{\phi}^{n+\theta}$, $\bar{\phi}^{n+1}$ are field functions,
while $r^{n+1}$, $\tilde{r}$ and $E[\bar{\phi}^{n+\theta}]$
are scalar variables.
Define
\begin{equation}
  \begin{split}
    &
    b^n = \frac{h(\bar{\phi}^{n+\theta})}{\sqrt{E[\bar{\phi}^{n+\theta}]}}, \qquad
    z^{n+1} = \int_{\Omega} b^n\phi^{n+1}, 
  \end{split}
  \label{equ:def_bn_dn}
\end{equation}
Note that $b^n$ is a field function, and
$z^{n+1}$ is a scalar variable.
Equation \eqref{equ:CH_disc_1} is then transformed into
\begin{multline}
  \nabla^2(\nabla^2\phi^{n+1})
  - \frac{S}{\lambda\omega_0}\nabla^2\phi^{n+1}
  + \frac{\gamma_0}{\lambda\omega_0 m\Delta t}\phi^{n+1} \\
  = \frac{1}{\lambda\omega_0 m}\left[
    g^{n+\theta} + \frac{\hat{\phi}}{\Delta t}
    \right]
  - \frac{S}{\lambda\omega_0}\nabla^2\bar{\phi}^{n+1}
  -\frac{1}{\omega_0}\nabla^2(\nabla^2\tilde{\phi})
  + \frac{\omega_0r^{n+1}+\tilde{r}}{\lambda\omega_0}\nabla^2 b^n.
  \label{equ:CH_disc_2}
\end{multline}

Equation \eqref{equ:alg_disc_3} can be written as
$ 
  \gamma_0r^{n+1}-\hat{r}
  =\frac{1}{2}\int_{\Omega}b^n( \gamma_0\phi^{n+1}-\hat{\phi}),
$ 
from which we get
\begin{equation}
  r^{n+1} = \frac{1}{\gamma_0}\left(\hat{r} - \frac{1}{2}\int_{\Omega}b^n\hat{\phi}  \right) + \frac{z^{n+1}}{2}
  \label{equ:r_expr}
\end{equation}
where $z^{n+1}$ is defined in \eqref{equ:def_bn_dn}
and involves the unknown field function $\phi^{n+1}$.
In light of this expression, equation \eqref{equ:CH_disc_2}
is transformed into
\begin{equation}
  \nabla^2(\nabla^2\phi^{n+1})
  - \frac{S}{\lambda\omega_0}\nabla^2\phi^{n+1}
  + \frac{\gamma_0}{\lambda\omega_0 m\Delta t}\phi^{n+1}
  = Q^n + \frac{z^{n+1}}{2\lambda}\nabla^2 b^n,
  \label{equ:CH_disc_3}
\end{equation}
where
\begin{equation}
  Q^n = \frac{1}{\lambda\omega_0 m}\left[
    g^{n+\theta} + \frac{\hat{\phi}}{\Delta t}
    \right]
  - \frac{S}{\lambda\omega_0}\nabla^2\bar{\phi}^{n+1}
  -\frac{1}{\omega_0}\nabla^2(\nabla^2\tilde{\phi})
  + \left[
    \frac{1}{\lambda\gamma_0}\left(
    \hat{r} - \frac{1}{2}\int_{\Omega} b^n\hat{\phi}
    \right)
    + \frac{\tilde{r}}{\lambda\omega_0}
    \right]\nabla^2b^n.
  \label{equ:Q_expr}
\end{equation}

Barring the unknown scalar variable $z^{n+1}$,
equation \eqref{equ:CH_disc_3} is a fourth-order
equation about $\phi^{n+1}$. The left-hand-side (LHS)
of this equation can be reformulated into two de-coupled
Helmholtz type equations (see e.g.~\cite{YueFLS2004,DongS2012,Dong2012}).
By adding/subtracting a term $\alpha\nabla^2\phi^{n+1}$ ($\alpha$
denoting a constant to be determined) on the LHS
of \eqref{equ:CH_disc_3}, we get
\begin{equation}
  \nabla^2\left[\nabla^2\phi^{n+1}+\alpha\phi^{n+1}  \right]
  - \left(\alpha + \frac{S}{\lambda\omega_0}  \right)\left[
    \nabla^2\phi^{n+1} - \frac{\gamma_0}{\lambda\omega_0m\Delta t\left(\alpha+\frac{S}{\lambda\omega_0}  \right)}\phi^{n+1}
    \right]
  = Q^n + \frac{z^{n+1}}{2\lambda}\nabla^2b^n.
  \label{equ:CH_disc_4}
\end{equation}
By requiring
$
\alpha = - \frac{\gamma_0}{\lambda\omega_0m\Delta t\left(\alpha+\frac{S}{\lambda\omega_0}  \right)},
$
we can determine the constant $\alpha$,
\begin{equation}
  \alpha = -\frac{S}{2\lambda\omega_0}\left[
    1 - \sqrt{1 - \frac{4\gamma_0}{\lambda\omega_0m\Delta t}\left(
    \frac{\lambda\omega_0}{S}
    \right)^2}
    \right]
  \label{equ:alpha_expr}
\end{equation}
with the requirement
\begin{equation}
  S \geqslant \lambda\omega_0\sqrt{\frac{4\gamma_0}{\lambda\omega_0m\Delta t}}
  = \sqrt{\frac{4\gamma_0\lambda\omega_0}{m\Delta t}}.
  \label{equ:S_condition}
\end{equation}
This is the condition the chosen constant $S$ must satisfy.

Therefore, equation \eqref{equ:CH_disc_4} can be transformed into
the following equivalent form
\begin{subequations}
  \begin{equation}
    \nabla^2\psi^{n+1} - \left(\alpha+\frac{S}{\lambda\omega_0} \right)\psi^{n+1}
    = Q^n + \frac{z^{n+1}}{2\lambda}\nabla^2b^n,
    \label{equ:CH_psi}
  \end{equation}
  \begin{equation}
    \nabla^2\phi^{n+1} + \alpha\phi^{n+1} = \psi^{n+1},
    \label{equ:CH_phi}
  \end{equation}
\end{subequations}
where $\psi^{n+1}$ is an auxiliary field variable defined by
\eqref{equ:CH_phi}.
Note that $\alpha<0$ and $\alpha+\frac{S}{\lambda\omega_0}>0$
under the condition \eqref{equ:S_condition} for $S$.
It can also be noted that, if $z^{n+1}$ is known,
then the two equations \eqref{equ:CH_psi} and \eqref{equ:CH_phi}
are not coupled. One can first solve \eqref{equ:CH_psi}
for $\psi^{n+1}$, and then solve \eqref{equ:CH_phi}
for $\phi^{n+1}$.
The unknown variable $z^{n+1}$, which depends on
$\phi^{n+1}$, causes a complication in the solution of this system.

Let us now turn the attention to the boundary conditions.
Equation \eqref{equ:alg_disc_5} can be written as
\begin{equation}
  \mathbf{n}\cdot\nabla\phi^{n+1}
  = \frac{1}{\omega_0}\left(
  g_b^{n+\theta}-
  \mathbf{n}\cdot\nabla\tilde{\phi}
  \right)
  \quad \text{on} \ \partial\Omega.
  \label{equ:phi_bc}
\end{equation}
Equation \eqref{equ:alg_disc_4} can be re-written as
\begin{equation}
  -\lambda\mathbf{n}\cdot\nabla
    \left(\omega_0\nabla^2\phi^{n+1}+\nabla^2\tilde{\phi}  \right)
  + S\mathbf{n}\cdot\nabla\left(\phi^{n+1}-\bar{\phi}^{n+1}  \right)
  + \left(\omega_0r^{n+1}+\tilde{r} \right)\mathbf{n}\cdot\nabla b^n
  = g_a^{n+\theta},
  \quad \text{on} \ \partial\Omega.
  \label{equ:psi_bc_trans_1}
\end{equation}
In light of the equations \eqref{equ:CH_phi} and \eqref{equ:r_expr},
we can transform this equation into
\begin{multline}
  \mathbf{n}\cdot\nabla\psi^{n+1}
  = \left(\alpha+\frac{S}{\lambda\omega_0} \right)\mathbf{n}\cdot\nabla\phi^{n+1}
  - \frac{1}{\lambda\omega_0}\left(
   g_a^{n+\theta} + S\mathbf{n}\cdot\nabla\bar{\phi}^{n+1}
  \right)
  - \frac{1}{\omega_0}\mathbf{n}\cdot\nabla(\nabla^2\tilde{\phi}) \\
  + \left[
  \frac{\tilde{r}}{\lambda\omega_0} + \frac{1}{\lambda\gamma_0}\left(
  \hat{r} - \frac{1}{2}\int_{\Omega} b^n\hat{\phi}
  \right)
  +\frac{z^{n+1}}{2\lambda}
  \right]\mathbf{n}\cdot\nabla b^n,
  \quad \text{on} \ \partial\Omega.
\end{multline}
Substitute equation \eqref{equ:phi_bc} into
the above equation, and we have the boundary
condition about $\psi^{n+1}$:
\begin{equation}
  \mathbf{n}\cdot\nabla\psi^{n+1} =
  T^n + \frac{z^{n+1}}{2\lambda}\mathbf{n}\cdot\nabla b^n,
  \quad \text{on} \ \partial\Omega,
  \label{equ:psi_bc}
\end{equation}
where
\begin{multline}
  T^n = \left(\alpha+\frac{S}{\lambda\omega_0} \right)
  \frac{1}{\omega_0}
  \left(
  g_b^{n+\theta}
  -\mathbf{n}\cdot\nabla\tilde{\phi}
  \right)
  - \frac{1}{\lambda\omega_0}\left(g_a^{n+\theta}
  +S\mathbf{n}\cdot\nabla\bar{\phi}^{n+1}\right) \\
  - \frac{1}{\omega_0}\mathbf{n}\cdot\nabla(\nabla^2\tilde{\phi}) 
  + \left[
  \frac{\tilde{r}}{\lambda\omega_0} + \frac{1}{\lambda\gamma_0}\left(
  \hat{r} - \frac{1}{2}\int_{\Omega} b^n\hat{\phi}
  \right)
  \right]\mathbf{n}\cdot\nabla b^n.
  \label{equ:Tn_expr}
\end{multline}

Therefore, the original system of
equations \eqref{equ:alg_disc_1}--\eqref{equ:alg_disc_5}
has been reduced to the system
consisting of equations \eqref{equ:CH_psi}--\eqref{equ:CH_phi},
\eqref{equ:psi_bc} and \eqref{equ:phi_bc}. 
After $\phi^{n+1}$ is solved from this system,
$r^{n+1}$ can be computed
based on equation \eqref{equ:r_expr}.

To solve the system consisting of
equations \eqref{equ:CH_psi}--\eqref{equ:CH_phi},
\eqref{equ:psi_bc} and \eqref{equ:phi_bc}, 
it is critical to note that the unknown variable
$z^{n+1}$  involved therein is a scalar, not a field function,
and that the equations are linear with respect
to $\phi^{n+1}$, $\psi^{n+1}$,
and  $z^{n+1}$.
In what follows we present an efficient algorithm for
solving this system.
We define two sets of field variables,
$(\psi_i^{n+1},\phi_i^{n+1})$ ($i=1,2$),
as follows: \\
\noindent\underline{For $\psi_1^{n+1}$:}
\begin{subequations}
  \begin{equation}
    \nabla^2\psi_1^{n+1} - \left(\alpha+\frac{S}{\lambda\omega_0} \right)\psi_1^{n+1} = Q^n,
    \label{equ:psi_1_equ}
  \end{equation}
  \begin{equation}
    \mathbf{n}\cdot\nabla\psi_1^{n+1} = T^n, \quad \text{on} \ \partial\Omega.
    \label{equ:psi_1_bc}
  \end{equation}
\end{subequations}
\noindent\underline{For $\phi_1^{n+1}$:}
\begin{subequations}
  \begin{equation}
    \nabla^2\phi_1^{n+1} + \alpha\phi_1^{n+1} = \psi_1^{n+1}, \label{equ:phi_1_equ}
  \end{equation}
  \begin{equation}
    \mathbf{n}\cdot\nabla\phi_1^{n+1} =
    \frac{1}{\omega_0}\left(g_b^{n+\theta} - \mathbf{n}\cdot\nabla\tilde{\phi}  \right),
    \quad \text{on} \ \partial\Omega.
    \label{equ:phi_1_bc}
  \end{equation}
\end{subequations}
\noindent\underline{For $\psi_2^{n+1}$:}
\begin{subequations}
  \begin{equation}
    \nabla^2\psi_2^{n+1} - \left(\alpha+\frac{S}{\lambda\omega_0} \right)\psi_2^{n+1}
    = \frac{1}{2\lambda}\nabla^2 b^n,
    \label{equ:psi_2_equ}
  \end{equation}
  \begin{equation}
    \mathbf{n}\cdot\nabla\psi_2^{n+1} =
    \frac{1}{2\lambda}\mathbf{n}\cdot\nabla b^n, \quad \text{on} \ \partial\Omega.
    \label{equ:psi_2_bc}
  \end{equation}
\end{subequations}
\noindent\underline{For $\phi_2^{n+1}$:}
\begin{subequations}
  \begin{equation}
    \nabla^2\phi_2^{n+1} + \alpha\phi_2^{n+1} = \psi_2^{n+1}, \label{equ:phi_2_equ}
  \end{equation}
  \begin{equation}
    \mathbf{n}\cdot\nabla\phi_2^{n+1} = 0, \quad \text{on} \ \partial\Omega.
    \label{equ:phi_2_bc}
  \end{equation}
\end{subequations}
Then we have the following result.
\begin{theorem}
\label{thm:thm_2}

For given scalar value $z^{n+1}$,
the following field functions solve the system consisting of equations
\eqref{equ:CH_psi}--\eqref{equ:CH_phi},
\eqref{equ:psi_bc} and \eqref{equ:phi_bc}:
\begin{subequations}
  \begin{equation}
    \psi^{n+1} = \psi_1^{n+1} + z^{n+1}\psi_2^{n+1}, 
    \label{equ:psi_expr}
  \end{equation}
  \begin{equation}
    \phi^{n+1} = \phi_1^{n+1} + z^{n+1}\phi_2^{n+1}, 
    \label{equ:phi_expr}
  \end{equation}
\end{subequations}
where $(\psi_i^{n+1},\phi_i^{n+1})$ ($i=1,2$) are given by
equations \eqref{equ:psi_1_equ}--\eqref{equ:phi_2_bc}.

\end{theorem}
\noindent This theorem can be proved by straightforward substitutions
and verifications.

We still need to determine the value for $z^{n+1}$.
Substituting the expression $\phi^{n+1}$ in \eqref{equ:phi_expr}
into the definition for $z^{n+1}$  in \eqref{equ:def_bn_dn}
results in
\begin{equation}
    z^{n+1} =
    \frac{\int_{\Omega}b^n\phi_1^{n+1} }{1-\int_{\Omega}b^n\phi_2^{n+1}}.
  \label{equ:z_xi_1}
\end{equation}
%
%
We have the following result:
\begin{theorem}
\label{thm:thm-3}
  The $\phi_2^{n+1}$ defined by \eqref{equ:psi_2_equ}--\eqref{equ:phi_2_bc}
  satisfies the property,
  \begin{equation}
    \int_{\Omega} b^n\phi_2^{n+1} \leqslant 0.
  \end{equation}  
  
\end{theorem}
\noindent A proof of this theorem is provided in Appendix B.
Based on this theorem, $z^{n+1}$ given by \eqref{equ:z_xi_1}
is well defined for any $\Delta t$.

Combining the above discussions, we arrive at the solution
algorithm for solving the system consisting of equations
\eqref{equ:alg_disc_1}--\eqref{equ:alg_disc_5}.
Given $(\phi^{n},r^n)$, we compute
$(\phi^{n+1}, \nabla^2\phi^{n+1},r^{n+1})$ through the following steps:
\begin{enumerate}[(S-1):]

\item
  Solve equations \eqref{equ:psi_1_equ}--\eqref{equ:psi_1_bc} for $\psi_1^{n+1}$; \\
  Solve equations \eqref{equ:phi_1_equ}--\eqref{equ:phi_1_bc} for $\phi_1^{n+1}$.

\item
  Solve equations \eqref{equ:psi_2_equ}--\eqref{equ:psi_2_bc} for $\psi_2^{n+1}$; \\
  Solve equations \eqref{equ:phi_2_equ}--\eqref{equ:phi_2_bc} for $\phi_2^{n+1}$.

\item
  Solve equations \eqref{equ:z_xi_1} for $z^{n+1}$.

\item
  Compute $\psi^{n+1}$ and $\phi^{n+1}$ based on equations
  \eqref{equ:psi_expr} and \eqref{equ:phi_expr}, respectively. \\
  Compute $\nabla^2\phi^{n+1}$
  as follows in light of equation \eqref{equ:CH_phi}
  \begin{equation}
    \nabla^2\phi^{n+1} = \psi^{n+1} - \alpha\phi^{n+1}.
    \label{equ:laplace_phi}
  \end{equation}

\item
  Compute $r^{n+1}$ based on equation \eqref{equ:r_expr}.
  
\end{enumerate}


\begin{remark}

  This algorithm involves only the solution of four de-coupled Helmholtz
  type equations within a time step. These Helmholtz equations involve
  only two distinct coefficient matrices after spatial discretization,
  and they are both constant coefficient matrices and can be pre-computed.
  No nonlinear algebraic solver is involved in this algorithm.
  Thanks to these characteristics, the family of second-order energy-stable
  schemes represented by \eqref{equ:alg_disc_1}--\eqref{equ:alg_disc_5}
  can be implemented in a very efficient fashion.
  
\end{remark}


\subsection{Implementation with $C^0$ Spectral Elements}

The solution algorithm presented in the previous subsection
can be implemented using any commonly-used method for the spatial
discretization. In the current work we will employ $C^0$
continuous high-order
spectral elements~\cite{KarniadakisS2005,ZhengD2011}
for spatial discretizations.
We next consider the implementation of the algorithm
using $C^0$ spectral elements. The discussions in this subsection
without change also apply to implementations using
low-order finite elements.

We will first derive the weak forms about $(\psi_i^{n+1},\phi_i^{n+1})$
($i=1,2$) in the continuous space, and then specify the discrete
function space for the spectral element approximations.
In the process, certain terms involving derivatives of order two
or higher will be dealt with appropriately so that all quantities
involved in the weak formulation can be computed directly
using $C^0$ elements in the discrete function space.

Let $\varphi(\mathbf{x})$ denote a test function. Taking the $L^2$
inner product between $\varphi$ and the equation \eqref{equ:psi_1_equ}
leads to
\begin{equation}
  \int_{\Omega}\nabla\psi_1^{n+1}\cdot\nabla\varphi
  + \left(\alpha + \frac{S}{\lambda\omega_0} \right)\int_{\Omega}\psi_1^{n+1}\varphi
  = -\int_{\Omega}Q^n\varphi
  + \int_{\partial\Omega} T^n\varphi,
  \quad \forall \varphi,
  \label{equ:psi_1_weak_1}
\end{equation}
where we have used integration by part and
the equation \eqref{equ:psi_1_bc}.
By substituting the $Q^n$ expression from \eqref{equ:Q_expr}
and $T^n$ expression from \eqref{equ:Tn_expr}
into the RHS of the above equation and
integration by part, we obtain the weak form
about $\psi_1^{n+1}$,
\begin{equation}
  \begin{split}
    &
    \int_{\Omega}\nabla\psi_1^{n+1}\cdot\nabla\varphi
    + \left(\alpha + \frac{S}{\lambda\omega_0} \right)\int_{\Omega}\psi_1^{n+1}\varphi \\
    &= -\frac{1}{\lambda\omega_0m}\int_{\Omega}\left(
    g^{n+\theta} +
    \frac{\hat{\phi}}{\Delta t}
    \right)\varphi
    -\frac{S}{\lambda\omega_0}\int_{\Omega}\nabla\bar{\phi}^{n+1}\cdot\nabla\varphi
    -\frac{1}{\omega_0}\int_{\Omega}\nabla\left(\nabla^2\tilde{\phi} \right)\cdot\nabla\varphi
    -\frac{1}{\lambda\omega_0}\int_{\partial\Omega}g_a^{n+\theta}\varphi \\
    & \quad
    +\left[\frac{1}{\lambda\gamma_0}\left(\hat{r}-\frac{1}{2}\int_{\Omega}b^n\hat{\phi} \right)
      + \frac{\tilde{r}}{\lambda\omega_0}  \right]
    \int_{\Omega}\nabla b^n\cdot\nabla\varphi
    +
    \left(\alpha+\frac{S}{\lambda\omega_0} \right)
    \frac{1}{\omega_0}\int_{\partial\Omega}\left(
    g_b^{n+\theta}
    -\mathbf{n}\cdot\nabla\tilde{\phi}
    \right)\varphi.
  \end{split}
  \label{equ:psi_1_weakform}
\end{equation}
In the above equation note that the term $\nabla^2\tilde{\phi}$ is to be
computed by, in light of equation \eqref{equ:notation_3},
\begin{equation}
  \begin{split}
  \nabla^2\tilde{\phi} &=
  2(1-\theta)^2(\nabla^2\phi^n)
  + \left(\theta-\frac{1}{2} \right)(1-\theta)(\nabla^2\phi^{n-1}) \\
  &= 2(1-\theta)^2(\psi^n-\alpha\phi^n)
  + \left(\theta-\frac{1}{2} \right)(1-\theta)(\psi^{n-1}-\alpha\phi^{n-1})
  \end{split}
  \label{equ:laplace_phi_tilde}
\end{equation}
where $\nabla^2\phi^n$ has been computed based on
equation \eqref{equ:laplace_phi}.

Taking the $L^2$ inner product between $\varphi$
and the equation \eqref{equ:phi_1_equ} and integration
by part results in the weak form about $\phi_1^{n+1}$:
\begin{equation}
  \begin{split}
  &
  \int_{\Omega}\nabla\phi_1^{n+1}\cdot\nabla\varphi
  -\alpha\int_{\Omega}\phi_1^{n+1}\varphi 
  = -\int_{\Omega}\psi_1^{n+1}\varphi
  + \frac{1}{\omega_0}\int_{\partial\Omega}\left(
  g_b^{n+\theta}
  -\mathbf{n}\cdot\nabla\tilde{\phi}
  \right)\varphi,
  \quad \forall\varphi,
  \end{split}
  \label{equ:phi_1_weakform}
\end{equation}
where we have use equations \eqref{equ:phi_1_bc}.

By taking the $L^2$ inner product between $\varphi$
and the equation \eqref{equ:psi_2_equ} and integration
by part, we arrive at the weak form about $\psi_2^{n+1}$,
\begin{equation}
\int_{\Omega}\nabla\psi_2^{n+1}\cdot\nabla\varphi
+\left(\alpha+\frac{S}{\lambda\omega_0} \right)\int_{\Omega}\psi_2^{n+1}\varphi
=\frac{1}{2\lambda}\int_{\Omega}\nabla b^n\cdot\nabla\varphi,
\quad \forall \varphi,
\label{equ:psi_2_weakform}
\end{equation}
where we have used equation \eqref{equ:psi_2_bc}.
Taking the $L^2$ inner product between $\varphi$ and
equation \eqref{equ:phi_2_equ} and integration by part
leads to the weak form about $\phi_2^{n+1}$,
\begin{equation}
\int_{\Omega}\nabla\phi_2^{n+1}\cdot\nabla\varphi
-\alpha\int_{\Omega}\phi_2^{n+1}\varphi
= -\int_{\Omega}\psi_2^{n+1}\varphi,
\quad \forall\varphi,
\label{equ:phi_2_weakform}
\end{equation}
where the equation \eqref{equ:phi_2_bc} has been used.


We discretize the domain $\Omega$ using a mesh of $N_{el}$ spectral
elements. Let $K$ (positive integer) denote
the element order, which is a measure of the highest
polynomial degree in field expansions within an element.
Let $\Omega_h$ denote the discretized domain,
and $\Omega_h^e$ ($1\leqslant e\leqslant N_{el}$)
denote the element $e$, $\Omega_h=\cup_{e=1}^{N_{el}}\Omega_h^e$.
Define function space
\begin{equation*}
  H_{\phi}=\left\{\
  v\in H^1(\Omega_h) \ :\
  v\ \text{is a polynomial
    of degree characterized by}\ K\ \text{on}\ \Omega_h^e,
  \ \text{for}\ 1\leqslant e\leqslant N_{el}
  \ \right\}.
\end{equation*}
Let $(\cdot)_h$ denote the discretized version of
the variable $(\cdot)$ below.
The fully discretized equations for $(\psi_i^{n+1},\phi_i^{n+1})$
($i=1,2$) are the following. \\
\noindent\underline{For $\psi_{1h}^{n+1}$:} \ \
find $\psi_{1h}^{n+1}\in H_{\phi}$ such that
\begin{equation}
  \begin{split}
    &
    \int_{\Omega_h}\nabla\psi_{1h}^{n+1}\cdot\nabla\varphi_h
    + \left(\alpha + \frac{S}{\lambda\omega_0} \right)\int_{\Omega_h}\psi_{1h}^{n+1}\varphi_h \\
    &= -\frac{1}{\lambda\omega_0m}\int_{\Omega_h}\left(
    g_h^{n+\theta} + \frac{\hat{\phi}_h}{\Delta t}
    \right)\varphi_h
    -\frac{S}{\lambda\omega_0}\int_{\Omega_h}\nabla\bar{\phi}_h^{n+1}\cdot\nabla\varphi_h
    -\frac{1}{\omega_0}\int_{\Omega_h}\nabla\left(\nabla^2\tilde{\phi}_h \right)\cdot\nabla\varphi_h \\
    & \quad
    +\left[\frac{1}{\lambda\gamma_0}\left(\hat{r}-\frac{1}{2}\int_{\Omega_h}b_h^n\hat{\phi_h} \right)
      + \frac{\tilde{r}}{\lambda\omega_0}  \right]
    \int_{\Omega_h}\nabla b_h^n\cdot\nabla\varphi_h \\
    &\quad
    +\left(\alpha+\frac{S}{\lambda\omega_0} \right)
    \frac{1}{\omega_0}\int_{\partial\Omega_h}\left(
    g_{bh}^{n+\theta}
    -\mathbf{n}_h\cdot\nabla\tilde{\phi}_h
    \right)\varphi_h
    -\frac{1}{\lambda\omega_0}\int_{\partial\Omega_h}g_{ah}^{n+\theta}\varphi_h,
    \qquad \forall\varphi_h\in H_{\phi},
  \end{split}
  \label{equ:psi_1_weakform_disc}
\end{equation}
where note that $\nabla^2\tilde{\phi}_h$ is to be computed according to
equation \eqref{equ:laplace_phi_tilde}. \\
\noindent\underline{For $\phi_{1h}^{n+1}$:} \ \
find $\phi_{1h}^{n+1}\in H_{\phi}$ such that
\begin{multline}
  \int_{\Omega_h}\nabla\phi_{1h}^{n+1}\cdot\nabla\varphi_h
  -\alpha\int_{\Omega_h}\phi_{1h}^{n+1}\varphi_h 
  = -\int_{\Omega_h}\psi_{1h}^{n+1}\varphi_h \\
  +
  \frac{1}{\omega_0}\int_{\partial\Omega_h}\left(
  g_{bh}^{n+\theta}
  -\mathbf{n}_h\cdot\nabla\tilde{\phi}_h
  \right)\varphi_h, 
  \quad \forall\varphi_h\in H_{\phi},
  \label{equ:phi_1_weakform_disc}
\end{multline}
\noindent\underline{For $\psi_{2h}^{n+1}$:} \ \
find $\psi_{2h}^{n+1}\in H_{\phi}$ such that
\begin{equation}
\int_{\Omega_h}\nabla\psi_{2h}^{n+1}\cdot\nabla\varphi_h
+\left(\alpha+\frac{S}{\lambda\omega_0} \right)\int_{\Omega_h}\psi_{2h}^{n+1}\varphi_h
=\frac{1}{2\lambda}\int_{\Omega_h}\nabla b_h^n\cdot\nabla\varphi_h,
\qquad \forall \varphi_h\in H_{\phi}.
\label{equ:psi_2_weakform_disc}
\end{equation}
\noindent\underline{For $\phi_{2h}^{n+1}$:} \ \
find $\phi_{2h}^{n+1}\in H_{\phi}$ such that
\begin{equation}
\int_{\Omega_h}\nabla\phi_{2h}^{n+1}\cdot\nabla\varphi_h
-\alpha\int_{\Omega_h}\phi_{2h}^{n+1}\varphi_h
= -\int_{\Omega_h}\psi_{2h}^{n+1}\varphi_h,
\qquad \forall\varphi_h\in H_{\phi}.
\label{equ:phi_2_weakform_disc}
\end{equation}

The final discretized algorithm consists of the following within
each time step:
(i) Solve equations \eqref{equ:psi_1_weakform_disc}--\eqref{equ:phi_2_weakform_disc}
for $(\psi_i^{n+1},\phi_i^{n+1})$ ($i=1,2$), respectively.
(ii) Compute $z^{n+1}$ from equations
\eqref{equ:z_xi_1}.
(iii) Compute $\psi^{n+1}$, $\phi^{n+1}$, and $r^{n+1}$
based on equations \eqref{equ:psi_expr},
\eqref{equ:phi_expr}, and \eqref{equ:r_expr}, respectively.



\section{Representative Numerical Examples}
\label{sec:tests}

In this section we present several example problems in two dimensions
to test the performance of the family of energy-stable schemes
developed in the previous section.
In the numerical simulations we have non-dimensionalized
the physical
variables, the governing equations, and the boundary conditions.
As detailed in previous works (see e.g.~\cite{Dong2014}), 
the non-dimensional form of the governing equations and boundary conditions 
will remain the same, if the physical variables
are normalized consistently. Let $L$ denote a length scale,
$U_0$ a velocity scale, $\varrho_0$ a density scale, and
$d$ ($d=2$ or $3$) the spatial dimension.
The normalization constants for consistent non-dimensionalization of
different physical variables involved in the current work are listed
in Table~\ref{tab:normalization}.

\begin{table}
  \centering
  \begin{tabular}{ll | ll}
    \hline
    variable & normalization constant & variable & normalization constant \\
    $\mathbf{x}$, $\eta$ & $L$ & $t$, $\Delta t$ & $L/U_0$ \\
    $\phi$, $\phi_1$, $\phi_2$, $\phi_{in}$, $\theta$, $\gamma_0$, $\omega_0$, $z^{n+1}$ & $1$ & $\lambda$ & $\varrho_0U_0^2L^2$ \\
    $m$ & $L/(\varrho_0U_0)$ & $h(\phi)$, $F(\phi)$, $\mathscr{H}$, $S$ & $\varrho_0U_0^2$ \\
    $g(\mathbf{x})$ & $U_0/L$ & $g_a(\mathbf{x})$ & $\varrho_0U_0/L$ \\
    $g_b(\mathbf{x})$ & $1/L$ & $E[\phi]$, $C_0$ & $\varrho_0U_0^2L^d$ \\
    $r(t)$ & $\sqrt{\varrho_0U_0^2L^d}$ & $\alpha$, $\psi_1$, $\psi_2$, $\psi$ & $1/L^2$ \\
    \hline
  \end{tabular}
  \caption{Normalization constants for consistent non-dimensionalization
  of physical variables.}
  \label{tab:normalization}
\end{table}

\subsection{Convergence Rates}

We first use a manufactured analytic solution to the Cahn-Hilliard
equation to numerically demonstrate the rates of convergence
in space and time of the algorithm presented in Section \ref{sec:method}.

We consider the domain $0\leqslant x\leqslant 2$ and
$-1\leqslant y\leqslant 1$, and the following solution
to the Cahn-Hilliard equation \eqref{equ:CH} on this domain,
\begin{equation}
  \phi = \cos(\pi x)\cos(\pi y)\sin(t).
  \label{equ:anal_soln}
\end{equation}
The source term $g(\mathbf{x},t)$ in \eqref{equ:CH}
is chosen such that the analytic expression of \eqref{equ:anal_soln}
satisfies \eqref{equ:CH}.
The conditions \eqref{equ:wbc_1} and \eqref{equ:wbc_2}
are imposed on the domain boundary, and
the source terms $g_a(\mathbf{x},t)$ and $g_b(\mathbf{x},t)$
are chosen such that the analytic expression of
\eqref{equ:anal_soln} satisfies
\eqref{equ:wbc_1} and \eqref{equ:wbc_2} on the domain
boundary.

\begin{figure}
  \centerline{
    \includegraphics[width=3in]{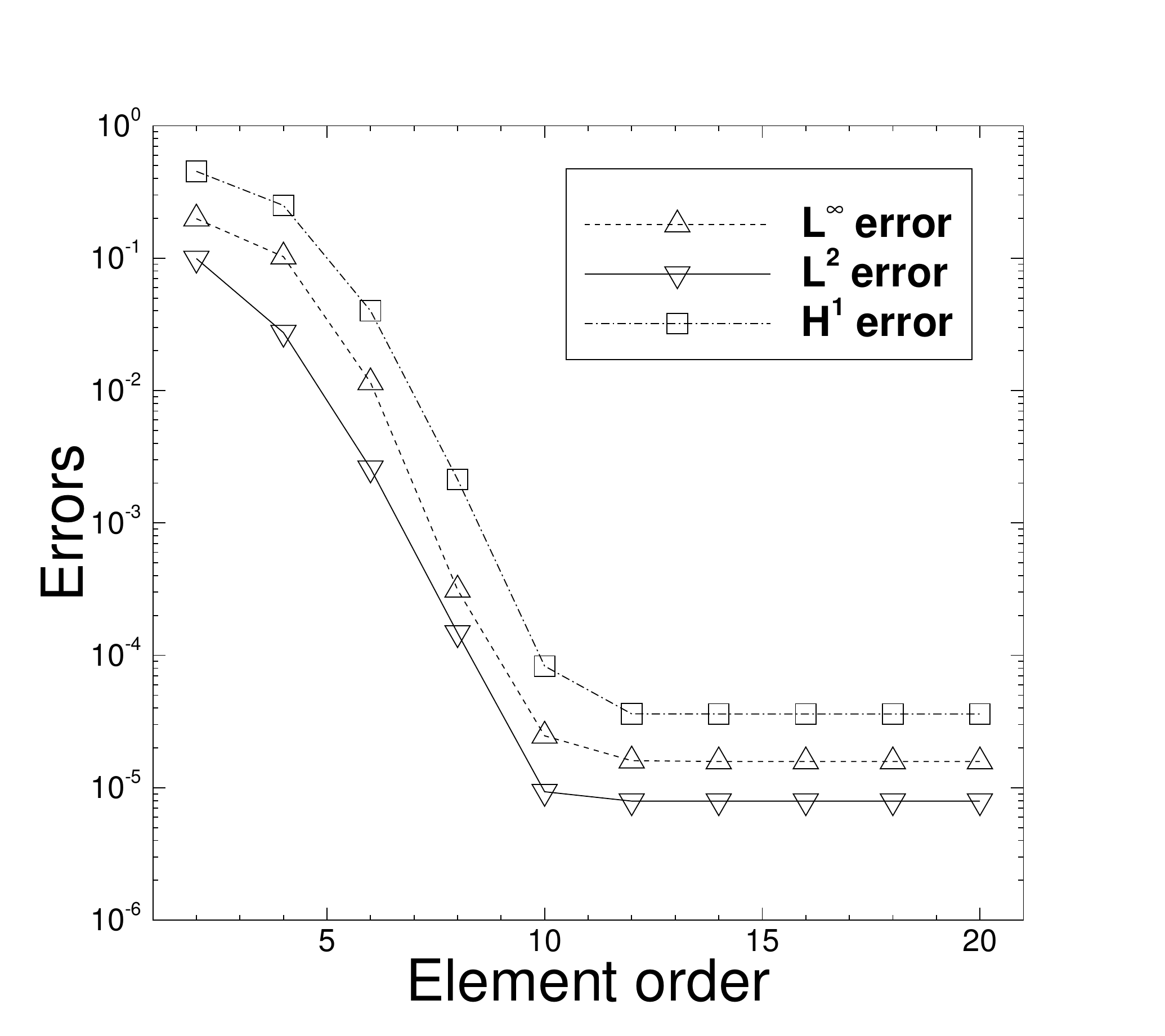}(a)
    \includegraphics[width=3in]{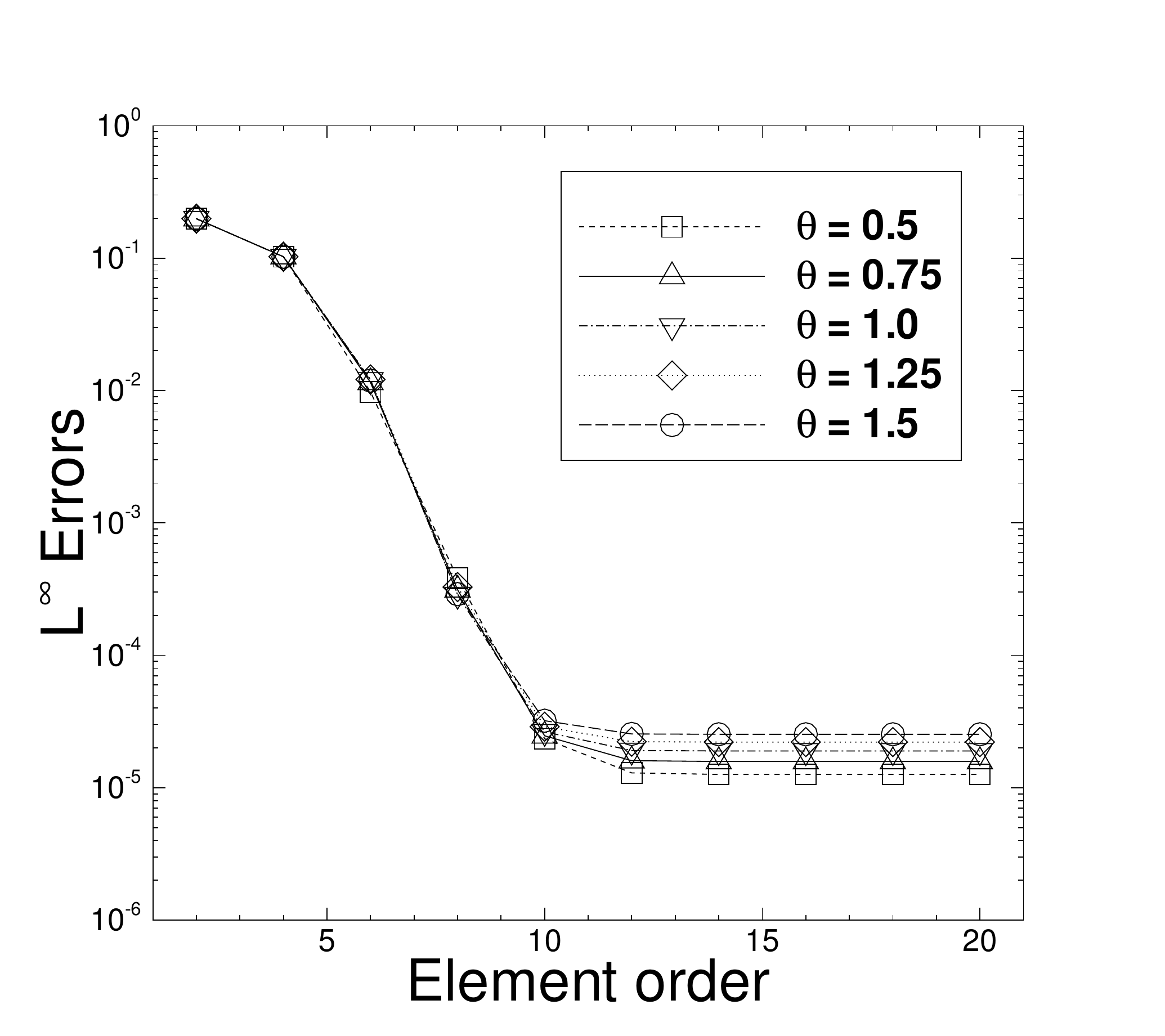}(b)
  }
  \caption{
    Spatial convergence tests: (a) $L^{\infty}$, $L^2$ and $H^1$ errors
    as a function of the
    element order (fixed $\Delta t=0.001$, $t_0=0.1$ and $t_f=0.2$)
    computed using the algorithm with $\theta=0.75$.
    (b) $L^{\infty}$ errors as a function of the element order computed
    using algorithms corresponding to several $\theta$ values.
  }
  \label{fig:spatial_conv}
\end{figure}

\begin{table}
  \centering
  \begin{tabular}{ll| ll}
    \hline
    parameter & value & parameter & value \\
    $C_0$ & $0$ & $\lambda$ & $0.01$  \\
    $m$ & $0.01$ & $\eta$ & $0.1$  \\
    $\theta$ & $1/2\sim 3/2$ & $\Delta t$ & (varied) \\
    $\Delta t_{\min}$ & $1e-4$ & $S$ & $\sqrt{\frac{4\gamma_0\lambda\omega_0}{m\Delta t}}$
    or $\sqrt{\frac{4\gamma_0\lambda\omega_0}{m\Delta t_{\min}}}$  \\
    $t_0$ & $0.1$ & $t_f$ & $0.2$ (spatial tests) or $0.3$ (temporal tests) \\
    Element order & (varied) & Elements & $2$ \\
    \hline
  \end{tabular}
  \caption{Simulation parameter values for convergence tests.}
  \label{tab:simu_param}
\end{table}


We discretize the domain using two equal-sized quadrilateral
elements, with the element order and the time step size
$\Delta t$ varied systematically in the spatial and
temporal convergence tests. The algorithm from Section \ref{sec:method}
is employed to numerically integrate the Cahn-Hilliard equation in
time from $t=t_0$ to $t=t_f$. The initial phase field function
$\phi_{in}$ is obtained by setting $t=t_0$ in the analytic expression
\eqref{equ:anal_soln}.
The numerical
solution at $t=t_f$ is then compared with the analytic
solution, and various norms of the errors are computed.
The values for the simulation parameters are summarized in
Table \ref{tab:simu_param} for this problem.

To test the spatial convergence rate, we employ a fixed
$\Delta t=0.001$, $t_0=0.1$ and $t_f=0.2$, and vary
the element order systematically between $2$ and $20$.
Then for each element order, the numerical errors of $\phi$ in
$L^{\infty}$, $L^2$ and $H^1$ norms at $t=t_f$ are obtained.
Figure \ref{fig:spatial_conv}(a) shows these errors
as a function of the element order using
the algorithm with $\theta=3/4$. We observe an exponential
decrease of the numerical errors with increasing element order
(for element orders $10$ and below),
and a level-off of the error curves beyond element
order $12$ due to
the saturation of temporal errors.
Figure \ref{fig:spatial_conv}(b) shows the $L^{\infty}$ errors
versus the element order obtained using the
algorithms corresponding several $\theta$ values
ranging from $1/2$ to $3/2$.
Some differences in the saturation error level (for orders
$12$ and above) can be observed with different algoirthms.
The saturation error appears larger for algorithms
with larger $\theta$ values.

\begin{figure}
  \centerline{
    \includegraphics[width=3in]{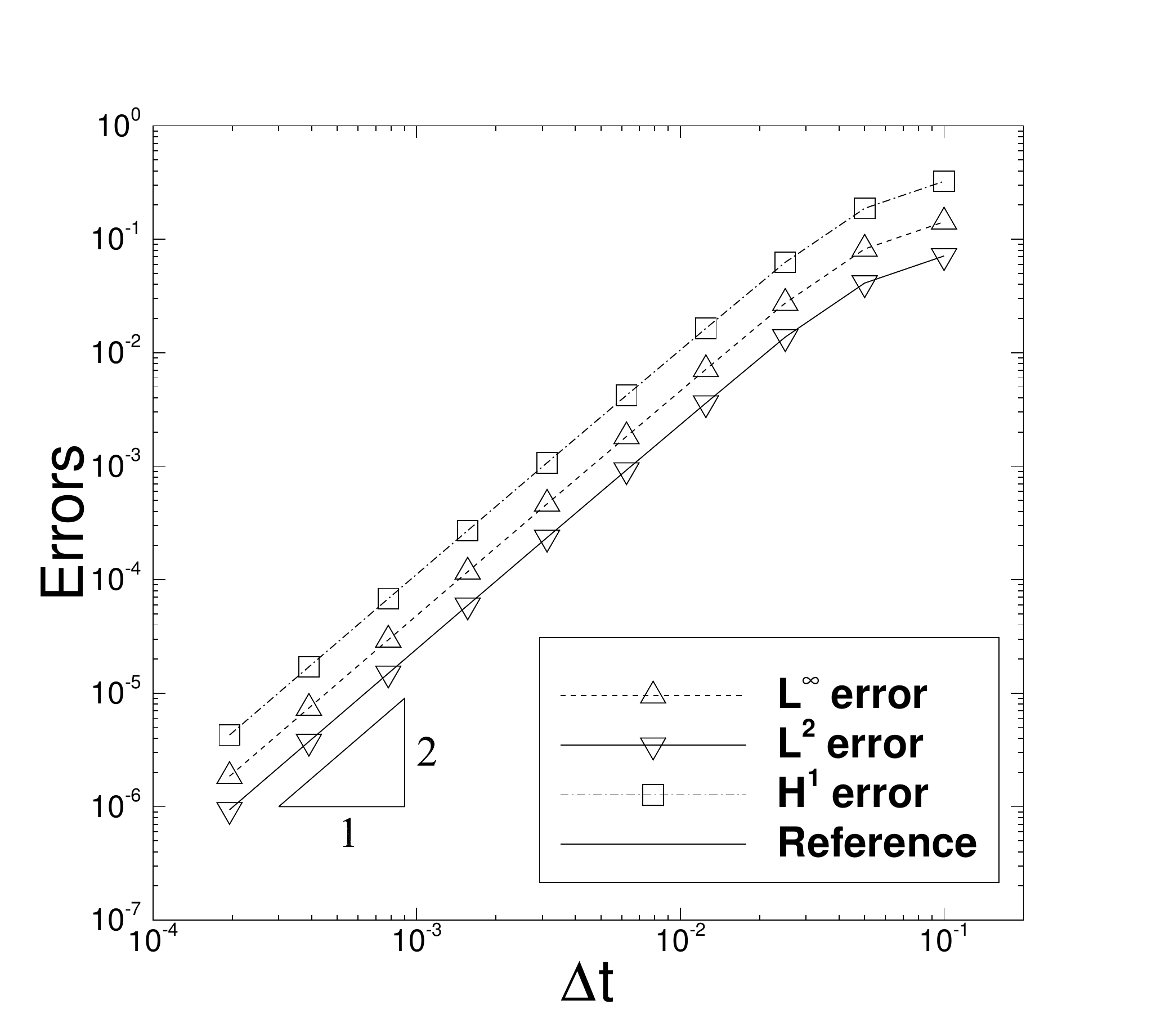}(a)
    \includegraphics[width=3in]{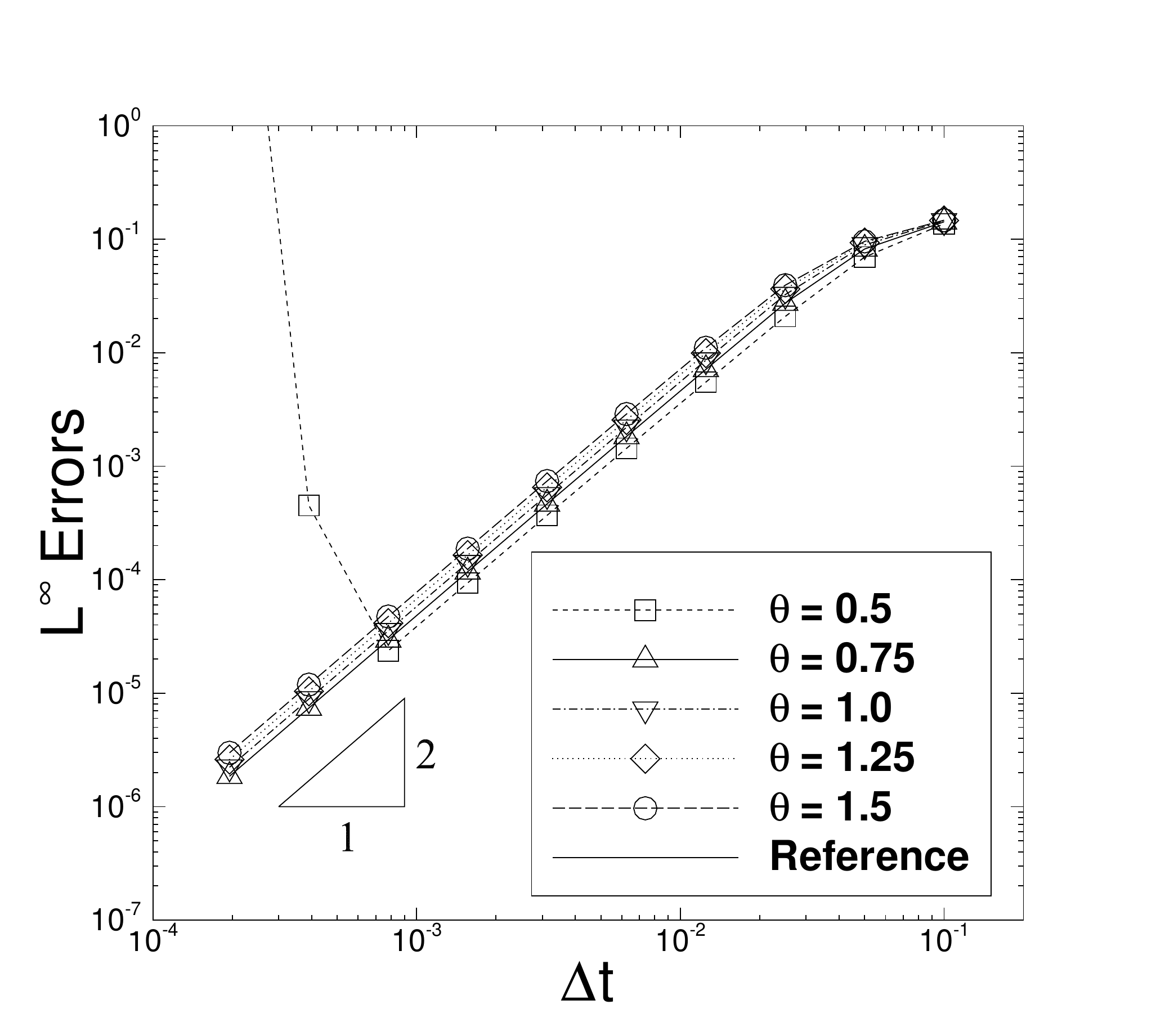}(b)
  }
  \caption{
    Temporal convergence tests: (a) $L^{\infty}$, $L^2$ and $H^1$ errors
    as a function of time step size $\Delta t$
    (fixed element order $16$, $t_0=0.1$ and $t_f=0.3$)
    computed using the algorithm corresponding to $\theta=0.75$.
    (b) $L^{\infty}$ errors as a function of $\Delta t$ computed
    using the algorithms corresponding to several $\theta$ values.
  }
  \label{fig:temporal_conv}
\end{figure}

To test the temporal convergence rate, we employ a fixed
element order $16$, $t_0=0.1$ and $t_f=0.3$, and
vary the time step size $\Delta t$ systematically
between $0.1$ and $1.953125\times 10^{-4}$.
For each $\Delta t$ the numerical errors in different norms
are computed at $t=t_f$.
Figure \ref{fig:temporal_conv}(a) shows the numerical
errors as a function of $\Delta t$ obtained using
the algorithm with $\theta=3/4$. A second-order
convergence rate is observed.
Figure \ref{fig:temporal_conv}(b) shows the $L^{\infty}$
errors versus $\Delta t$ obtained using the algorithms
corresponding to several $\theta$ values.
The errors are generally smaller with a smaller $\theta$
value in the algorithm. However, with
$\theta=1/2$ (the lower boundary for the $\theta$ range,
corresponding to
the Crank-Nicolson scheme)
we observe a weak instability when $\Delta t$ becomes small
(below $\Delta t=0.00078125$), which results in
larger errors with the two smallest $\Delta t$ values
in the test. Note that $t_f$ is fixed in the tests,
and a larger number of time steps are computed
with a smaller $\Delta t$.
The observation of poor performance of
the algorithm with $\theta=1/2$
will appear again with further numerical tests in
subsequent sections. 

The test results of this section indicate that the
family of algorithms presented in Section \ref{sec:method}
exhibits a spatial exponential convergence rate
and a temporal second-order convergence rate.

\subsection{Evolution and Coalescence of Drops}

In this section we consider the evolution of a drop
and the coalescence of two drops to illustrate the dynamics
of the Cahn-Hilliard equation and the energy stability
of the family of algorithms from Section \ref{sec:method}.

\paragraph{Evolution of a Drop}
We first look into the evolution of a material drop under
the Cahn-Hilliard dynamics.
Consider a square domain
$
\Omega = \{\ (x,y) \ |\ 0\leqslant x,y\leqslant 1  \ \},
$
and two materials contained in this domain.
It is assumed that the evolution of the material regions
is described by the Cahn-Hilliard equation, and that
$\phi=1$ and $-1$ correspond to the bulk of the first and second materials,
respectively.
At $t=0$, the first material is located in a smaller square region
with dimension $2h_0 \times 2h_0$ (where $h_0=0.2$)
around the center of the domain, and the second material fills
the rest of the domain.
The goal here is to study the evolution dynamics 
of these material regions.

\begin{figure}[tbp]
  \centerline{
 \subfigure[$t=0.001$]{ \includegraphics[scale=.21]{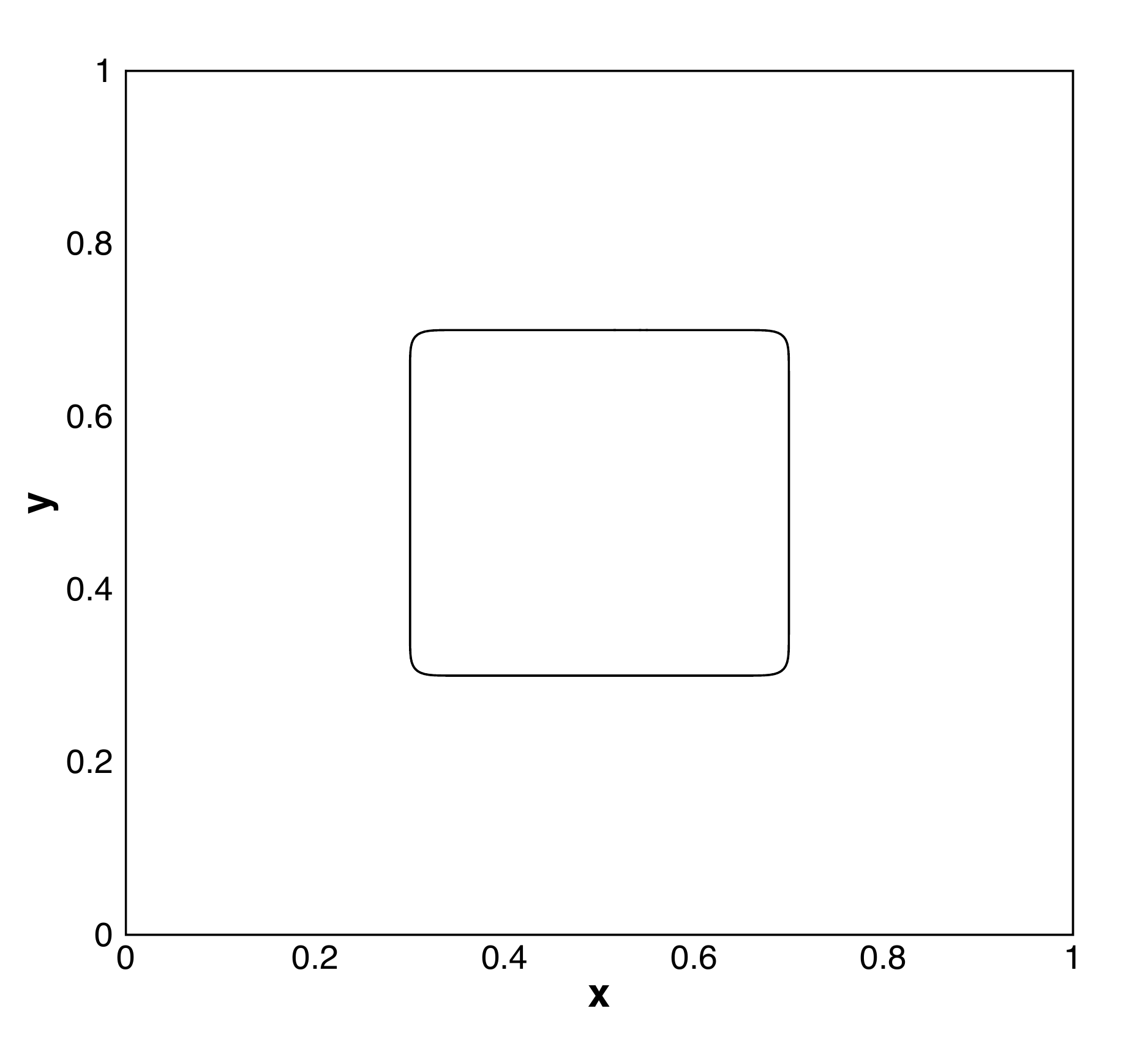}} 
 \subfigure[$t=1$]{ \includegraphics[scale=.21]{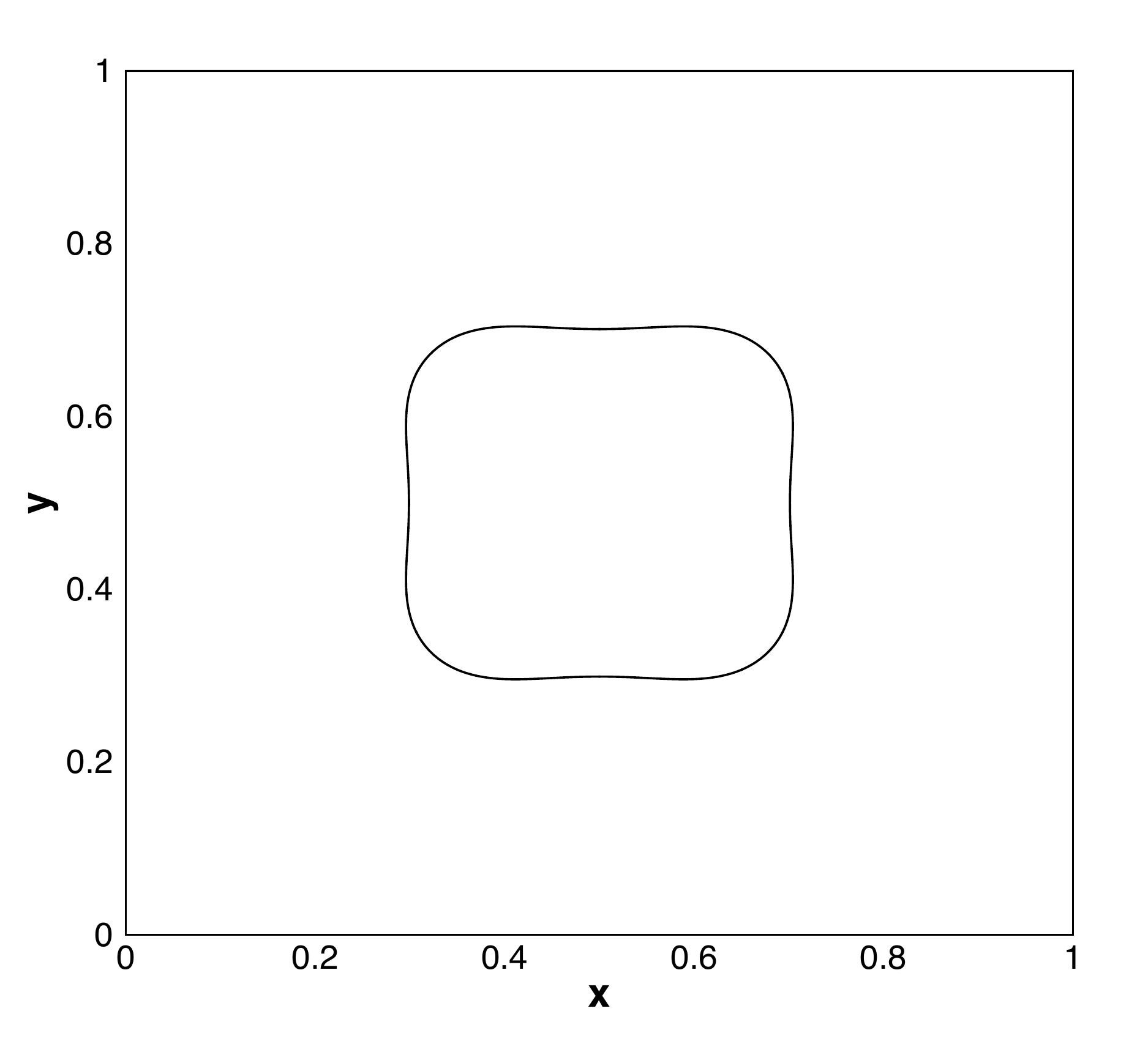}} 
  \subfigure[$t=4$]{ \includegraphics[scale=.21]{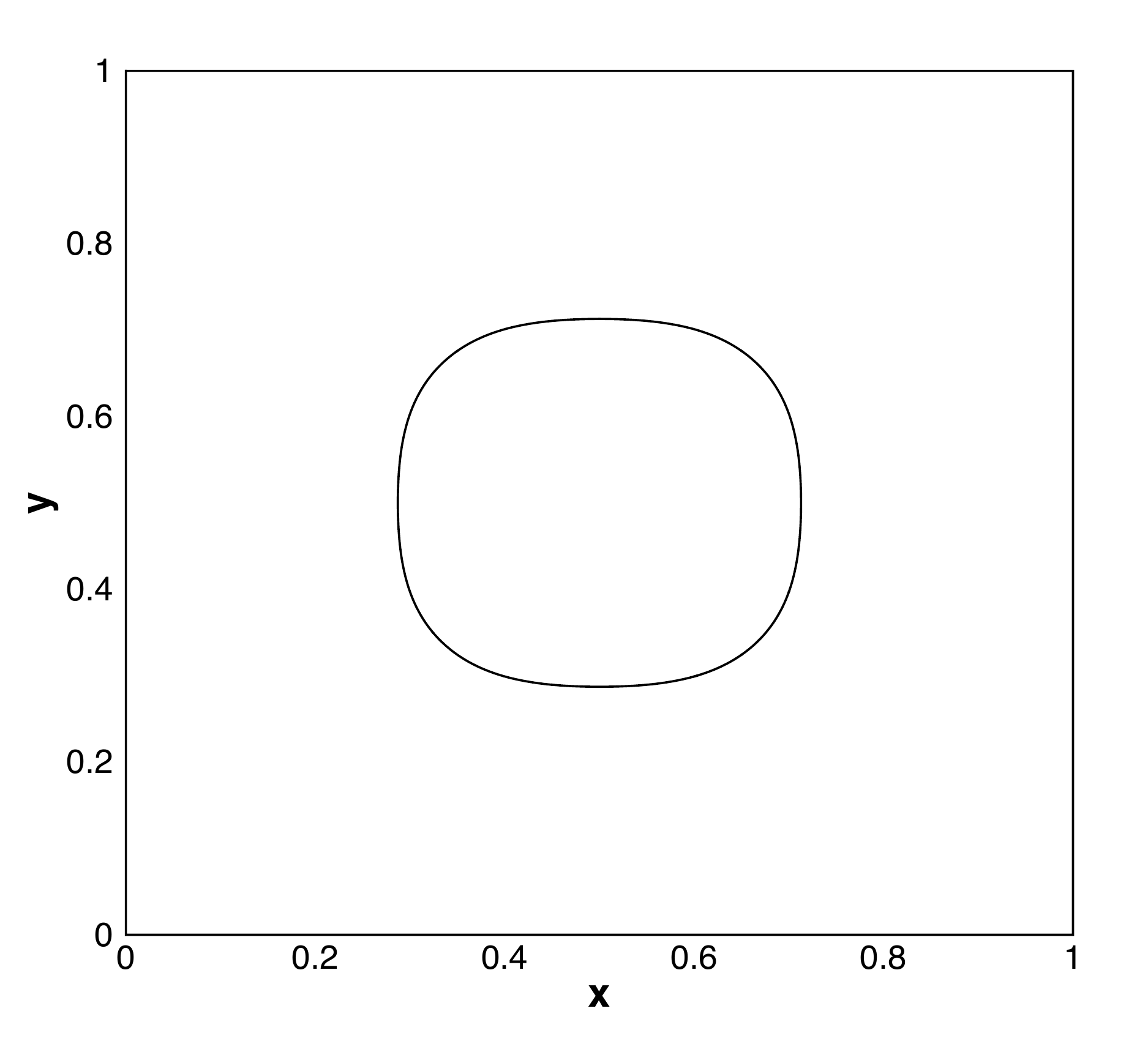}} 
  \subfigure[$t=10$]{ \includegraphics[scale=.21]{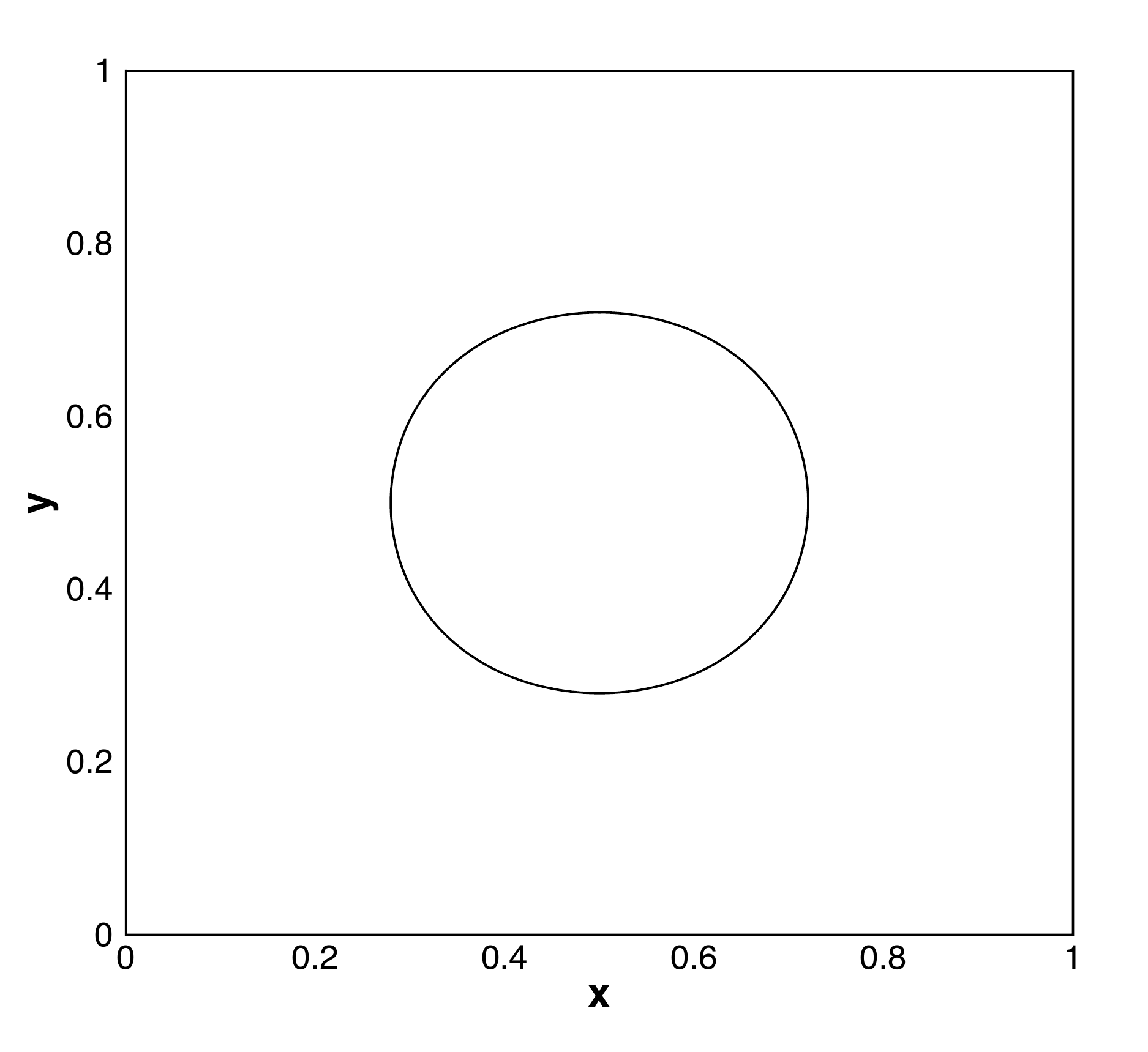}}
  }
  \caption{Temporal sequence of snapshots showing
    the evolution of a drop visualized by the contour level $\phi=0$.
    Results are obtained with the algorithm $\theta=0.75$,
    and time step size $\Delta t=10^{-3}$.
  }
\label{fig:evo0}
\end{figure}

\begin{figure}
  \centerline{
    \includegraphics[width=2.5in]{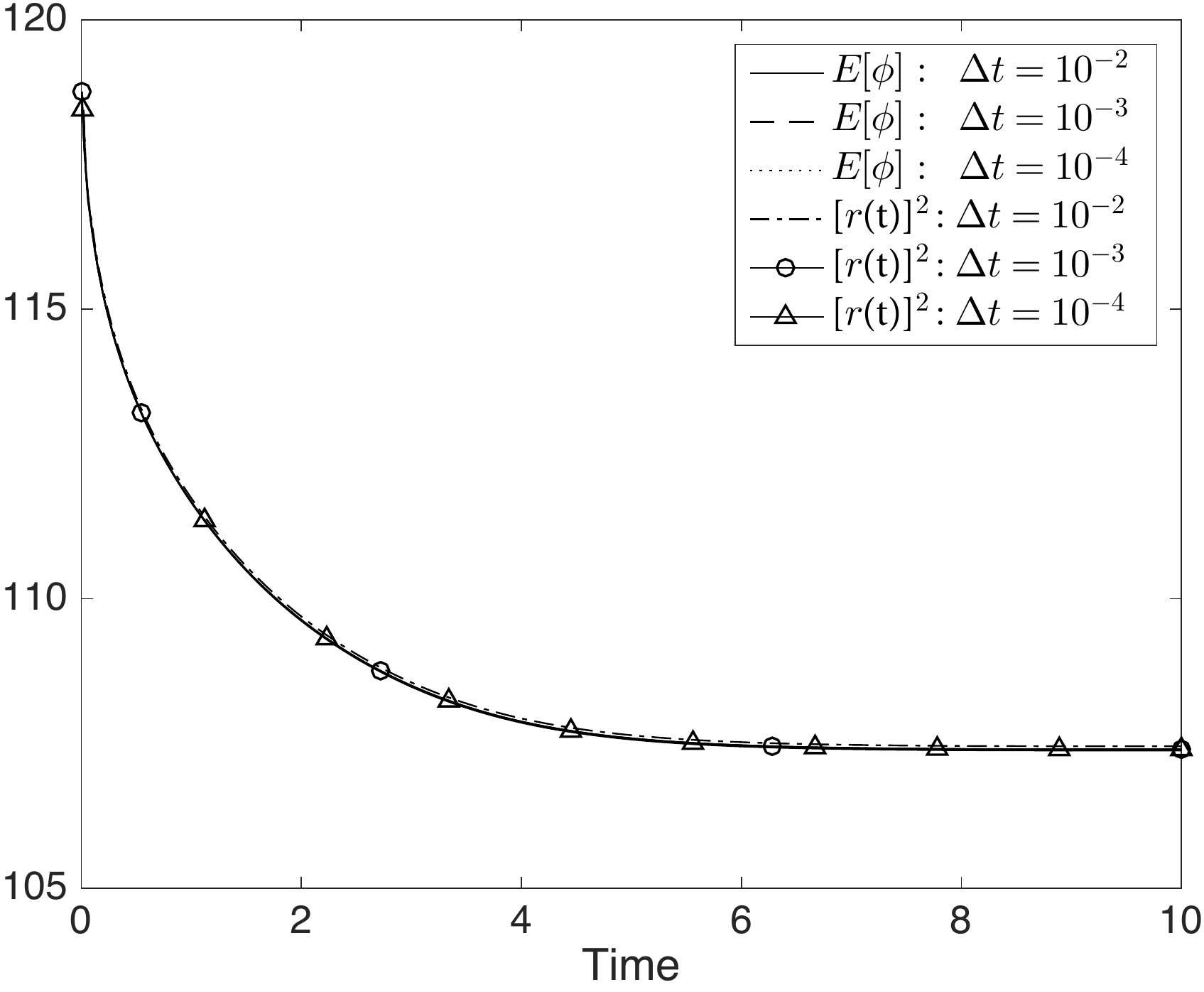}
  }
  \caption{
    Time histories of $E[\phi]$ and $[r(t)]^2$ computed using
    several $\Delta t$ values, with the algorithm $\theta=0.75$.
  }
  \label{fig:eng_hist_small_dt}
\end{figure}

We employ the algorithms from Section \ref{sec:method}
to numerically solve the Cahn-Hilliard equation with $g(\mathbf{x},t)=0$
in this domain.
We discretize the domain using
$400$ quadrilateral elements, with $20$ equal-sized elements
along both $x$ and $y$ directions.
The boundary conditions \eqref{equ:wbc_1} and \eqref{equ:wbc_2},
with $g_a=0$ and $g_b=0$, are imposed on the domain boundaries.
The initial distribution of the materials is given by
\begin{equation}
  \phi_{in}(\mathbf{x})=\frac{1}{2}\left[ \tanh \frac{x-x_0+h_0}{\sqrt{2}\eta}  - \tanh \frac{x-x_0-h_0}{\sqrt{2}\eta}   \right]\cdot
  \frac{1}{2} \left[ \tanh \frac{y-y_0+h_0}{\sqrt{2}\eta}  - \tanh \frac{y-y_0-h_0}{\sqrt{2}\eta}   \right],
\end{equation}
where $(x_0,y_0)=(0.5,0.5)$ is the center of the domain.
We employ the following (non-dimensional) parameter values for this problem:
\begin{equation}
  \left\{
  \begin{split}
    &
    \eta = 0.01, \quad \sigma = 151.15,\quad C_0=0, \\
    &
    \lambda = \frac{3}{2\sqrt{2}}\sigma\eta, \quad
    m = \frac{10^{-6}}{\lambda}, \quad
    S = \sqrt{\frac{4\gamma_0\lambda\omega_0}{m\Delta t}}, \\
    &
    \text{element order:} \ 8, \quad \text{number of elements:} \ 400, \\
    &
    \Delta t \ \text{varied}, \quad \theta \ \text{varied}.
  \end{split}
  \right.
  \label{equ:drop_evolve_param}
\end{equation}


Figure \ref{fig:evo0} shows the evolution of the material
regions with a temporal sequence of snapshots of
the interface (visualized by the contour level $\phi=0$)
between the two materials. These results are computed
with a time step size $\Delta t=0.001$
using the algorithm corresponding to $\theta=0.75$.
The initially square region of the first material
evolves gradually into a circular region under the
Cahn-Hilliard dynamics.
Figure \ref{fig:eng_hist_small_dt} shows the time histories
of the potential energy $E[\phi]$ defined in \eqref{equ:def_E1_E2}
and the quantity $[r(t)]^2$ computed from equation \eqref{equ:r_equ},
obtained using time step sizes $\Delta t=10^{-4}$, $10^{-3}$ and $10^{-2}$
with the algorithm $\theta = 0.75$.
Both $E[\phi]$ and $[r(t)]^2$ decrease over time, and they gradually
level off at a certain level.
In particular, we observe that the history curves for $E[\phi]$ and 
$[r(t)]^2$ essentially overlap with one another,
which is consistent with the fact that $r(t)$ computed based on
equation \eqref{equ:r_equ} is an approximation of $\sqrt{E[\phi]}$
(see equation \eqref{equ:def_r_q}).

\begin{figure}[tbp]
  \centering
 \subfigure[$t=10^4$]{ \includegraphics[scale=.24]{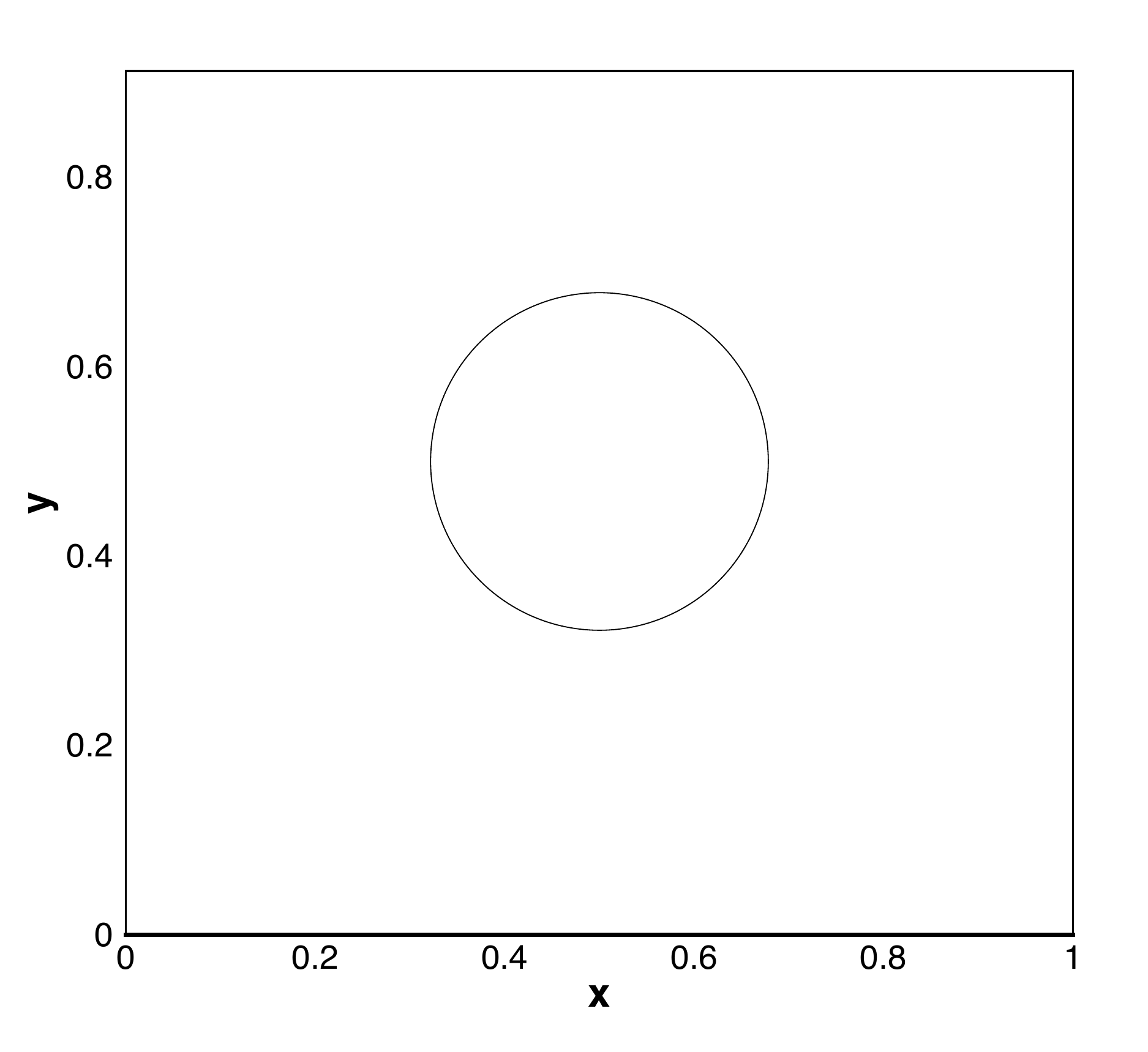}} 
 \subfigure[]{ \includegraphics[scale=.28]{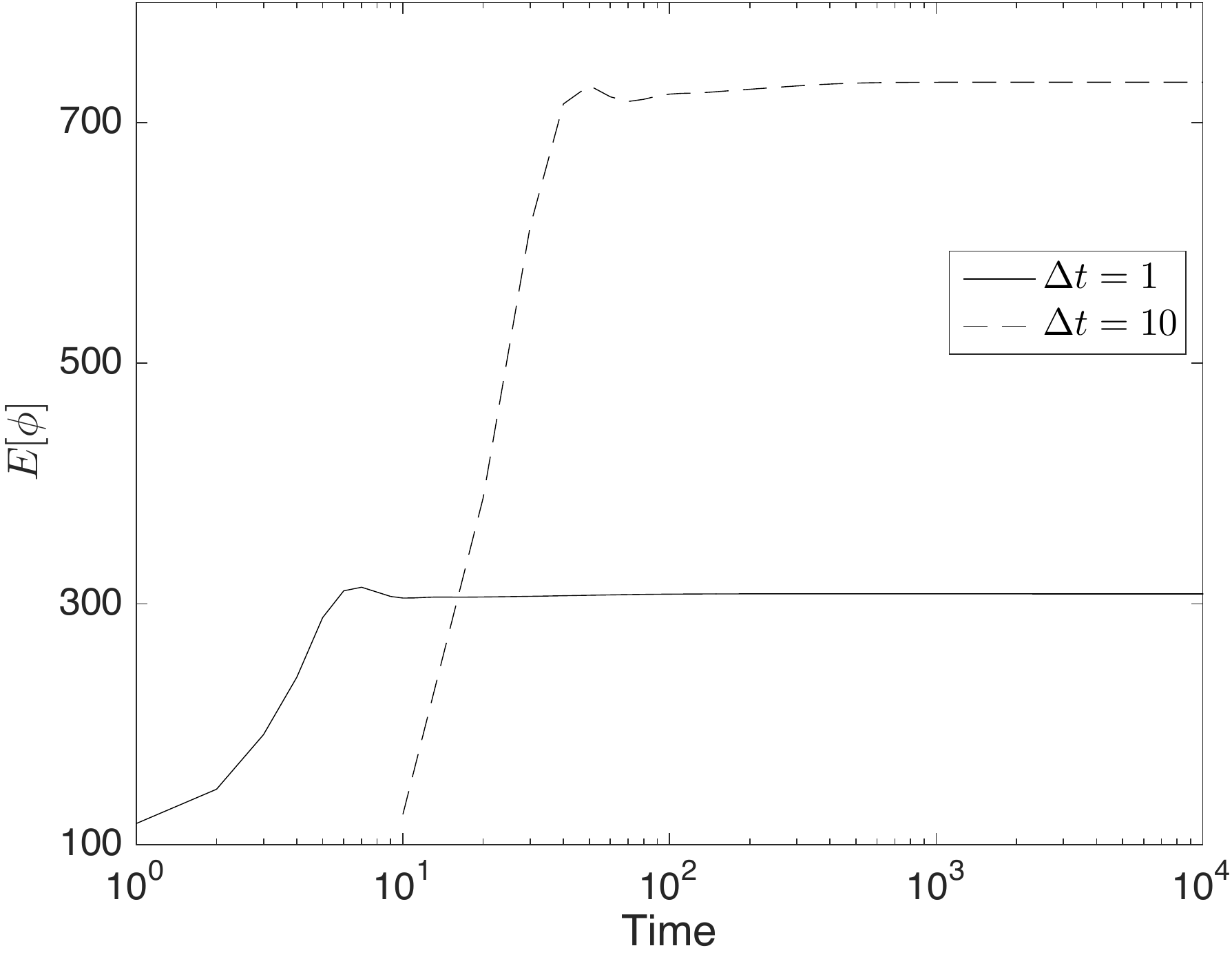}} 
  \subfigure[]{ \includegraphics[scale=.28]{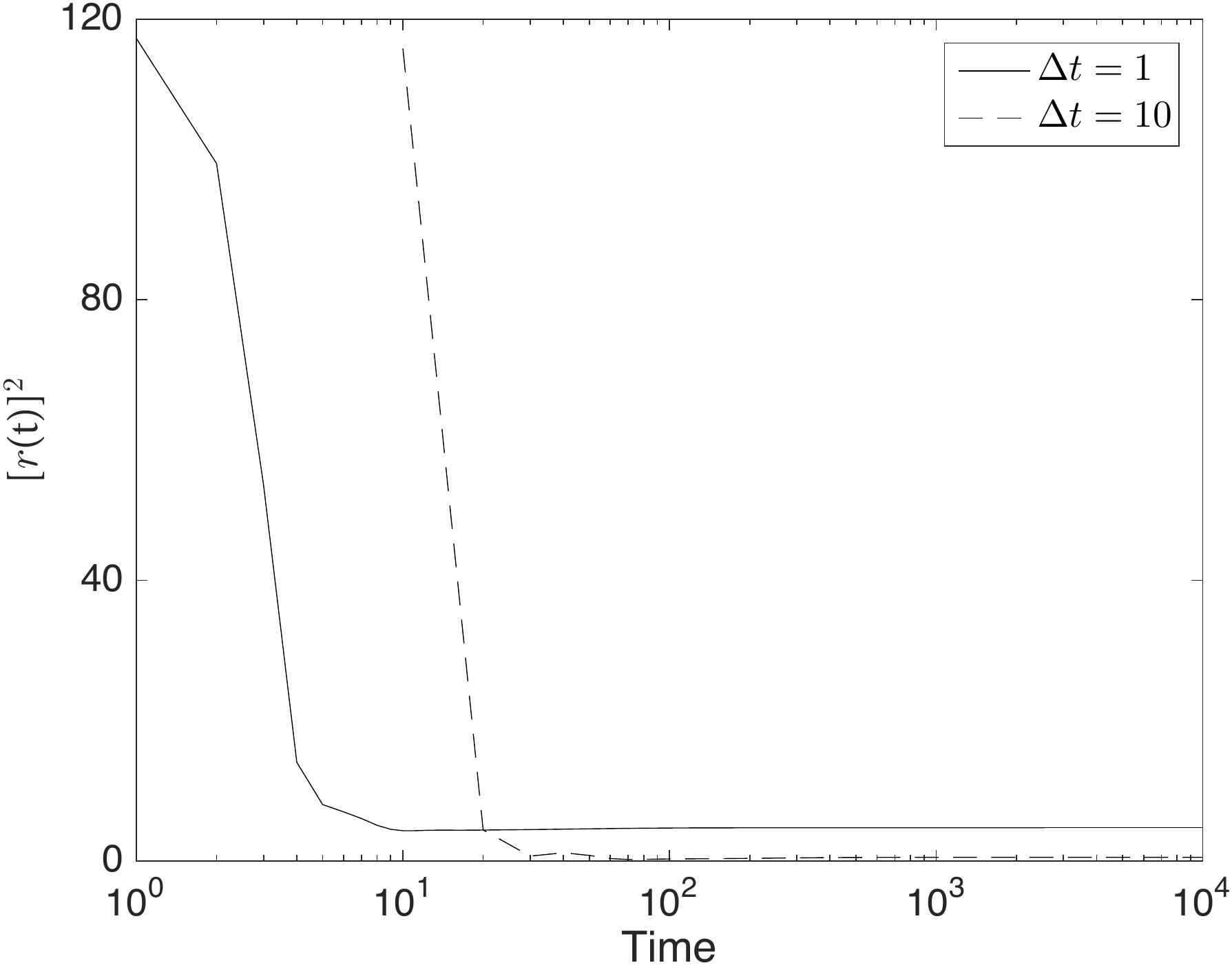}} 
  \caption{
    Drop evolution computed using large $\Delta t$ values:
    (a) Drop configuration at $t=10^4$ computed using $\Delta t=10$;
    (b)-(c) Time histories of $E[{\phi}]$ and $[r(t)]^2$ obtained with $\Delta t=1,10.$
    Simulation results are obtained with the algorithm $\theta=1$ (BDF2).
  }
\label{fig:evo1}
\end{figure}

The energy stable nature of the current
algorithms allows the use of large or fairly large
time step sizes in the computations.
This is demonstrated by the results in Figure \ref{fig:evo1}.
Figure \ref{fig:evo1}(a) shows the interface
between the two materials at a much later time $t=10^4$,
computed using a large time step size $\Delta t=10$
with the algorithm $\theta=1$.
Figures \ref{fig:evo1}(b) and (c) show the time histories
of  $E[\phi]$ and $[r(t)]^2$
corresponding to two large
time step sizes $\Delta t=1$ and $\Delta t=10$
with the algorithm $\theta=1$.
The long time histories demonstrate that
the computations with these large $\Delta t$ values
are indeed stable using the current algorithm.
On the other hand,
because these time step sizes are very large,
we do not expect that the computation results will be
accurate with these $\Delta t$ values.
It is observed that the $E[\phi]$ and $[r(t)]^2$ values
in Figures \ref{fig:evo1}(b)-(c) are quite different
from those obtained with small $\Delta t$
(see Figure \ref{fig:eng_hist_small_dt}).
In particular, we observe that $[r(t)]^2$ decreases sharply at
the beginning, 
and for a given $\Delta t$
its history curve gradually levels off at a certain value over time.
With a larger $\Delta t$, the $[r(t)]^2$ history
tends to level off at a smaller value. With very large $\Delta t$ values,
$[r(t)]^2$ will essentially tend to zero.
This seems to be a common characteristic to the family of
algorithms developed in this work.

\begin{figure}[tbp]
  \centering
 \subfigure[$E {[} \phi {]}$ with $\theta=0.75$]{ \includegraphics[scale=.38]{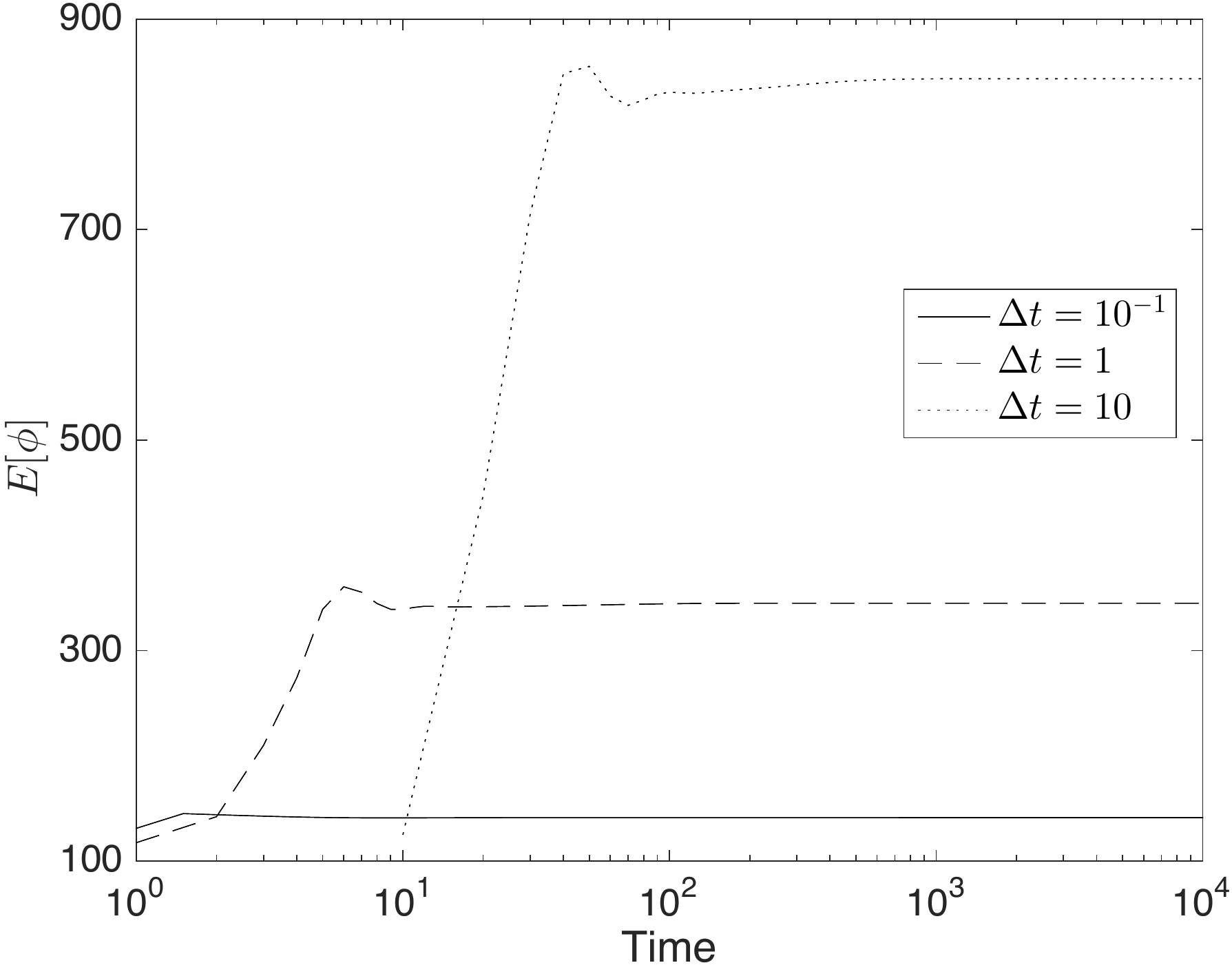}} 
 \subfigure[${[}r(t){]}^2$ with $\theta=0.75$]{\includegraphics[scale=.38]{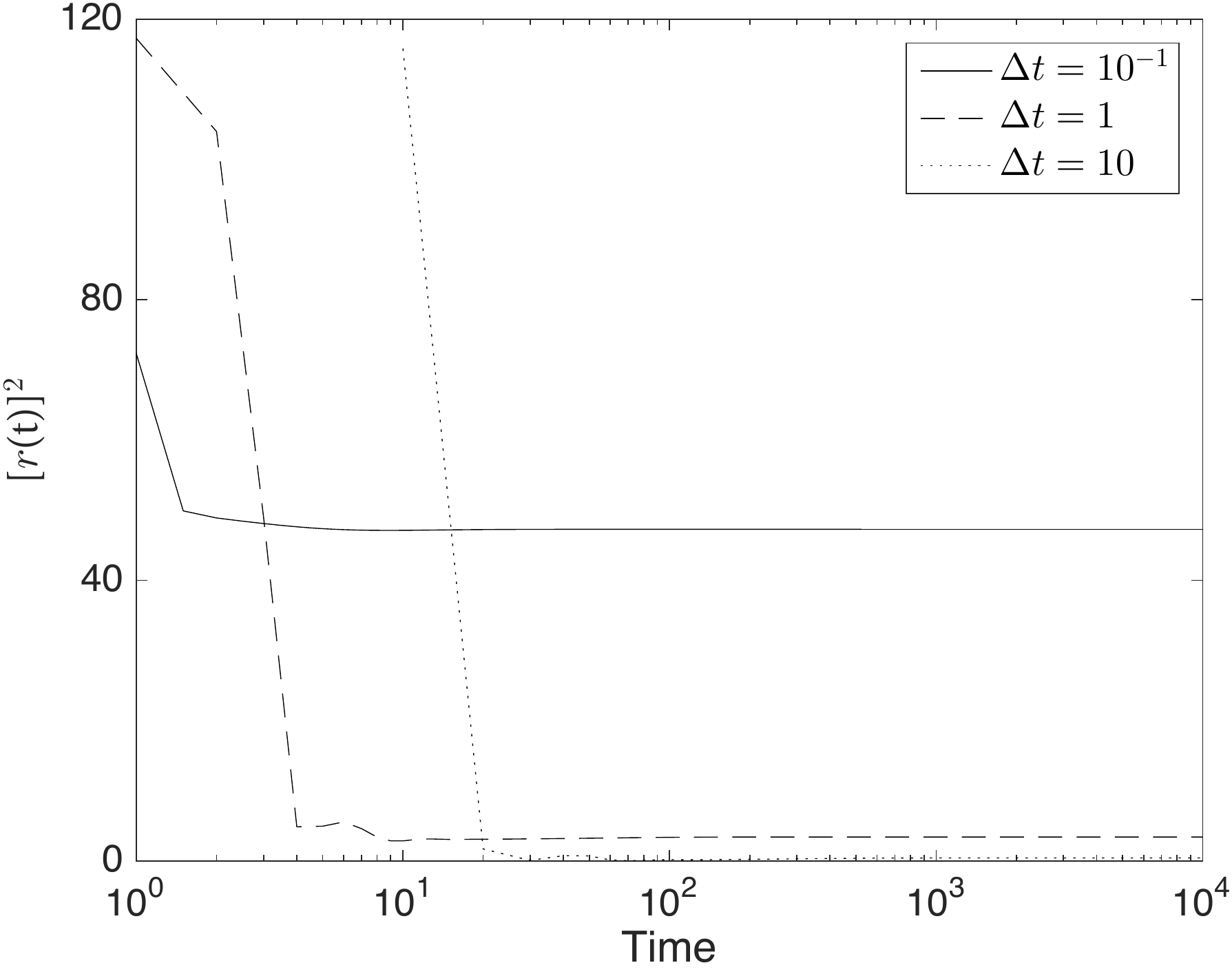}} 
 \subfigure[$E{[} \phi {]}$ with $\theta=1.25$]{ \includegraphics[scale=.38]{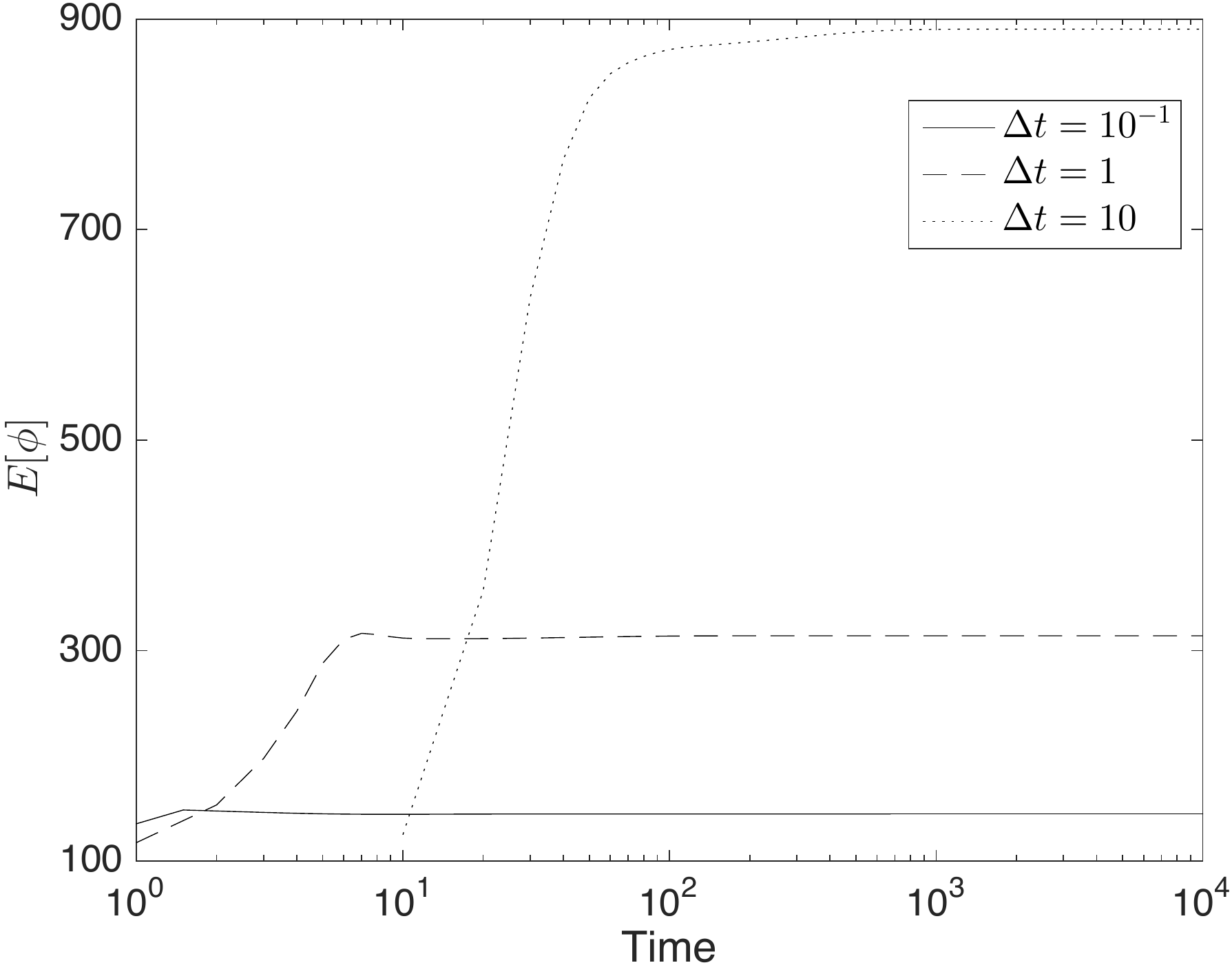}} 
 \subfigure[${[}r(t){]}^2$ with $\theta=1.25$]{ \includegraphics[scale=.38]{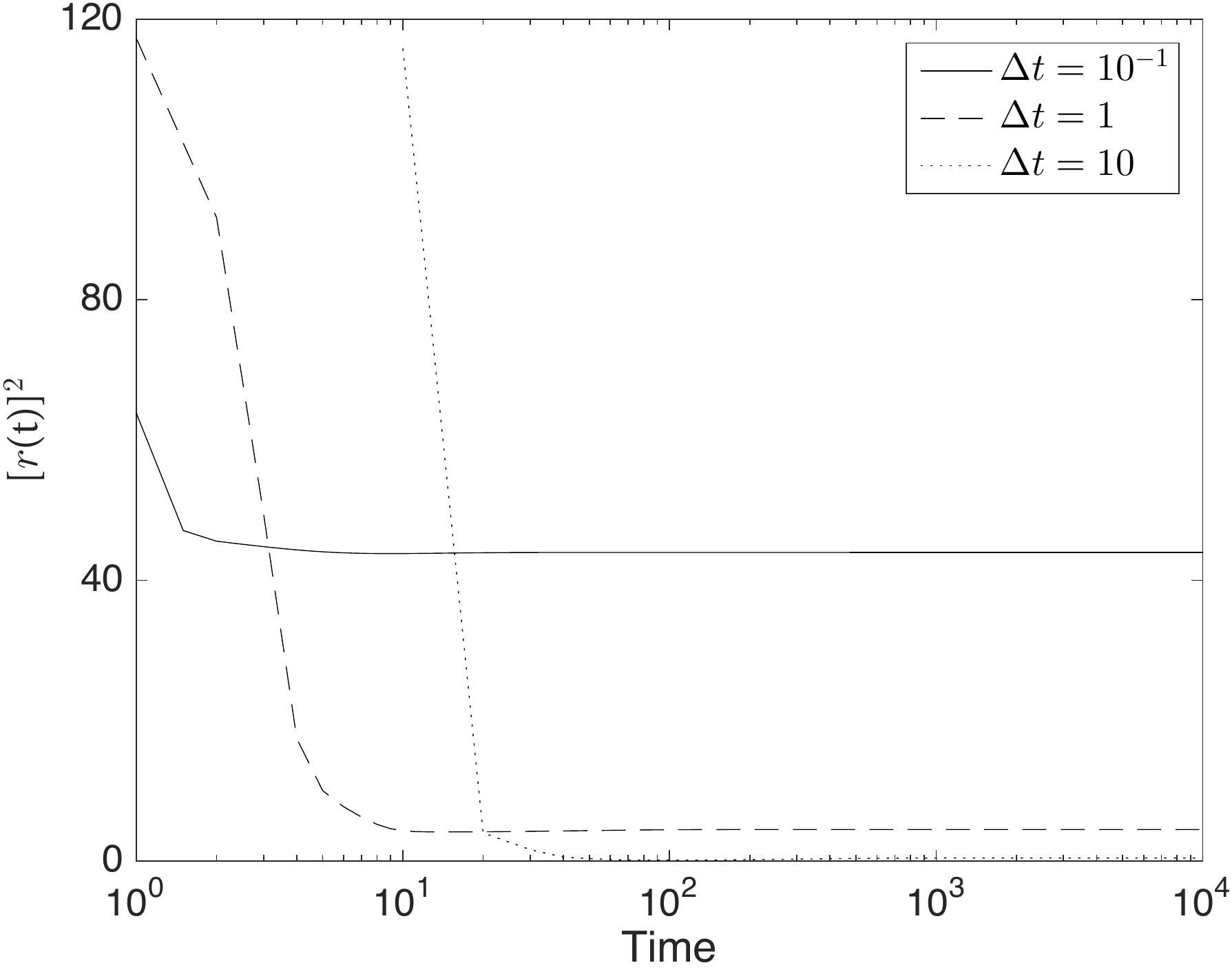}} 
 \caption{
   Time histories of $E[\phi]$ and $[r(t)]^2$ corresponding to
   several large $\Delta t$ values obtained with the algorithms
   $\theta=0.75$ and $\theta=1.25$.
 }
\label{fig:evo2}
\end{figure}

Similar behaviors have been observed with the
other members of this family of algorithms for
large $\Delta t$ values.
This is demonstrated by 
Figure \ref{fig:evo2}, in which we show time histories of
$E[\phi]$ and $[r(t)]^2$ corresponding to three
time step sizes $\Delta t=0.1$, $1$ and $10$,
computed using the algorithms $\theta=0.75$
and $\theta=1.25$.
The long-time histories signify the stability
of these computations.

\begin{figure}[tbp]
  \centering
 \subfigure[$\theta=0.5$]{ \includegraphics[scale=.38]{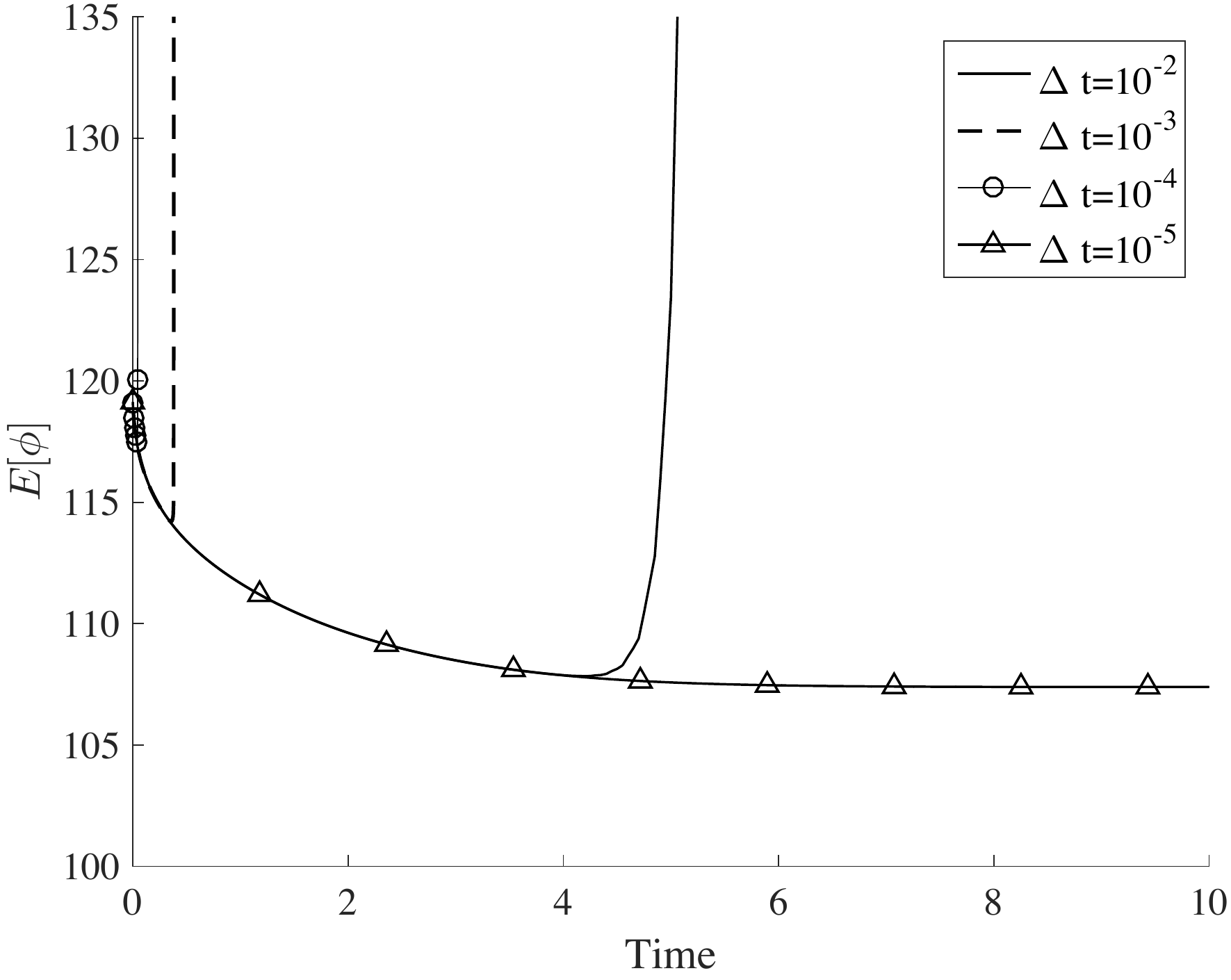}} 
 \subfigure[$\theta=1.5$]{ \includegraphics[scale=.38]{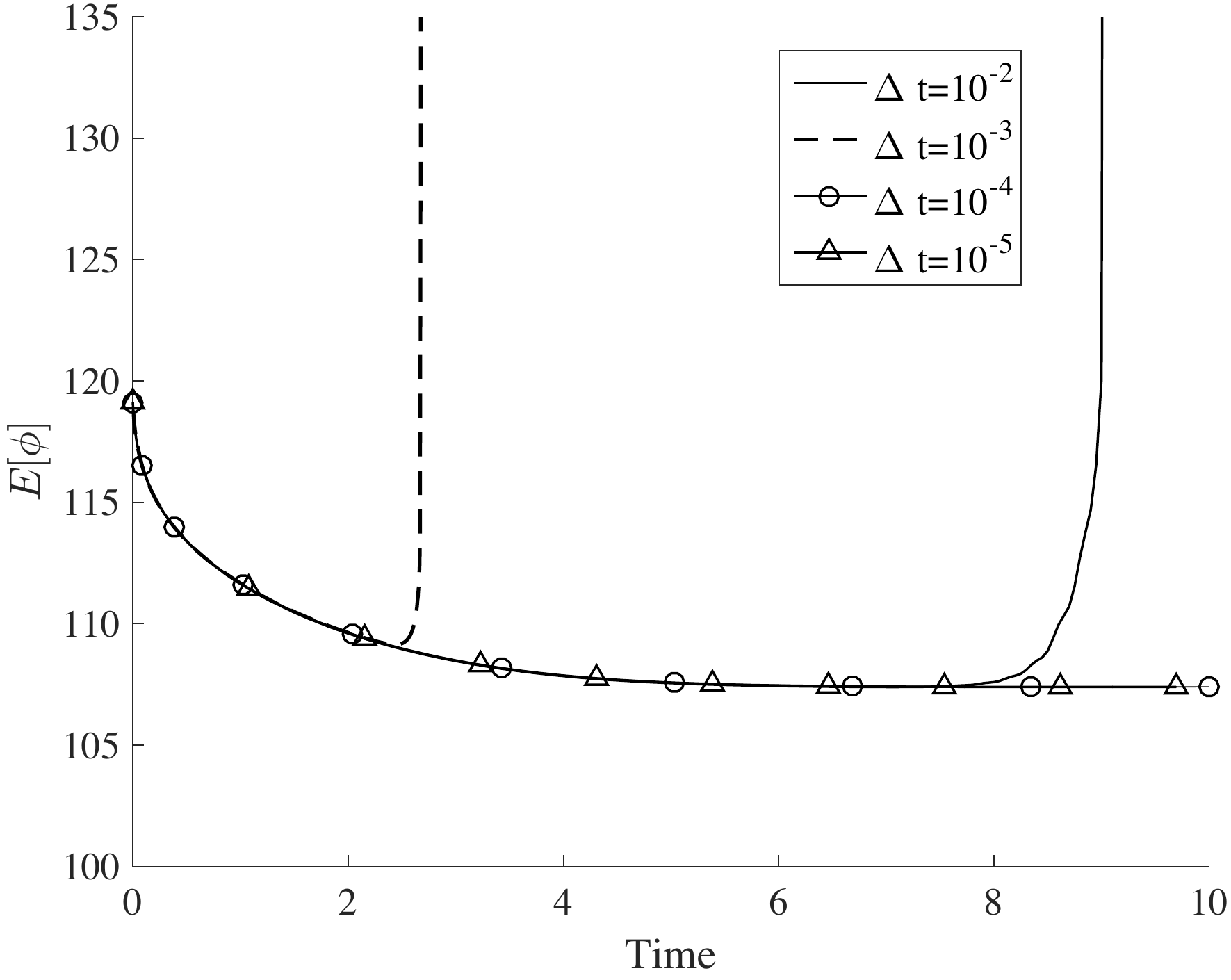}} 
 \caption{
   Time histories of $E[\phi]$ computed with the algorithms
   $\theta=0.5$ (a) and $\theta=1.5$ (b), using various $\Delta t$ values.
 }
\label{fig:evo3}
\end{figure}


Among the family of algorithms ($\frac{1}{2}\leqslant \theta\leqslant \frac{3}{2}$)
presented in Section \ref{sec:method},
we observe from numerical simulations that
the two members on the borders, $\theta=\frac{1}{2}$ and
$\theta=\frac{3}{2}$, seem to have a performance inferior to the rest of
this family. Figure \ref{fig:evo3} shows
the time histories of $E[\phi]$ corresponding to several
$\Delta t$ values, ranging from $10^{-2}$ to $10^{-5}$,
computed using the algorithms $\theta=\frac{1}{2}$
and $\theta=\frac{3}{2}$.
We observe that the computation using the algorithm $\theta=\frac{1}{2}$
becomes unstable with $\Delta t=10^{-2}$, $10^{-3}$ and $10^{-4}$,
and the computation using $\theta=\frac{3}{2}$ becomes
unstable with $\Delta t=10^{-2}$ and $10^{-3}$.
We do not observe such an instability with the other members of
this family of algorithms.
Note that the Theorem \ref{thm:thm_1} ensures the energy stability of
the semi-discretized algorithm (discrete in time, continuous in space)
given by \eqref{equ:alg_disc_1}--\eqref{equ:alg_disc_5}
for $\theta=\frac{1}{2}$ and $\theta = \frac{3}{2}$.
It is possible that the fully discretized
algorithm (in both space and time) may not preserve this
energy stability, which is
likely why we observe this instability
with the algorithms $\theta=\frac{1}{2}$ and $\theta = \frac{3}{2}$.


\paragraph{Coalescence of Two Drops}


We next consider the coalescence of two material drops governed
by the Cahn-Hilliard equation.
The computational domain and the setting follow those
for the single-drop case discussed above.
The difference lies in the initial distribution of the materials.
Here we assume that at $t=0$ the first material
occupies two circular regions that are right next to each other
and the second material fills the rest of the domain.
The two regions of the first material then coalesce with
each other to form a single drop 
under the Cahn-Hilliard dynamics.
The goal is to illustrate this process using the
current algorithms.

\begin{figure}[tbp]
  \centerline{
 \subfigure[$t=3\times 10^{-3}$]{ \includegraphics[scale=.25]{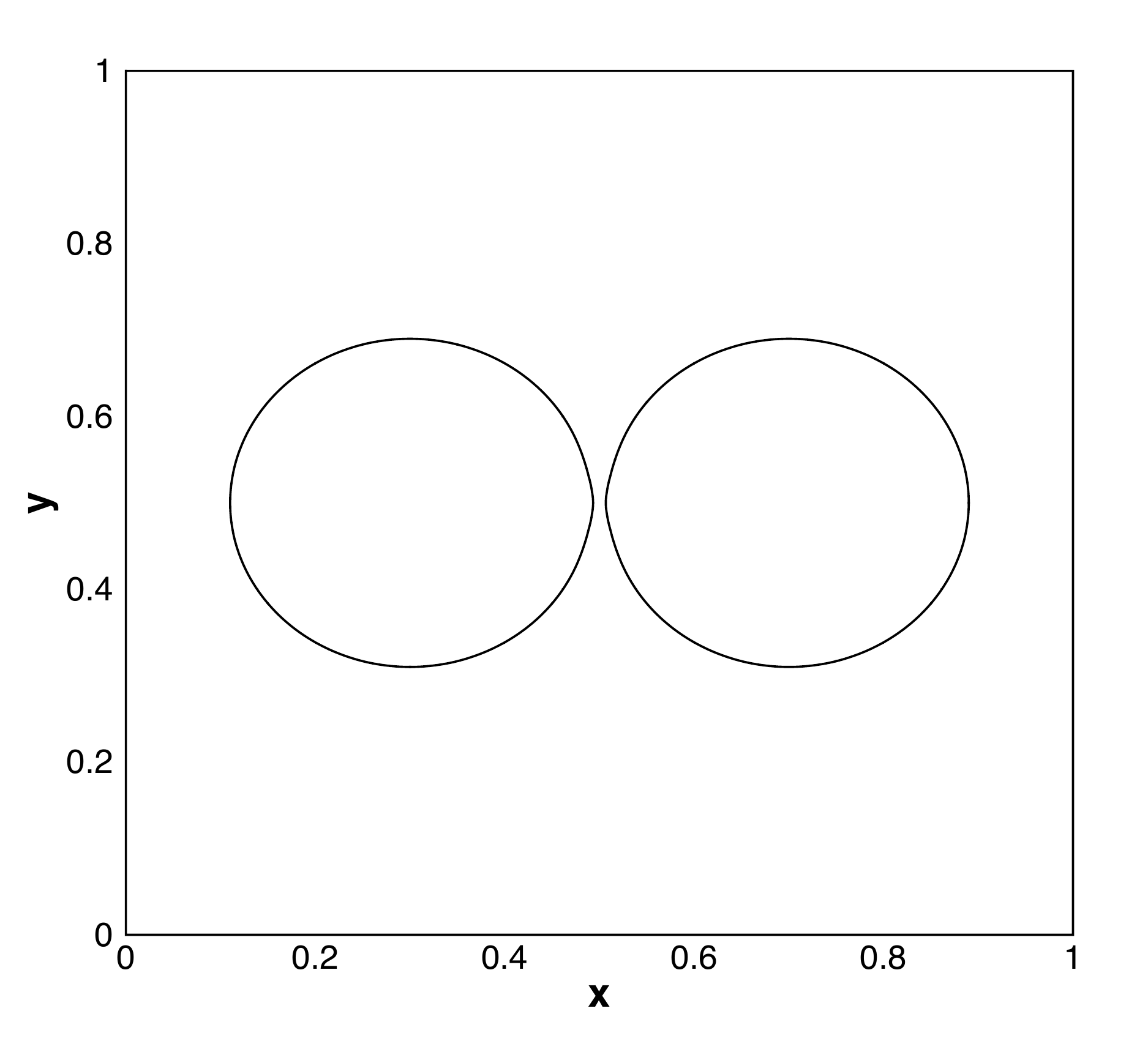}} 
 \subfigure[$t=0.25$]{ \includegraphics[scale=.25]{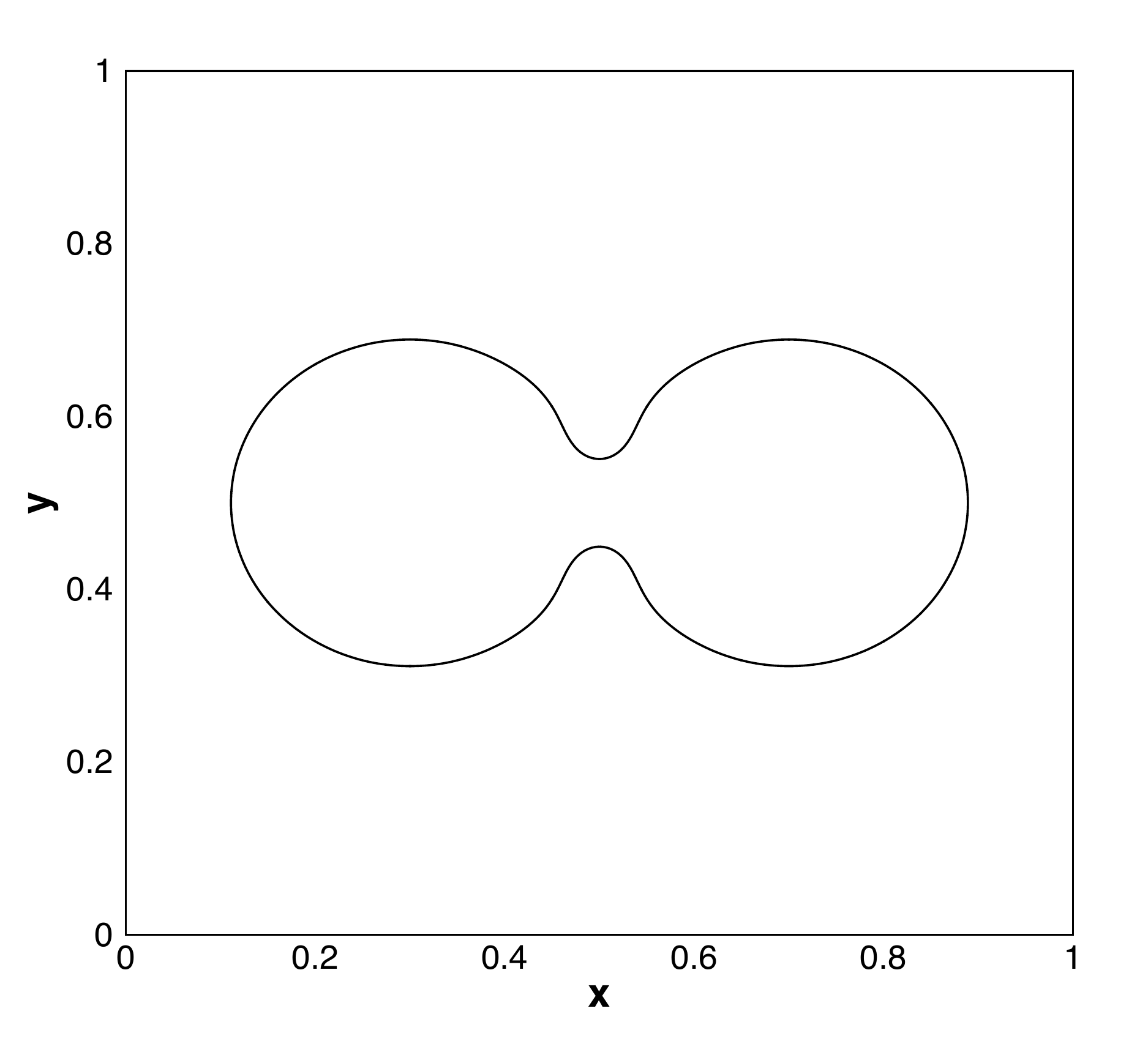}} 
 \subfigure[$t=2$]{ \includegraphics[scale=.25]{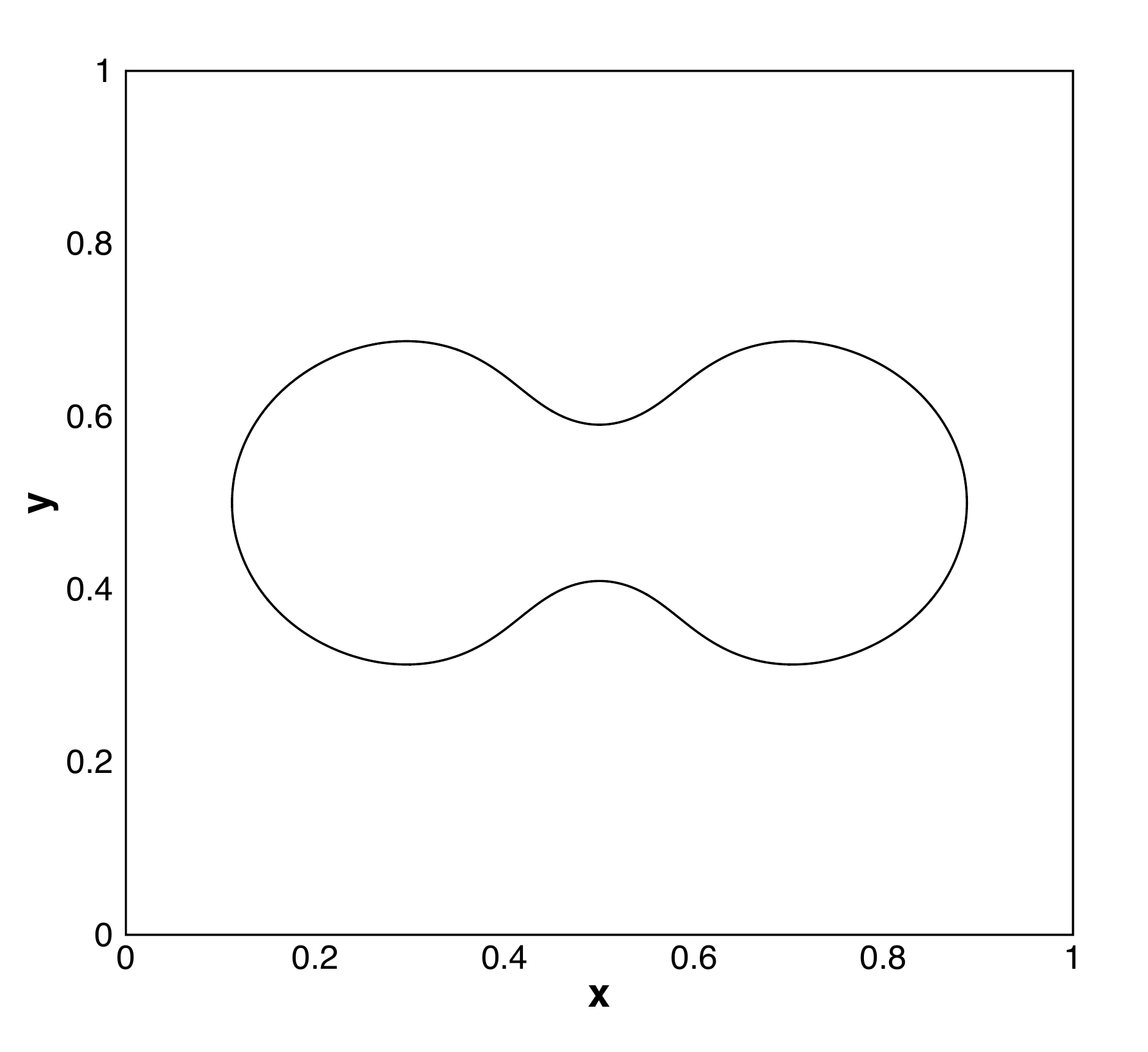}}
  }
  \centerline{
 \subfigure[$t=8$]{ \includegraphics[scale=.25]{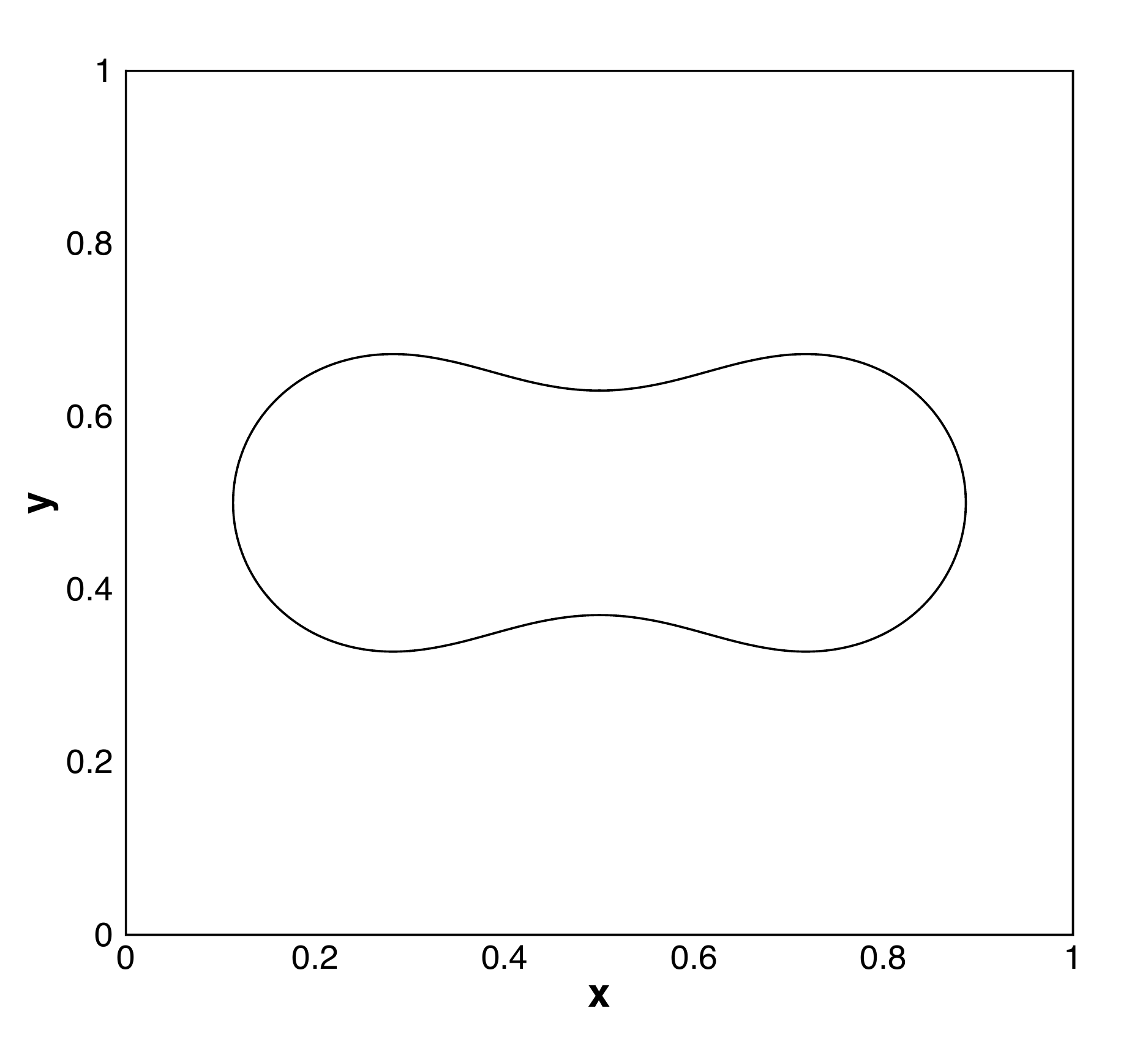}} 
 \subfigure[$t=20$]{ \includegraphics[scale=.25]{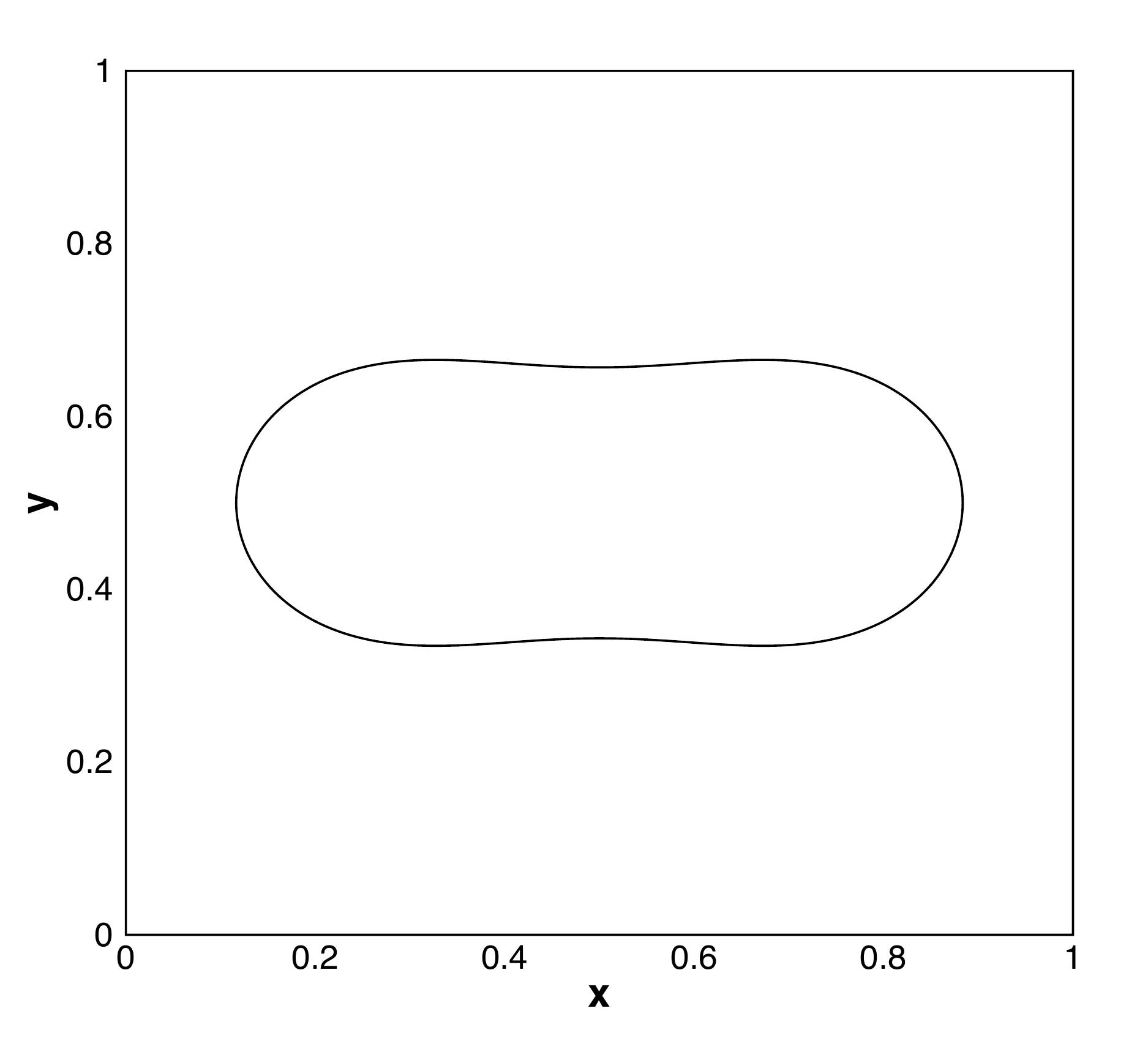}} 
 \subfigure[$t=40$]{ \includegraphics[scale=.25]{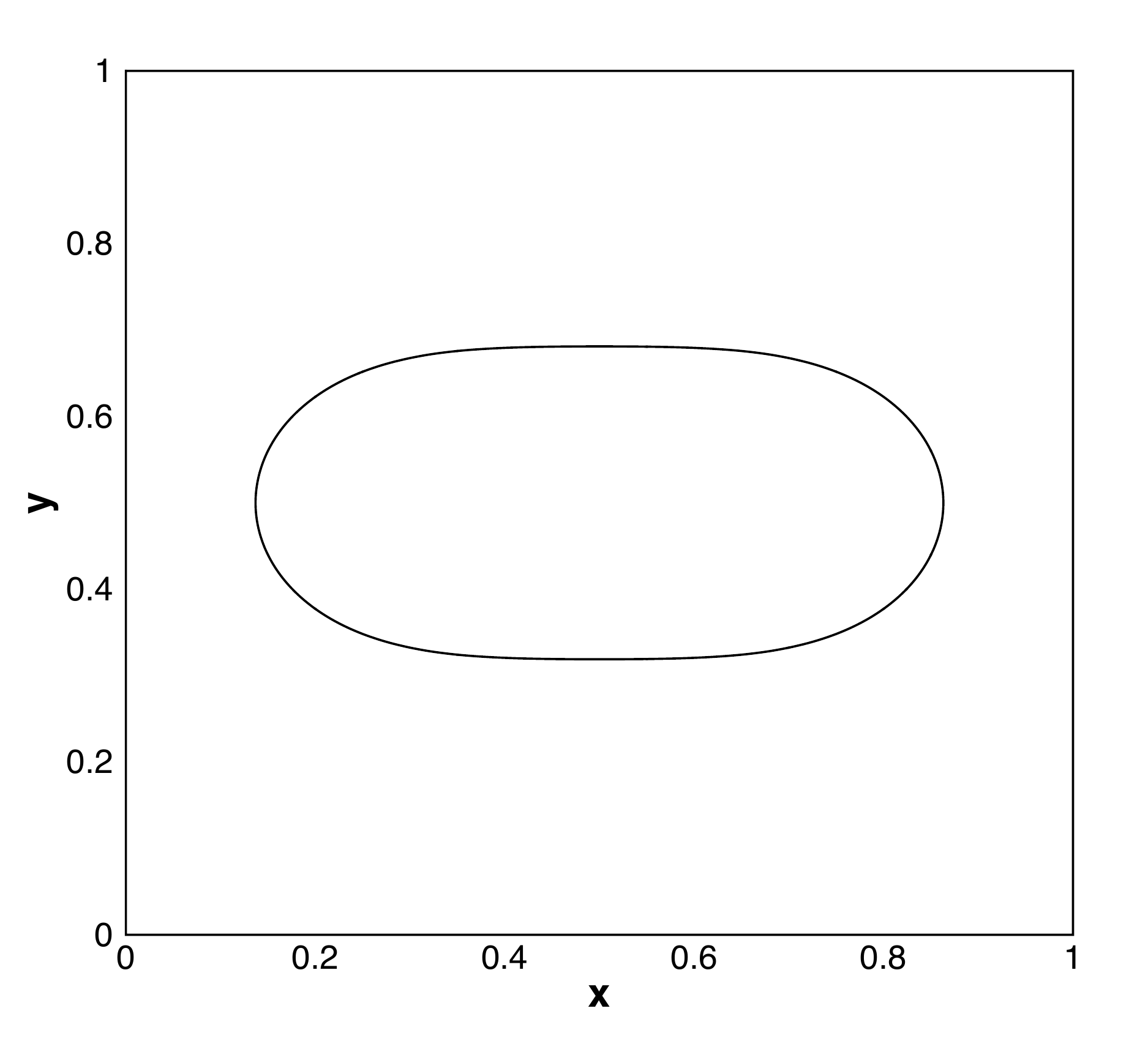}}
  }
  \centerline{
  \subfigure[$t=70$]{ \includegraphics[scale=.25]{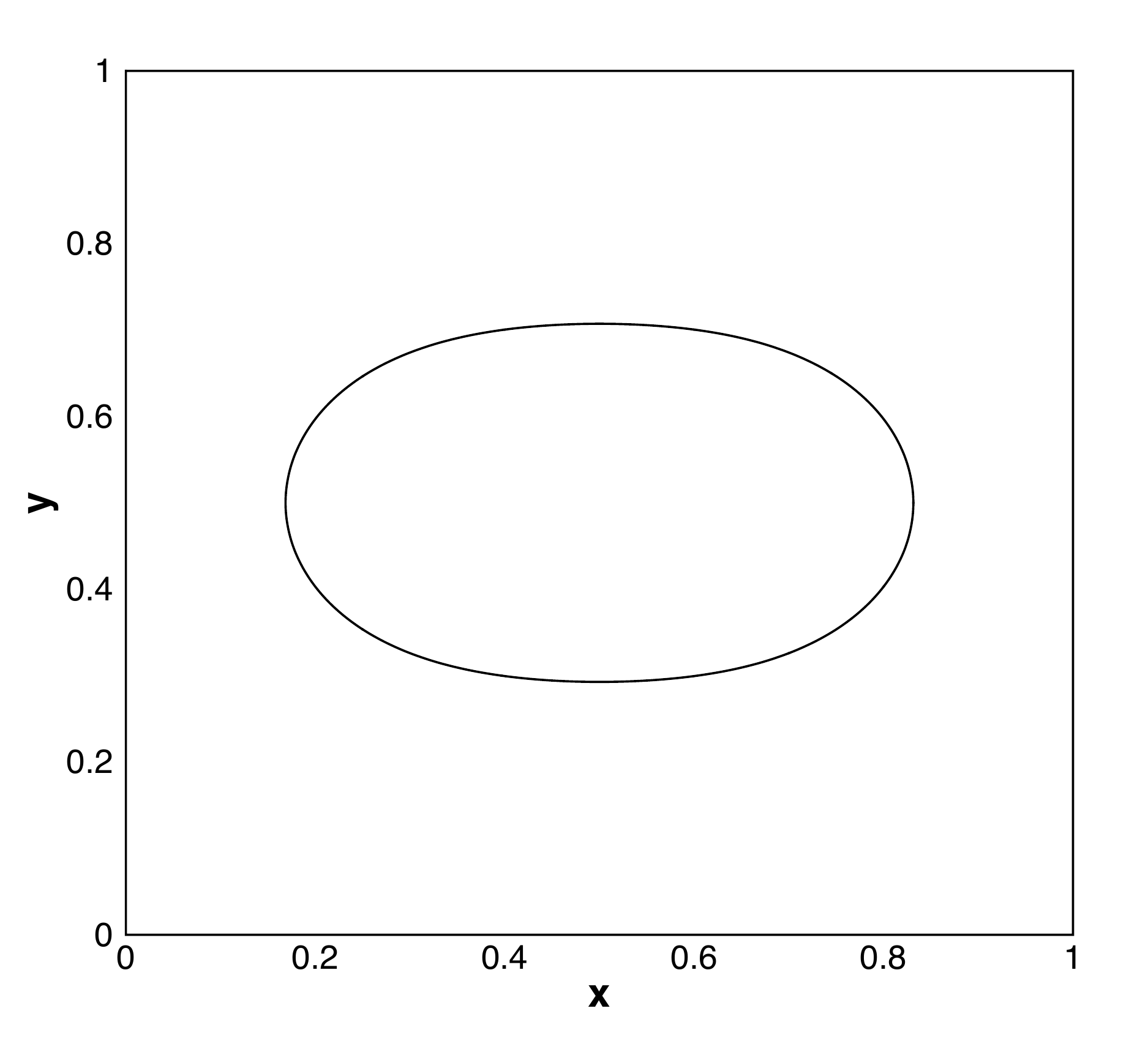}} 
    \subfigure[$t=100$]{ \includegraphics[scale=.25]{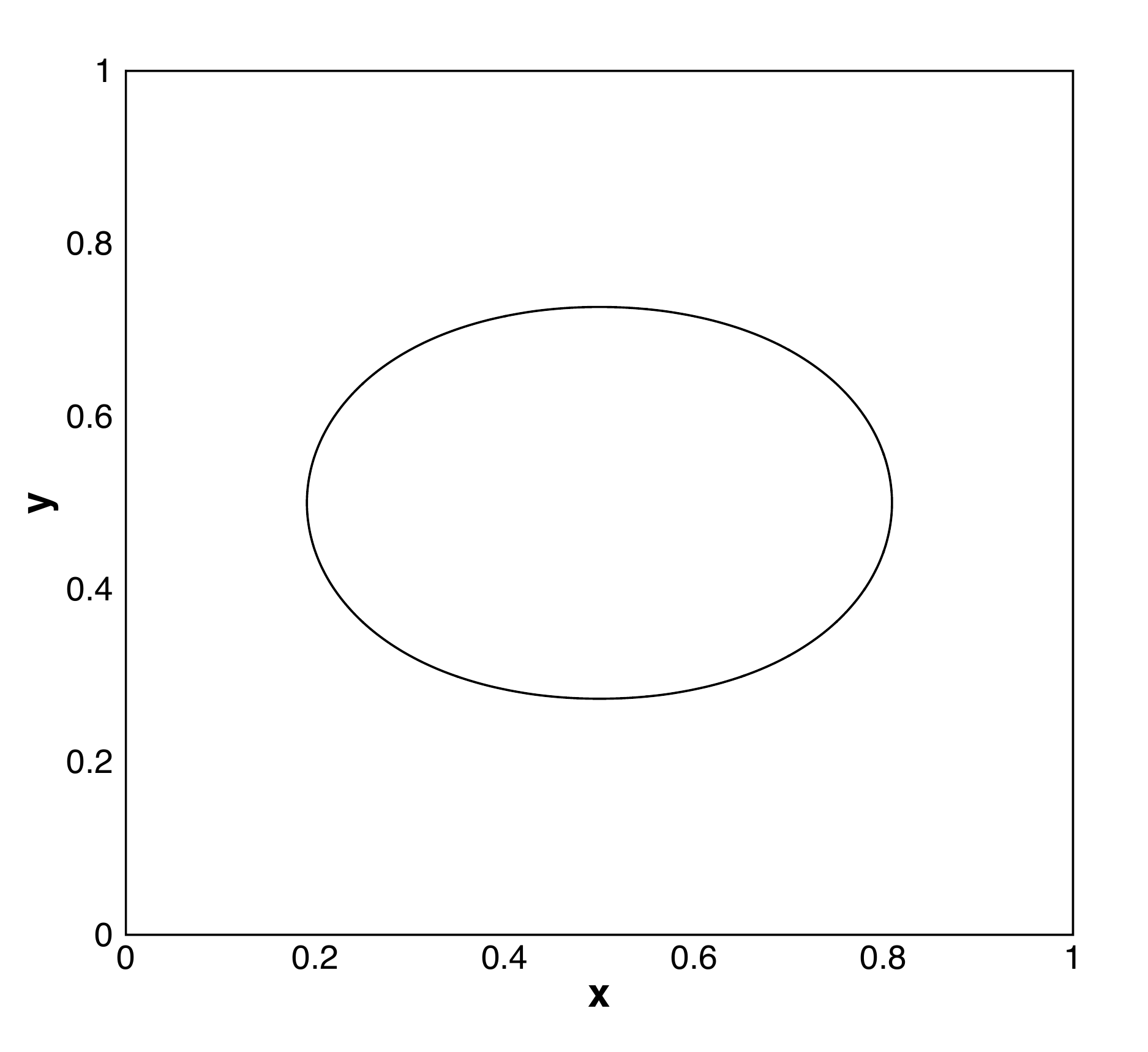}} 
    \subfigure[$t=200$]{ \includegraphics[scale=.25]{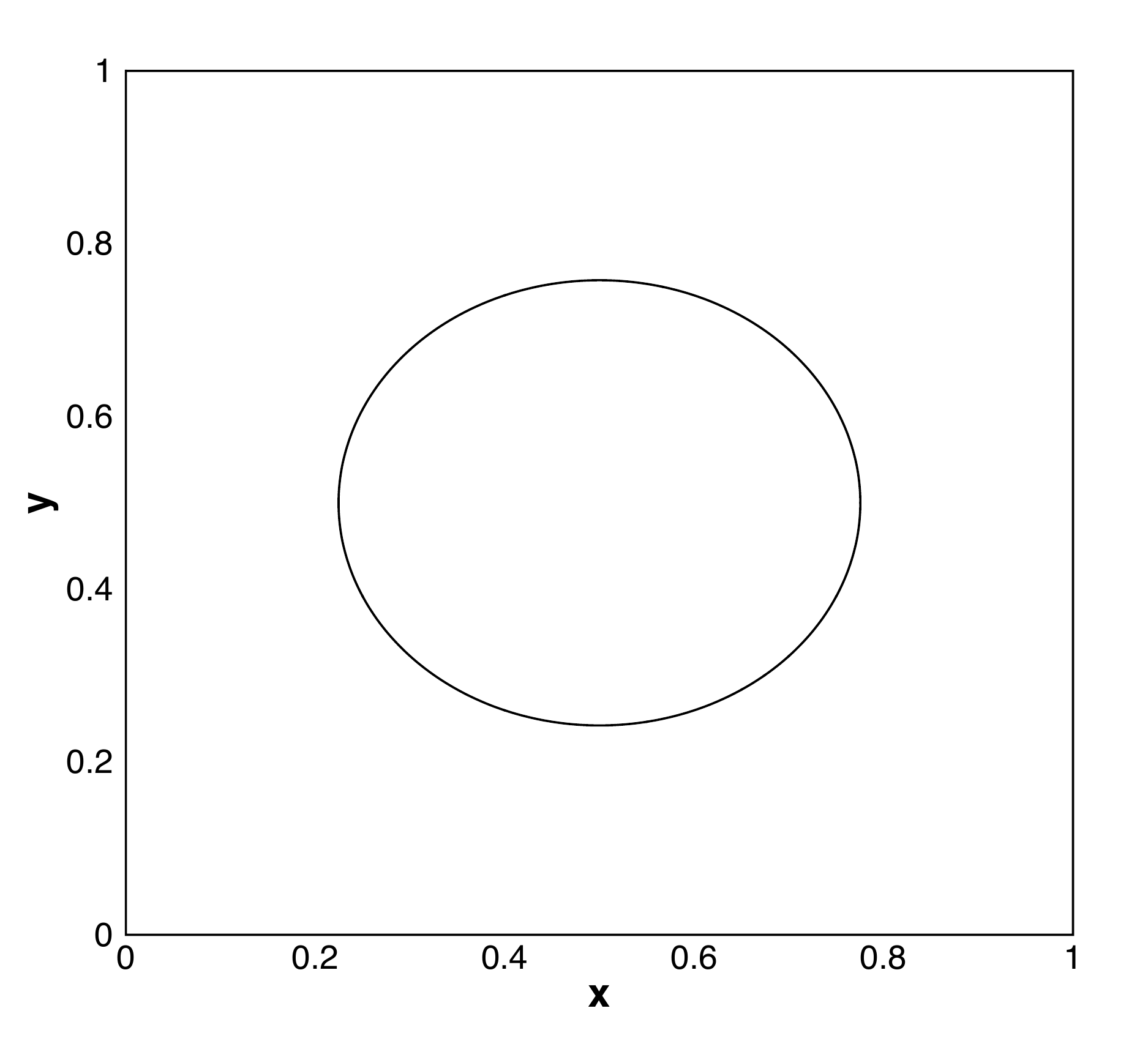}
    }
  }
  \caption{Temporal sequence of snapshots showing the coalescence
    of two circular drops. Results are obtained with the algorithm $\theta=0.75$.
  }
\label{fig:evo4}
\end{figure}

More specifically, we employ the following initial distribution for
the materials
\begin{equation}
\phi_{in}(\bs x)= 1-\tanh \frac{|\bs x-\bs x_0  |-R_0}{\sqrt{2}\eta}- \tanh \frac{|\bs x-\bs x_1  |-R_0}{\sqrt{2}\eta},
\end{equation}
where $\bs x_0=(x_0,y_0)=(0.3,0.5)$ and
$\bs x_1=(x_1,y_1)=(0.7,0.5)$ are the centers of the circular
regions for the first material,
and $R_0=0.19$ is the radius of these circles.
In the Cahn-Hilliard equation \eqref{equ:CH} we set $g=0$.
The boundary conditions \eqref{equ:wbc_1} and \eqref{equ:wbc_2}
with $g_a=0$ and $g_b=0$
are imposed on the domain boundaries.
The other simulation parameters follow those given
in \eqref{equ:drop_evolve_param}.

\begin{figure}[tbp]
  \centering
 \subfigure[$\theta=0.75$]{ \includegraphics[scale=.38]{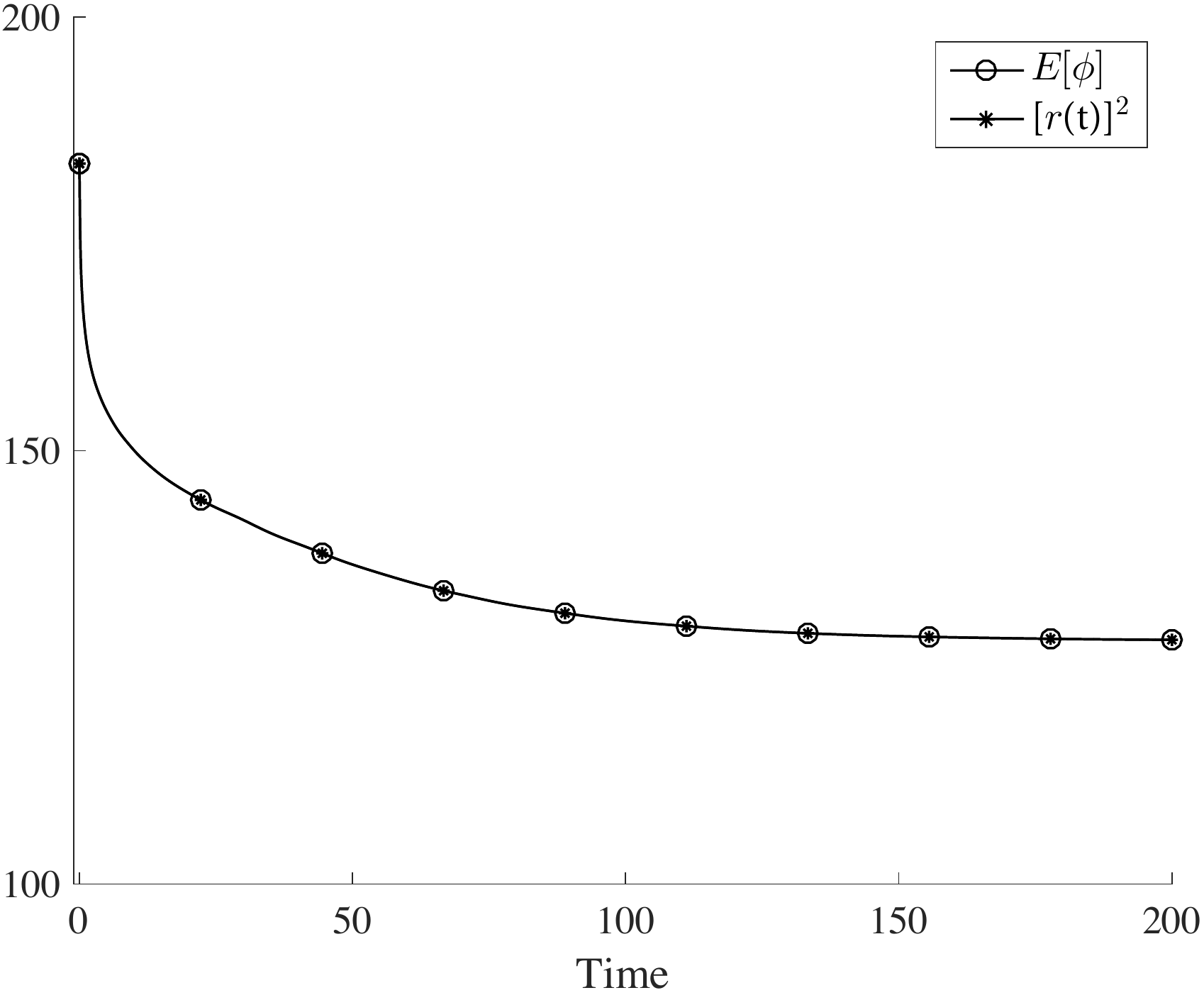}} 
 \subfigure[$\theta=1.25$]{ \includegraphics[scale=.38]{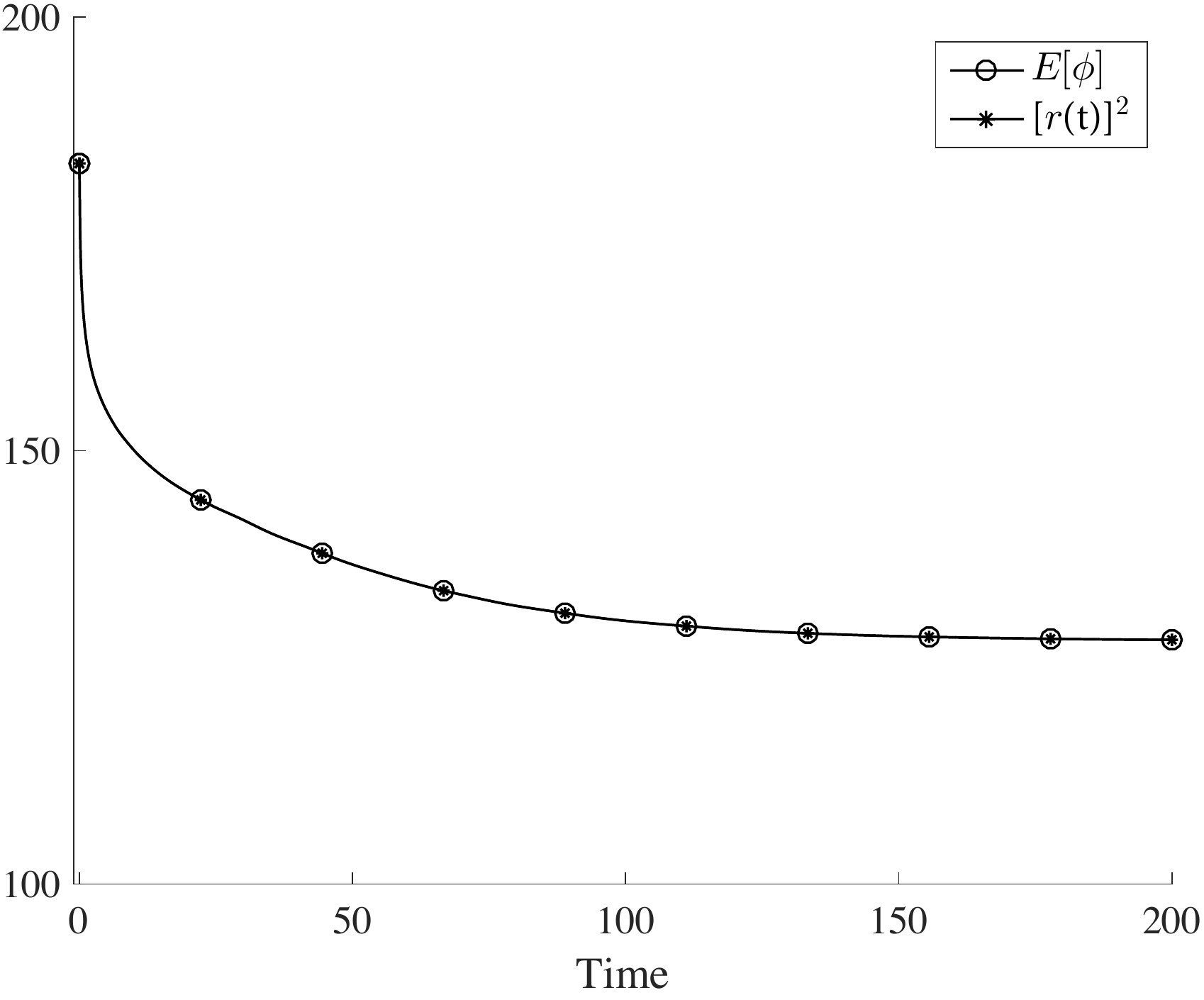}} 
 \caption{
   Time histories of $E[\phi]$ and $[r(t)]^2$ for the coalescence
   of two drops, computed using the algorithm
   (a) $\theta=0.75$, and (b) $\theta=1.25$.
 }
\label{fig:evo5}
\end{figure}

The process of coalescence of the
two drops is demonstrated in Figure \ref{fig:evo4}
with a temporal sequence of snapshots of the interfaces 
between the materials (visualized by the contour level $\phi=0$).
These results are computed using the algorithm $\theta=0.75$,
with a time step size $\Delta t=0.001$.
Figure \ref{fig:evo5} shows the time histories
of $E[\phi]$ and $[r(t)]^2$ computed using the algorithms
corresponding to $\theta=0.75$ and $\theta=1.25$.
The general characteristics for these history signals are similar to
those for the case with a single drop.
Both $E[\phi]$ and $[r(t)]^2$ are observed to decrease
over time, and their history curves overalp with
each other.


\subsection{Spinodal Decomposition}

In this section we consider the spinodal decomposition of
a homogeneous mixture into two coexisting phases governed
by the Cahn-Hilliard equation as another test of
the algorithms developed herein.

\begin{figure}[tbp]
  \centerline{
 \subfigure[$t=0$]{ \includegraphics[scale=.19]{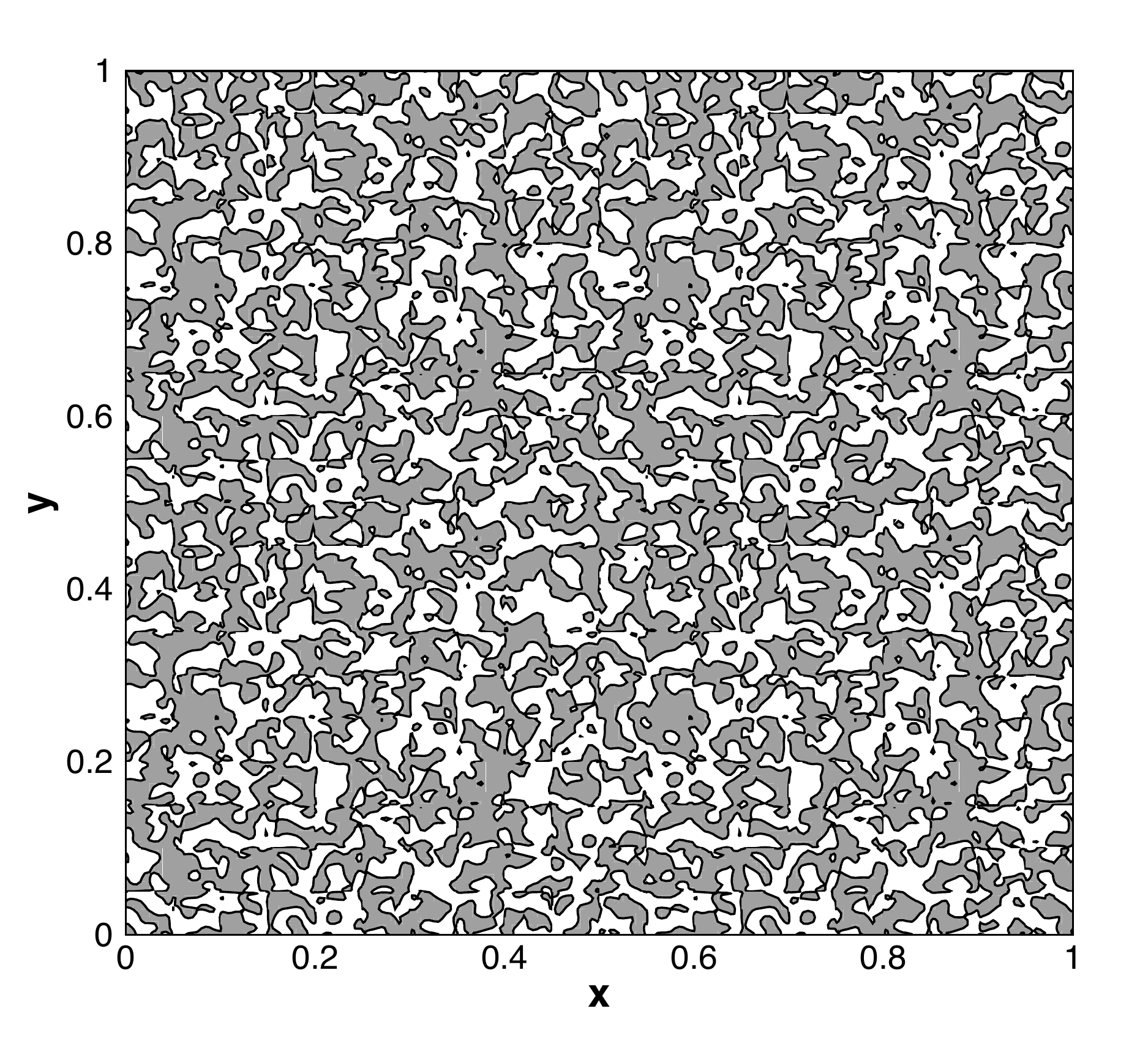}} 
 \subfigure[$t=0.1$]{ \includegraphics[scale=.19]{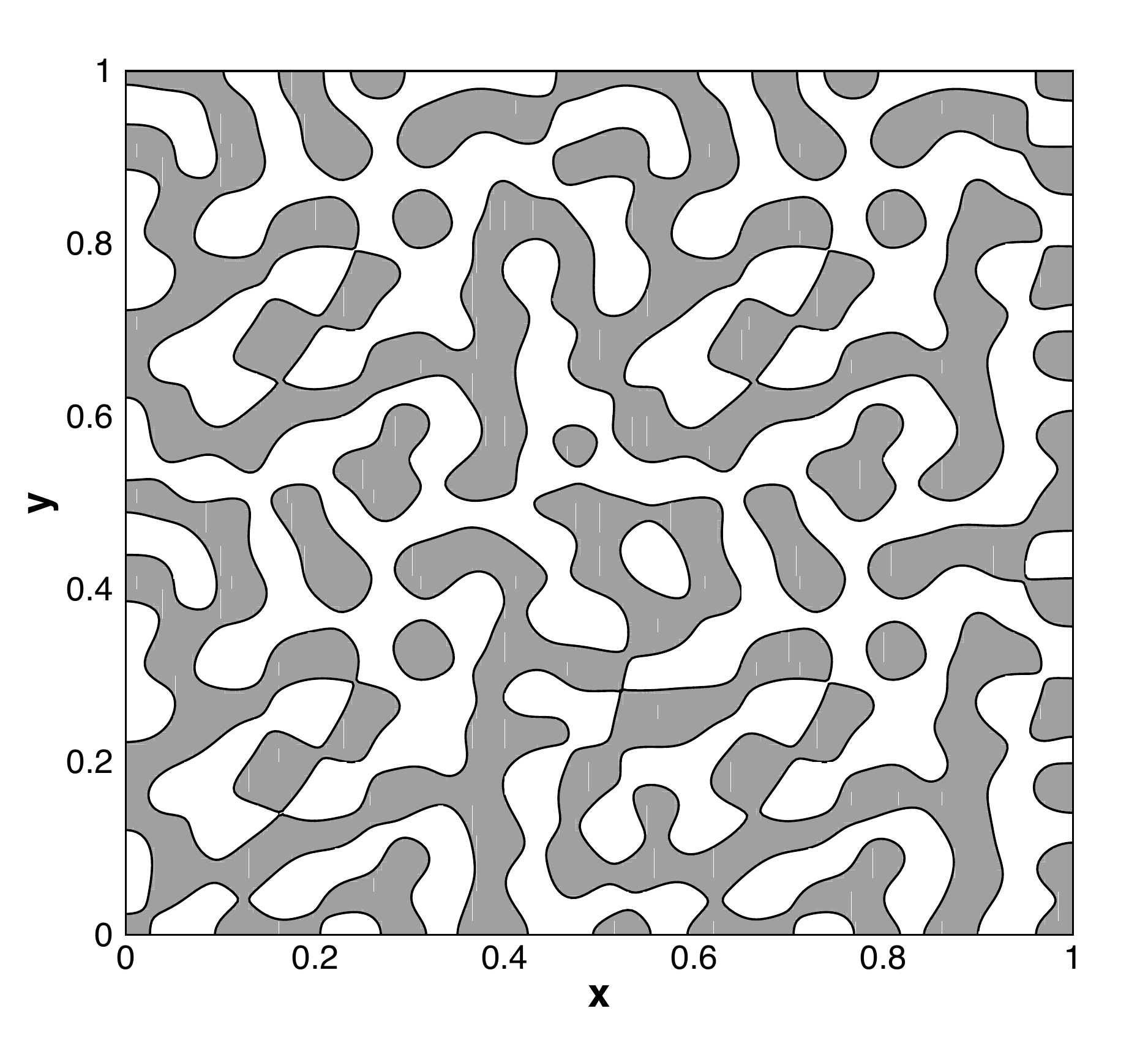}} 
 \subfigure[$t=0.2$]{ \includegraphics[scale=.19]{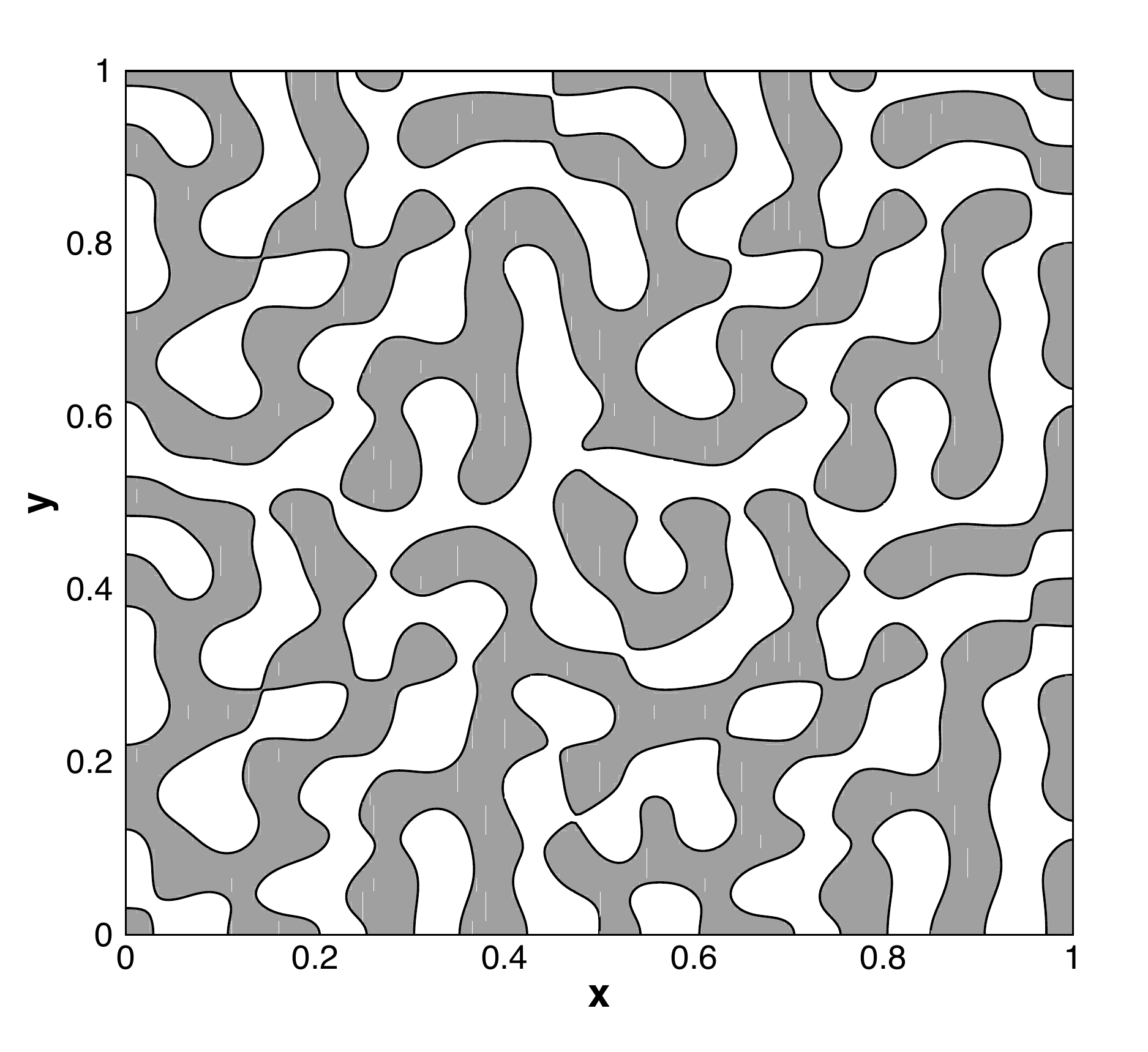}} 
 \subfigure[$t=0.5$]{ \includegraphics[scale=.19]{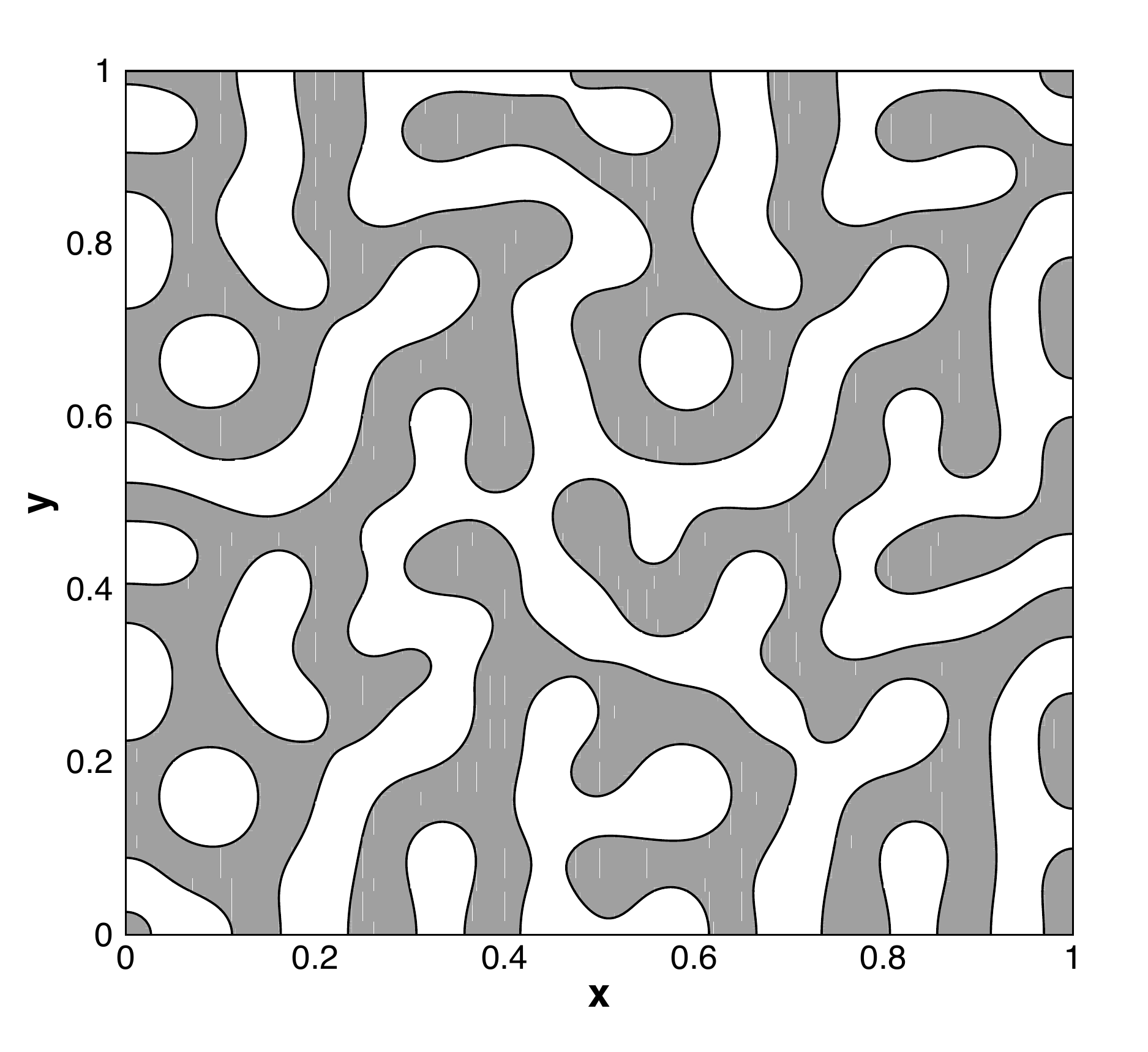}}
  }
  \centerline{
 \subfigure[$t=2$]{ \includegraphics[scale=.19]{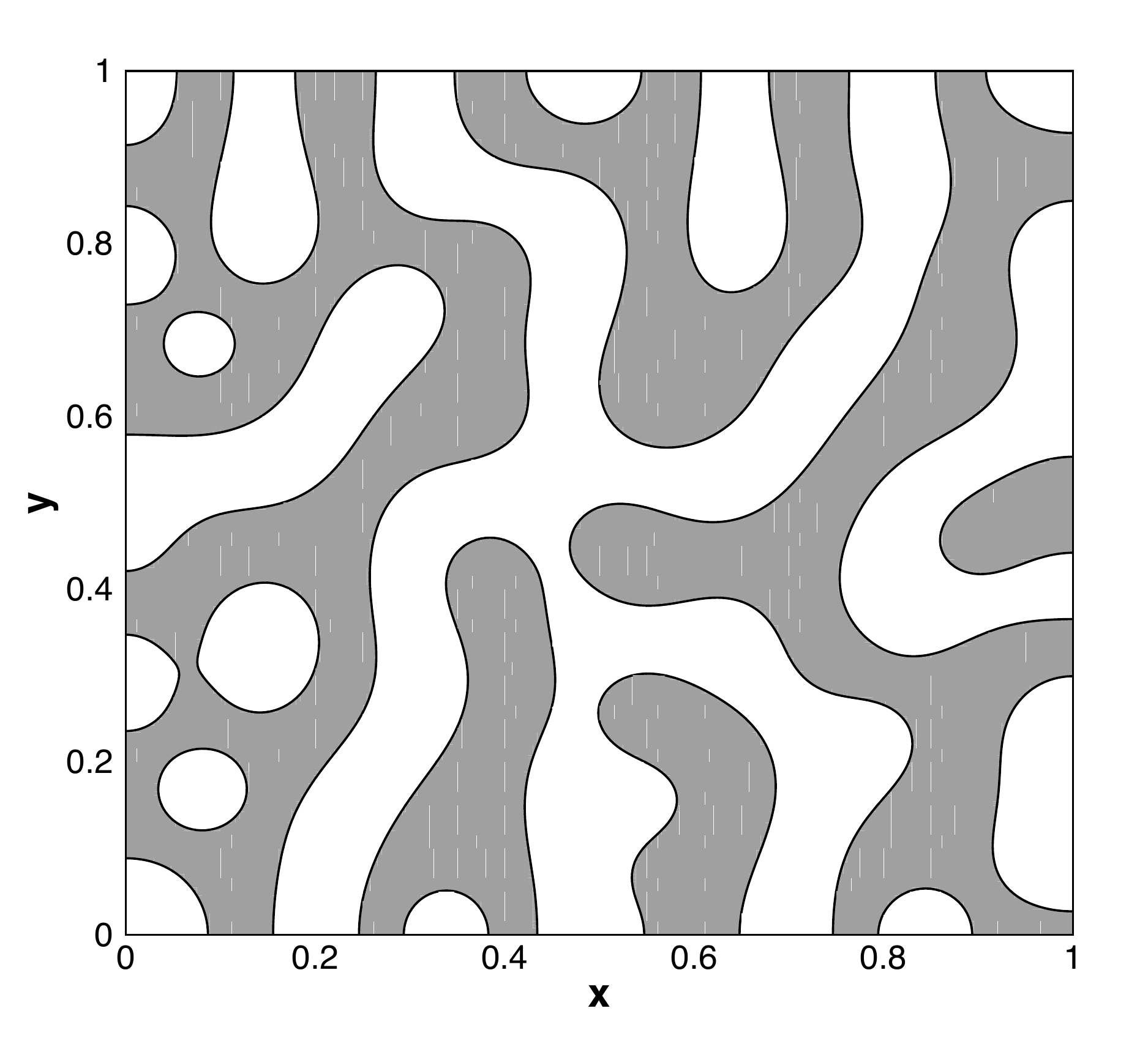}} 
 \subfigure[$t=4$]{ \includegraphics[scale=.19]{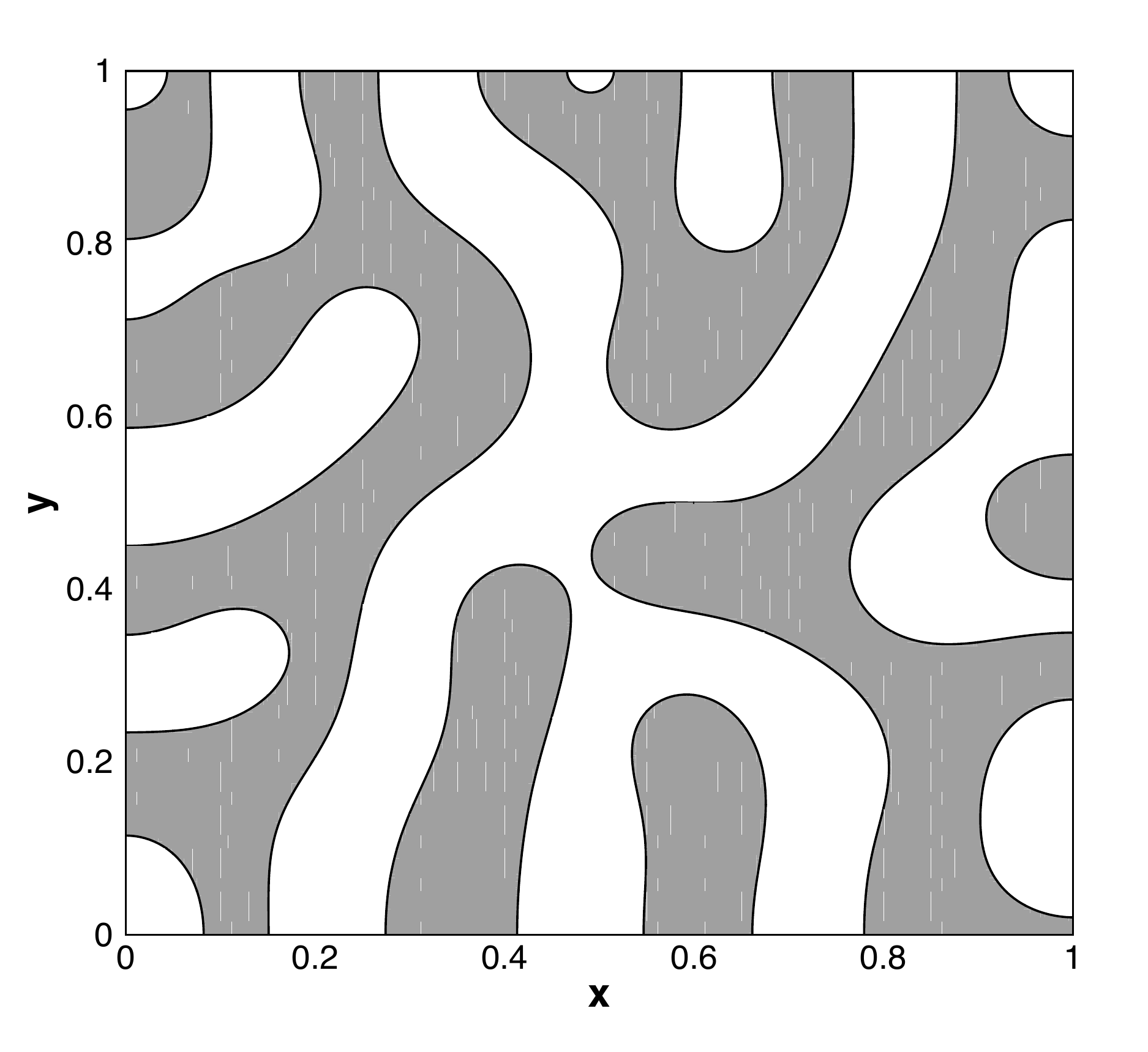}} 
 \subfigure[$t=7$]{ \includegraphics[scale=.19]{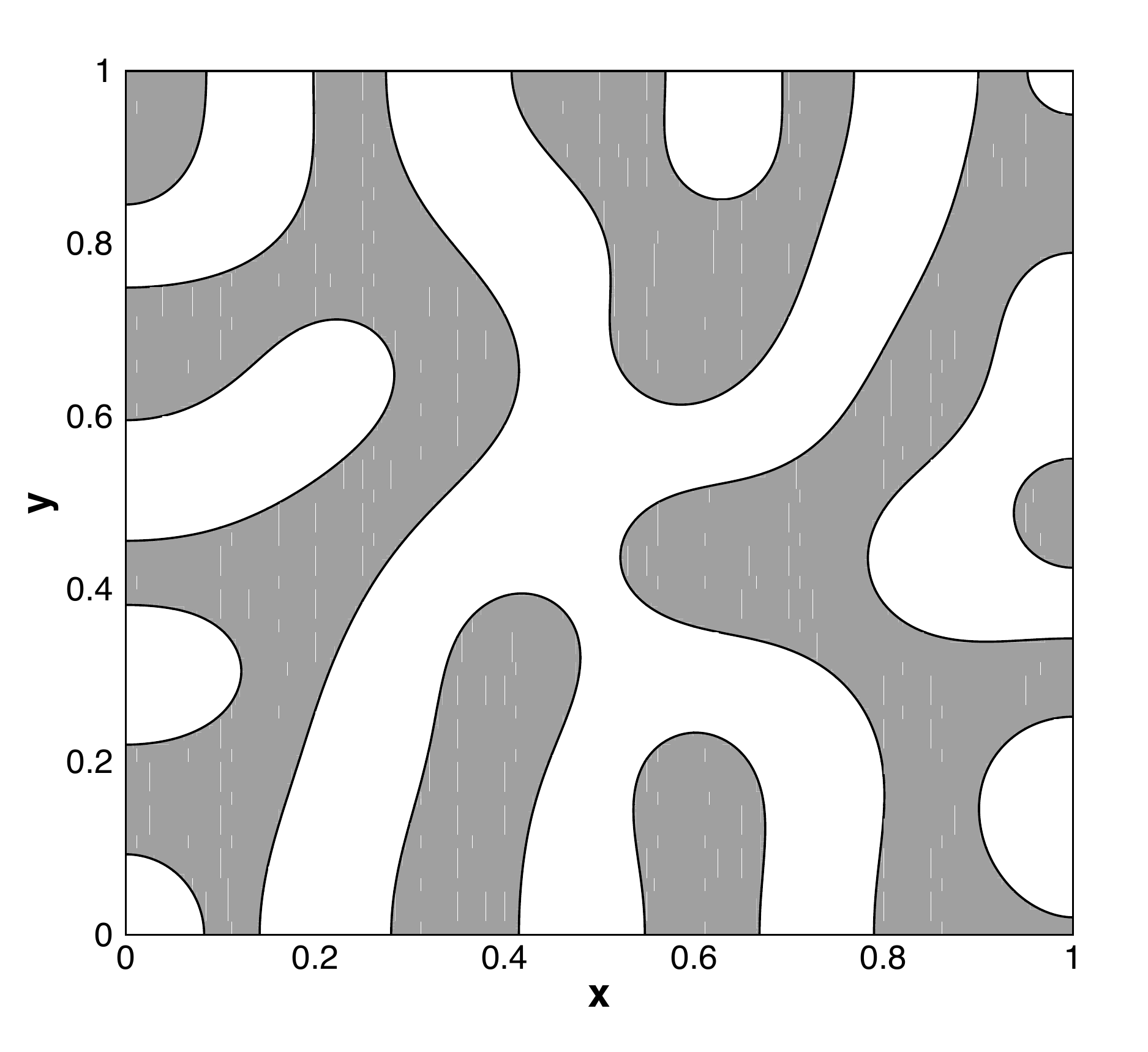}} 
 \subfigure[$t=10$]{ \includegraphics[scale=.19]{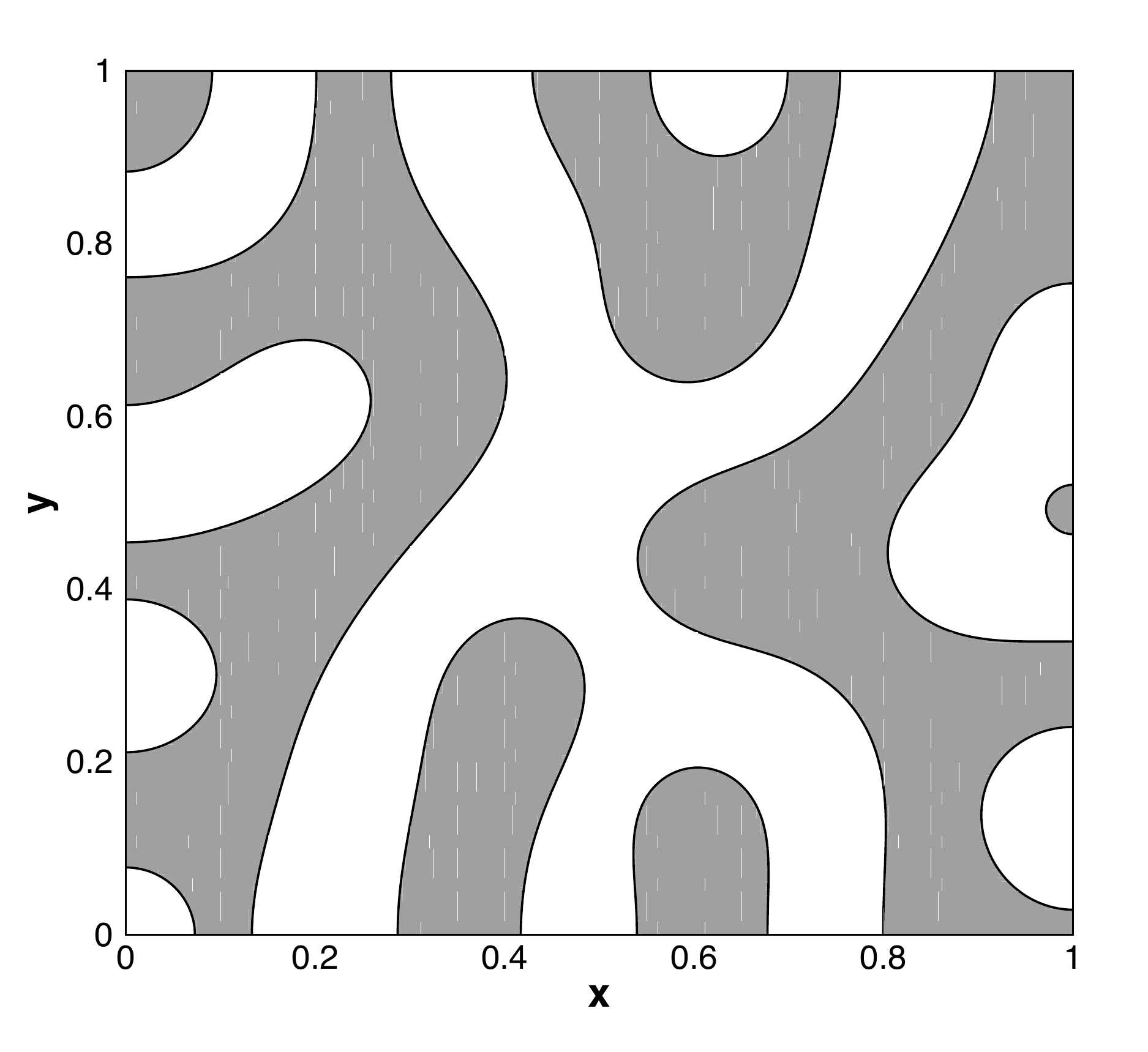}}
 }
  \caption{Spinodal decomposition: temporal sequence of snapshots of
    the interfaces visualized by $\phi=0$.
    Simulation results are obtained using the algorithm $\theta=1.25$.
  }
\label{fig:evo6}
\end{figure}

Consider the domain $\Omega=\{\ (x,y)\ |\ 0\leqslant x,y \leqslant 1\  \}$, and
a homogeneous mixture of two materials with a random initial
distribution (see Figure \ref{fig:evo6}(a)).
The evolution of the materials is assumed be described by
the Cahn-Hilliard equation \eqref{equ:CH} with $g(\mathbf{x},t)=0$,
and the goal is to simulate this evolution process.

We simulate this problem using the algorithms from Section \ref{sec:method}.
The domain is discretized using $400$ quadrilateral elements,
with $20$ uniform elements along both $x$ and $y$ directions.
The boundary conditions \eqref{equ:wbc_1} and \eqref{equ:wbc_2}
with $g_a=0$ and $g_b=0$
are imposed on the domain boundaries.
The initial random distribution of $\phi_{in}(\mathbf{x})$ is generated
using a random number generator from the
standard library of C language (see Figure \ref{fig:evo6}(a)).
The following simulation parameter values are employed
for this problem:
\begin{equation}
  \left\{
  \begin{split}
    &
    \eta = 0.01, \quad \lambda=0.001, \quad m=0.001, \\
    &
    C_0=0, \quad
    \Delta t = 0.001, \quad
    S = \sqrt{\frac{4\gamma_0\lambda\omega_0}{m\Delta t}}, \\
    &
    \text{element order:} \ 8, \quad \text{number of elements:} \ 400, \\
    &
    \theta= \ 0.75,\ 1.0,\ 1.25.
  \end{split}
  \right.
  \label{equ:spin_param}
\end{equation}

Figure \ref{fig:evo6} shows the typical evolution process of
the mixture with a temporal sequence of snapshots of
the interfaces formed between the two phases. The interface
is visualized by the contour level $\phi(\mathbf{x},t)=0$.
The lighter regions represents the first phase and
the darker region represents the second phase.
These results are obtained using the algorithm $\theta=1.25$.
It can be observed that two phases emerge from the initially
homogeneous distribution of the mixture. Over time
the grains of the two phases become increasingly coarser,
and a certain pattern can be observed from the two
regions.

\begin{figure}[tbp]
  \centering
 \subfigure[$\theta=0.75$]{ \includegraphics[scale=.27]{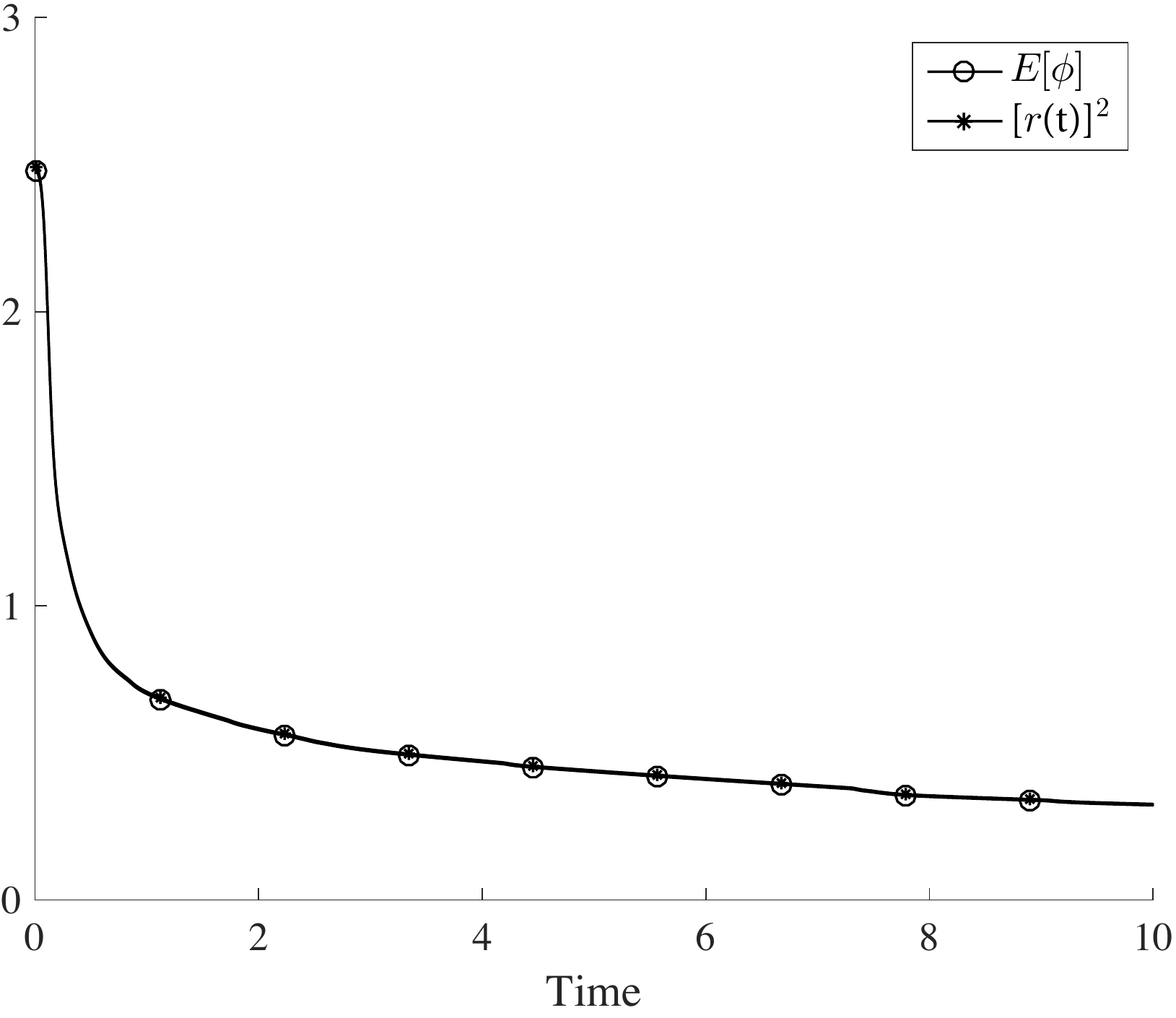}} 
 \subfigure[$\theta=1$]{ \includegraphics[scale=.27]{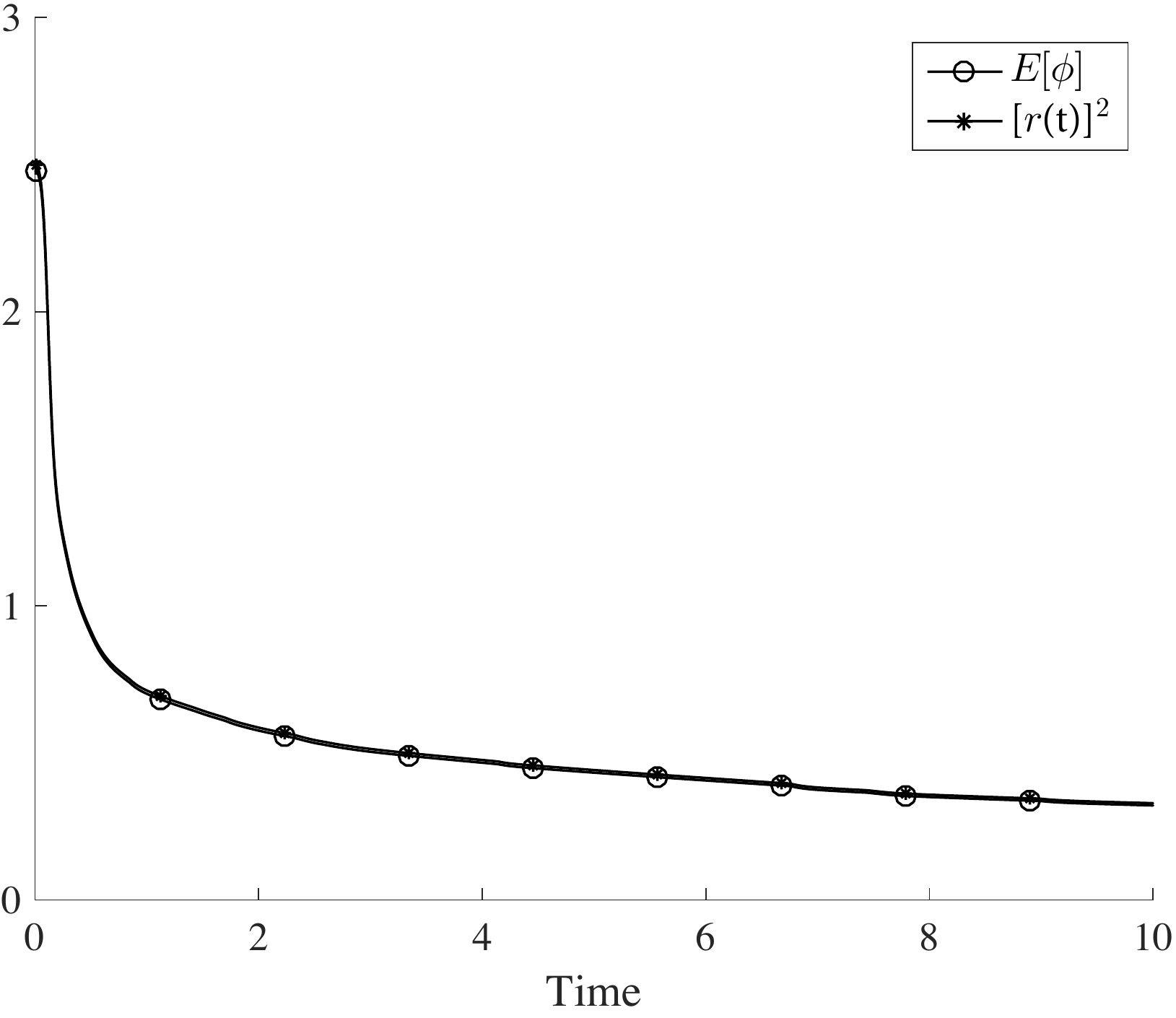}} 
  \subfigure[$\theta=1.25$]{ \includegraphics[scale=.27]{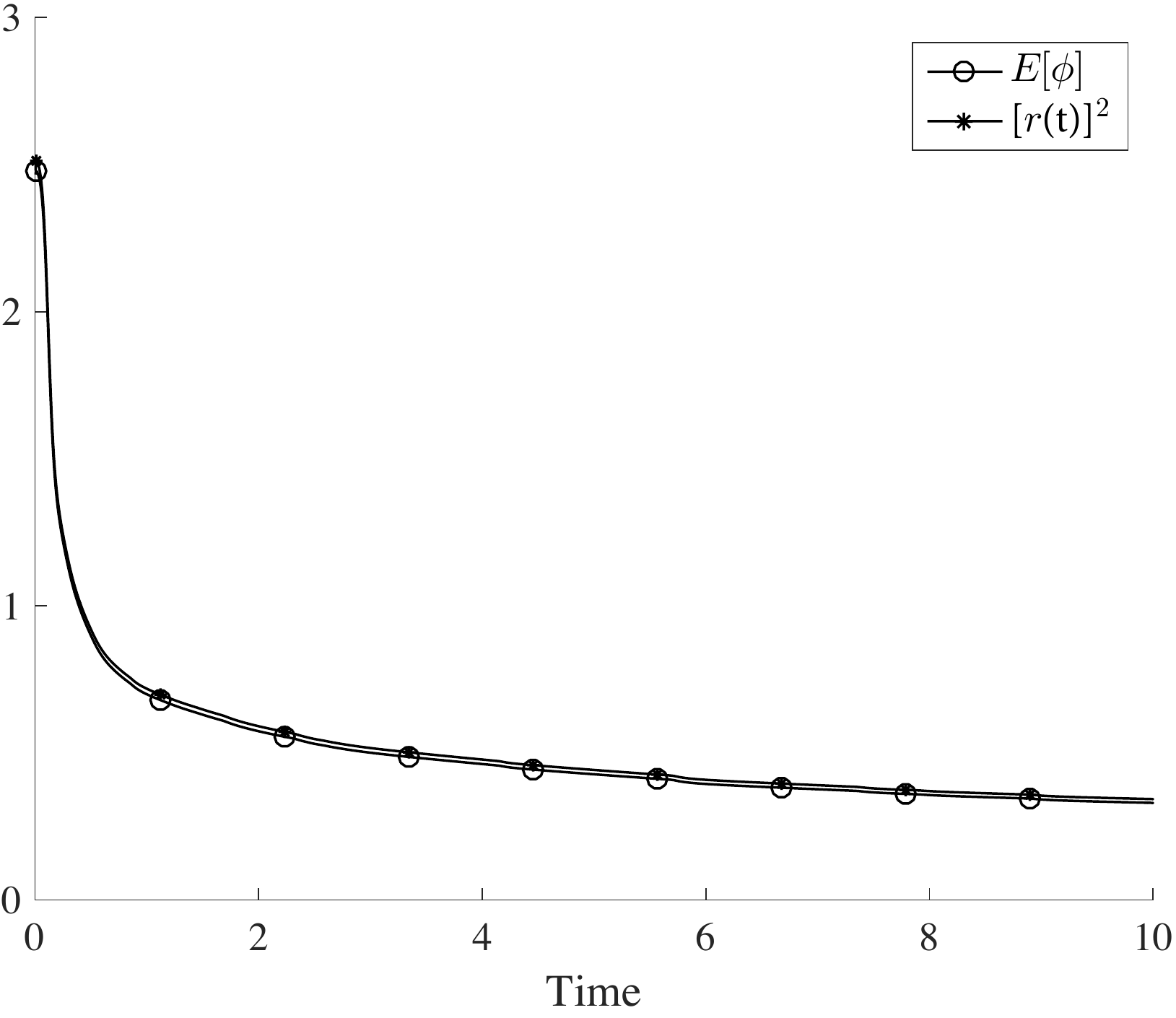}} \\
  \caption{
    Time histories of $E[\phi]$ and $[r(t)]^2$ for spinodal decomposition
    obtained with several algorithms corresponding to (a) $\theta=0.75,$ (b) $\theta=1,$ (c) $\theta=1.25.$
  }
\label{fig:evo7}
\end{figure}

Figure \ref{fig:evo7} shows the time histories of
the potential free energy $E[\phi]$ and the variable
$[r(t)]^2$ of the system obtained using three algorithms,
corresponding to $\theta=0.75$, $\theta=1$ and
$\theta=1.25$. We note that the history curves for
$E[\phi]$ and $[r(t)]^2$ essentially overlap with each other,
and that the results obtained with different algorithms
agree well with one another.



\subsection{Two-Phase Flow: Rising Air Bubble in Water}

As another test case, in this section we demonstrate the performance of
the algorithm developed herein in the context of a two-phase
flow solver, and simulate the two-phase flow of an air bubble
rising through the water. 

Following \cite{DongS2012,Dong2012}, we consider a system consisting
of two immiscible incompressible fluids, and combine
the Cahn-Hilliard equation and the incompressible Navier-Stokes equations
with variable density and variable viscosity
to model such a system; see \cite{DongS2012,DongW2016} for details.
We then combine the family of algorithms presented in Section \ref{sec:method}
for the Cahn-Hilliard equation and the algorithm from \cite{DongS2012}
for the momentum equations to form an overall method for numerically
solving the coupled system of Cahn-Hilliard and Navier-Stokes
equations. Note that the algorithm for the momentum equation
employed here is a semi-implicit type scheme and is only conditionally
stable~\cite{DongS2012}. So the overall algorithm for the two-phase
flows is not energy stable.

\begin{table}[tbp]
\centering 
\begin{tabular}{l l l } 
\hline 
Density [$kg/m^3$]:& air - $\rho_1=1.204$ & water - $\rho_2= 998.207$\\
Dynamic viscosity [$kg/(m\cdot s)$]: &air  - $\mu_1=1.78\times 10^{-5}$   & water - $\mu_2=1.002\times 10^{-3}$   \\
Surface tension [$kg/s^2$]:& air/water - $\sigma=7.28\times 10^{-2}$ &   \\
Gravity [$m/s^2$]:&$g_r=9.8$&\\
\hline 
\end{tabular}
\caption{Physical property values of air and water.}
\label{table:airbubble} 
\end{table}

\begin{figure}[tbp]
  \centerline{
 \subfigure[$t=0.05$]{ \includegraphics[scale=.21]{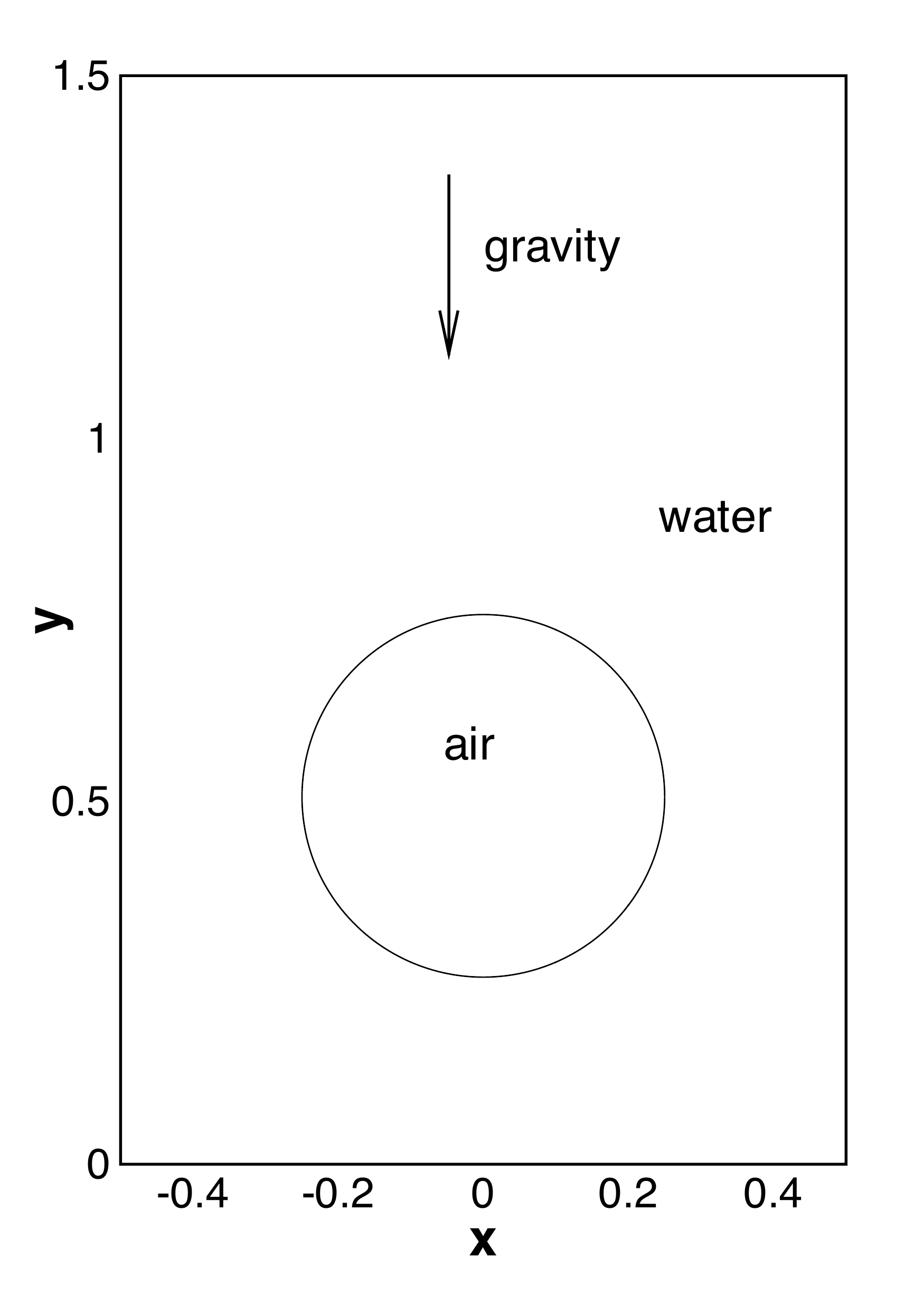}} 
  \subfigure[$t=0.4$]{ \includegraphics[scale=.21]{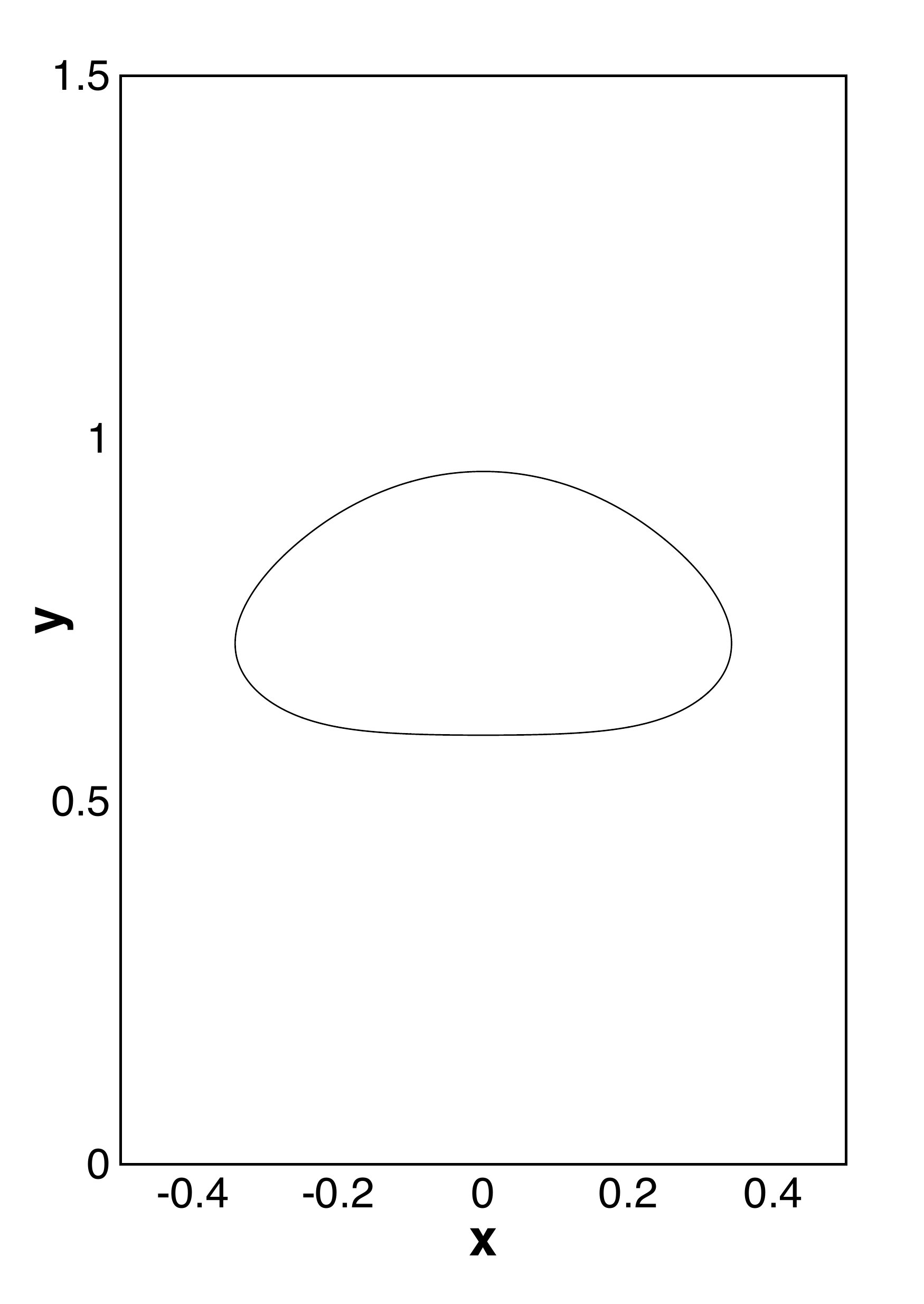}} 
   \subfigure[$t=0.8$]{ \includegraphics[scale=.21]{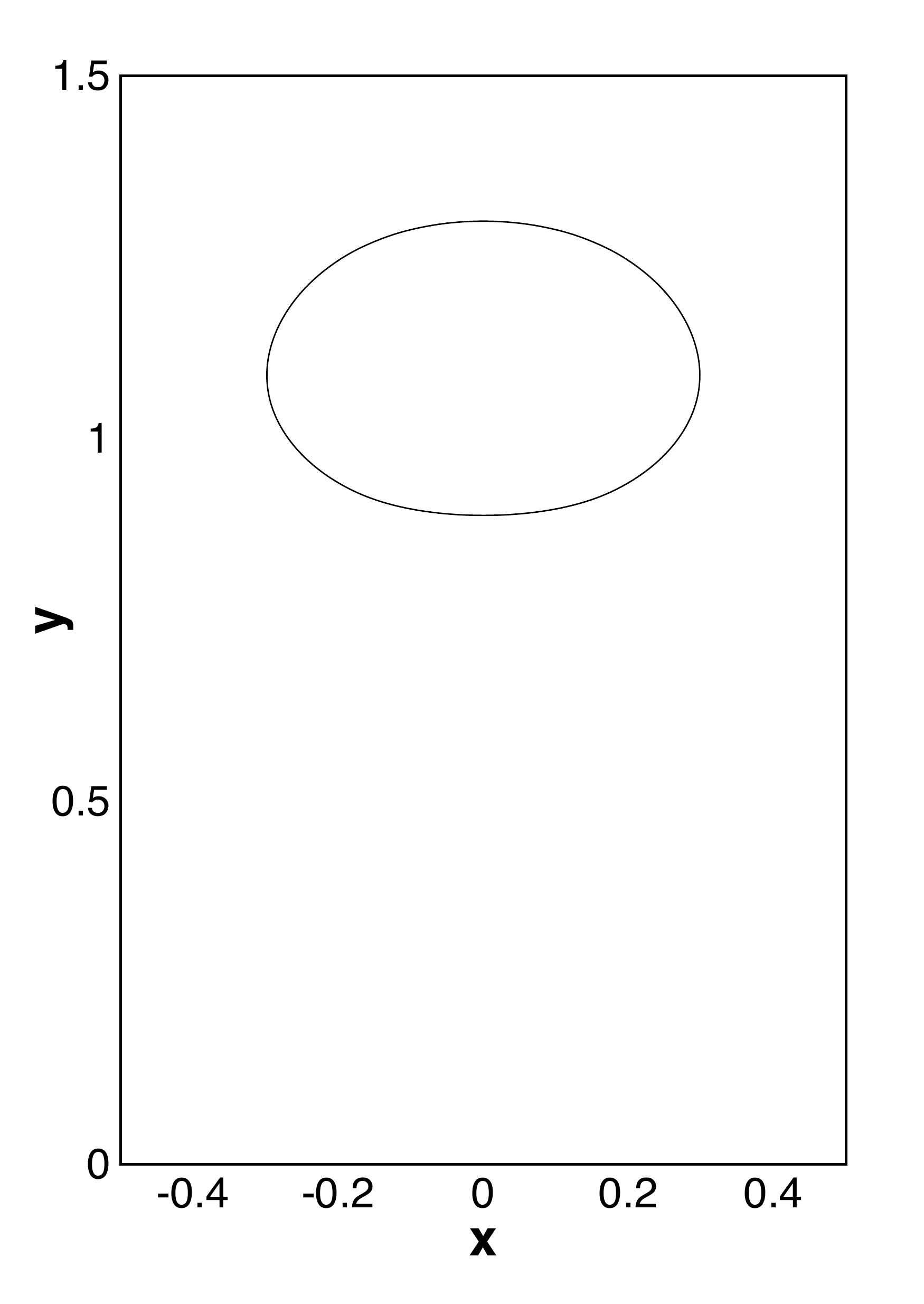}} 
   \subfigure[$t=1.15$]{ \includegraphics[scale=.21]{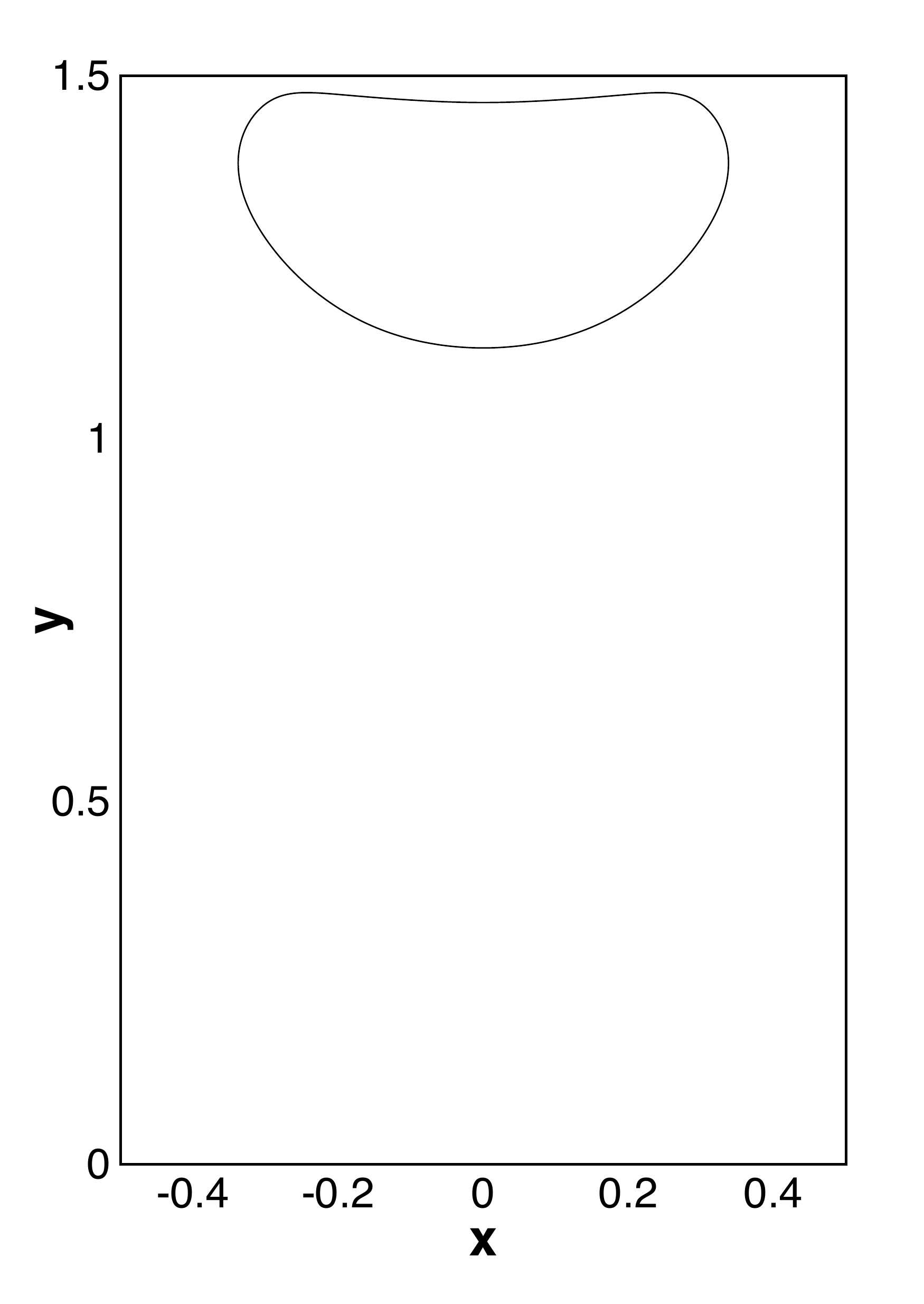}}
  }
  \centerline{
    \subfigure[$t=1.4$]{ \includegraphics[scale=.21]{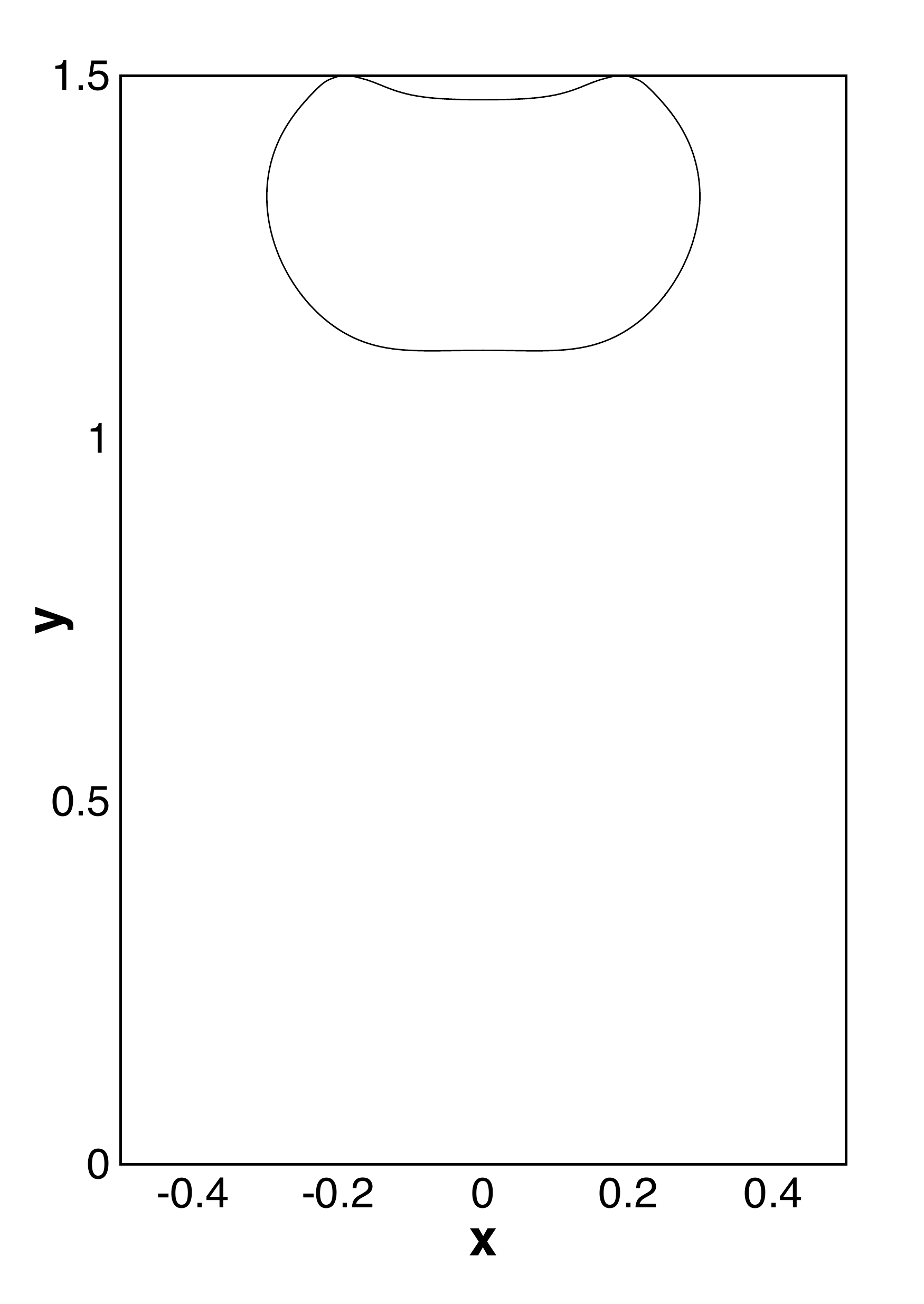}}
     \subfigure[$t=1.45$]{ \includegraphics[scale=.21]{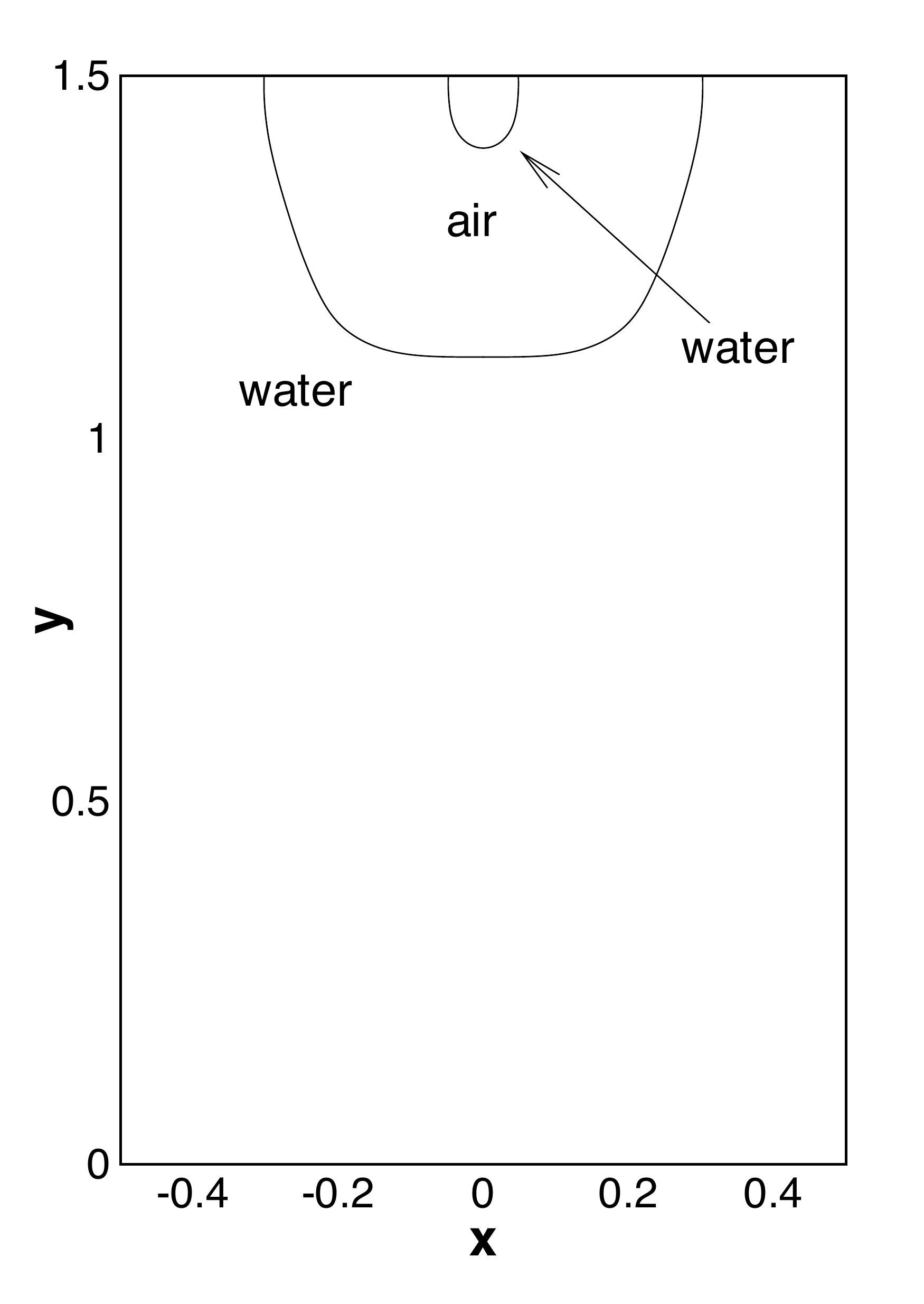}}
     \subfigure[$t=1.7$]{ \includegraphics[scale=.21]{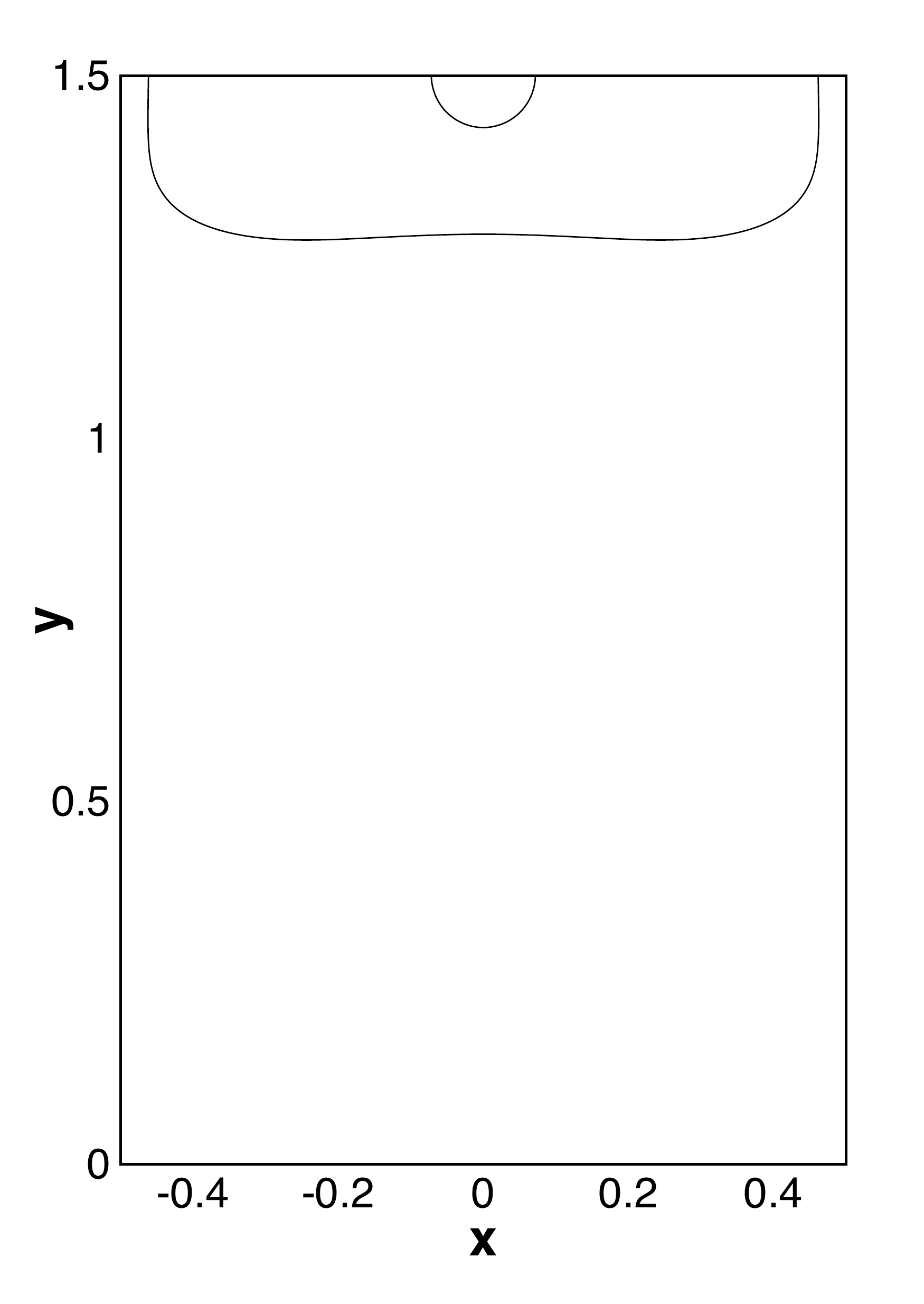}}
     \subfigure[$t=2.05$]{ \includegraphics[scale=.21]{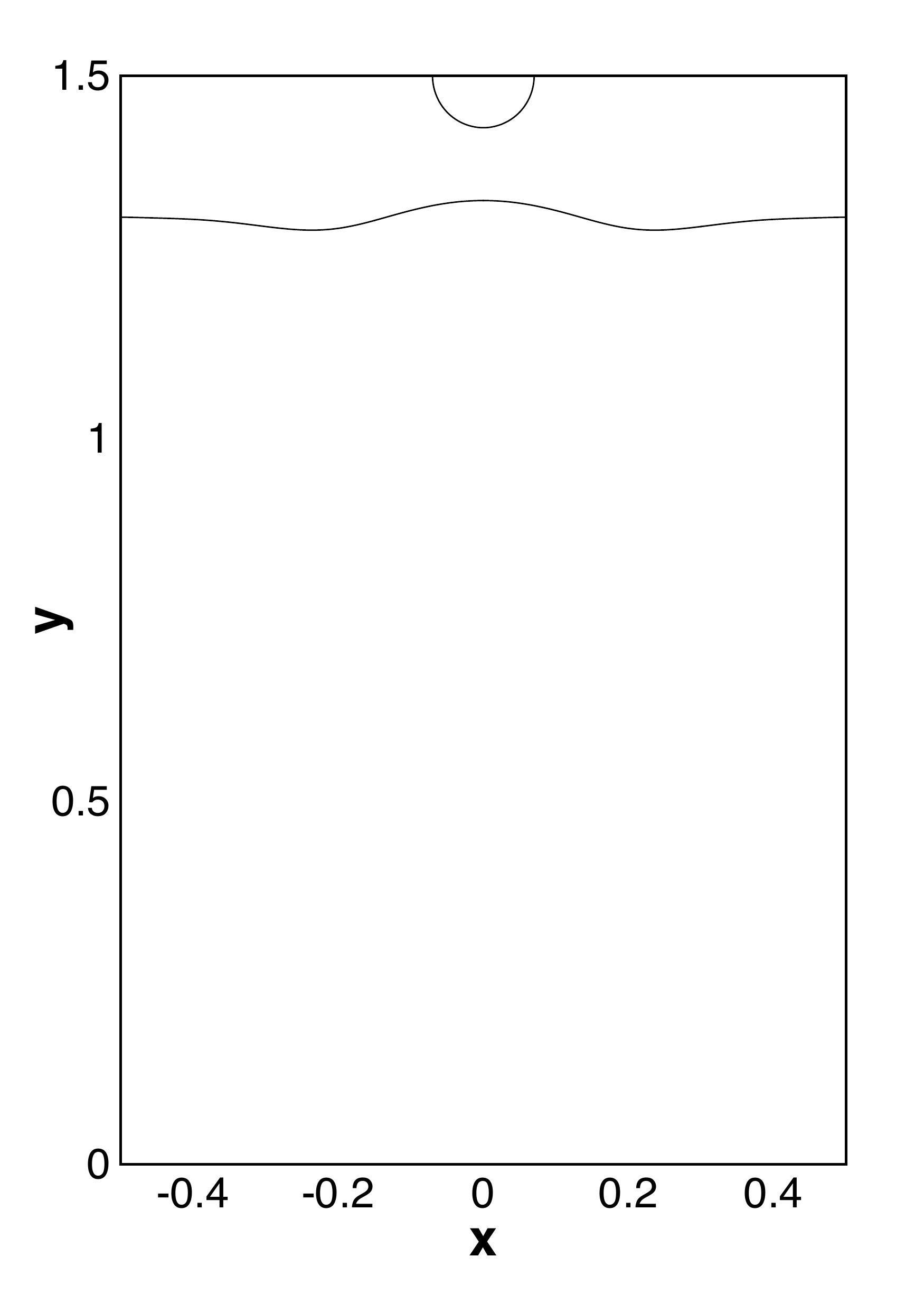}}
  }
  \caption{
    Time sequence of snapshots of an air bubble (initially circular)
    rising in water within a solid container,
    showing the bubble breakup on the upper wall and the formation
    of an air cavity.
}
\label{fig:evo8}
\end{figure}


We consider a solid container occupying the domain
$-\frac{L}{2}\leqslant x\leqslant \frac{L}{2}$
and $0\leqslant y\leqslant\frac{3}{2}L$, where $L=1cm$
(see Figure \ref{fig:evo8}(a)).
The container is filled with
water, and an air bubble is trapped inside the water.
The air bubble is initially circular with a
diameter $2R_0=0.5L=0.5cm$ and its center located at
$\mathbf{x}_0=(x_0,y_0)=(0,0.5L)$,
and it is held at rest.
The container walls are assumed to have a neutral wettability
($90$-degree contact-angle), and
the gravity is assumed to point downward.
At $t=0$, the air bubble is released, and starts to rise
through the water due to buoyancy. The bubble eventually
breaks up on the upper wall and forms an air cavity
at the top of the container.
The goal is to simulate this process.

The physical parameters for the air and water are summarized
in Table \ref{table:airbubble}. 
We choose $L$ as the characteristic length scale,
the air density as the characteristic density scale $\varrho_0$,
and $U_0=\sqrt{g_{r0}L}$ ($g_{r0}=1m/s^2$) as the characteristic
velocity scale. All the physical parameters are then
normalized according to Table \ref{tab:normalization}.

In the simulations the domain is discretized using
$600$ quadrilateral elements,
with $20$ and $30$ uniform elements in $x$ and $y$
directions, respectively. An element order $8$
is employed in the simulations.
We impose the boundary conditions \eqref{equ:wbc_1}
and \eqref{equ:wbc_2} with $g_a=0$ and $g_b=0$ for
the phase field function $\phi$ and the no-slip condition
for the velocity on the container walls.
The initial velocity is assumed to be zero, and the initial
distribution of the phase field function is given by
\begin{equation}
\phi_{in}(\bs x)=-\tanh\frac{|\bs x-\bs x_0|-R_0}{\sqrt{2}\eta}.
\end{equation}
The values for the simulation parameters in this problem
are given by
\begin{equation}
  \left\{
  \begin{split}
    &
    \frac{\eta}{L} = 0.01, \quad \frac{\sigma}{\varrho_0U_0^2L}=604.6\ (\text{surface tension}), \quad
    \lambda = \frac{3}{2\sqrt{2}}\sigma\eta, \quad
    \frac{m}{L/(\varrho_0U_0)}=\frac{10^{-7}}{\lambda/(\varrho_0U_0^2L^2)}, \\
    &
    C_0=0, \quad
    \frac{\Delta t}{L/U_0} = 2.5e-5, \quad
    S = \sqrt{\frac{4\gamma_0\lambda\omega_0}{m\Delta t}}, \\
    &
    \text{element order:} \ 8, \quad \text{number of elements:} \ 600, \\
    &
    \theta= 0.75.
  \end{split}
  \right.
  \label{equ:2p_param}
\end{equation}


The dynamics of this two-phase flow is illustrated by
Figure \ref{fig:evo8}, in which we have shown a temporal
sequence of snapshots of the air-water
interface in this system. The interface is visualized by
the contour level $\phi(\mathbf{x},t)=0$.
As the system is released the air bubble
rises through the water and experiences significant
deformations (Figure \ref{fig:evo8}(b)-(d)).
The bubble exhibits the typical shape of a circular
``cap'' (Figure \ref{fig:evo8}(b)) when it is still
far away from the upper wall.
But as the bubble approaches the upper wall,
its shape is affected by the presence of
the wall significantly (Figures \ref{fig:evo8}(c)-(d)).
After the bubble touches the upper wall, it traps
a layer of water between the wall and its upper
side (Figure \ref{fig:evo8}(e)). The trapped water
becomes a water drop sitting on the upper wall
over time (Figure \ref{fig:evo8}(g)).
The interface formed between the bulk of air
and the bulk of water moves sideways on the wall,
and the air forms a cavity at the top of the container
(Figures \ref{fig:evo8}(f)-(h)).
Our method has captured this process
and the interaction between the air-water interface
and the wall.


\section{Concluding Remarks}
\label{sec:summary}


In this paper we have developed a family of second-order 
energy-stable schemes for the Cahn-Hilliard type
equations. We start with the reformulated  system of equations
based on the scalar auxiliary variable approach, and approximate this
system at time
step $(n+\theta)$, and then develop corresponding
numerical approximations that are second-order accurate
and unconditionally energy stable.
This family of approximations contains the often-used Crank-Nicolson scheme
and the second-order backward differentiation formula 
as particular cases.
We have also developed an efficient solution algorithm
to overcome the difficulty caused by the unknown scalar
auxiliary variable in the discrete system of equations resulting from
this family of schemes.
Our overall computation algorithm only requires the solution
of four de-coupled individual Helmholtz type equations within
each time step, and these equations only involve constant
and time-independent coefficient matrices that can be
pre-computed.

While the current paper is only concerned with the numerical
approximation of the Cahn-Hilliard equation, the family of
second-order energy-stable approximations is readily applicable to
other types of equations resulting from gradient flows.
When combined with the invariant energy quadratization
or scalar auxiliary variable approach, they can be
readily used to design energy-stable schemes for other
gradient flows.



\section*{Appendix A: Proof of Theorem \ref{thm:thm_1}}

We note first the following useful relation ($\chi$ denoting a 
generic scalar variable):
%
\begin{multline}
\left(\chi^{n+1}-2\chi^n+\chi^{n-1}\right)\left[
\left(\theta +\frac{1}{2} \right)\chi^{n+1}
-2\theta\chi^{n}
+\left(\theta -\frac{1}{2} \right)\chi^{n-1}
\right] \\
= \theta\left|\chi^{n+1}-2\chi^n+\chi^{n-1}  \right|^2
+ \frac{1}{2}\left(
\left|\chi^{n+1}-\chi^n  \right|^2
- \left|\chi^{n}-\chi^{n-1} \right|^2
\right),
\label{equ:relation_2}
\end{multline}
%
%
This relation can be verified by elementary operations.

Assume that $g=0$, $g_a=0$, and $g_b=0$.
Multiply $\Delta t\mathscr{H}^{n+\theta}$ to equation \eqref{equ:alg_disc_1}
and integrate over $\Omega$, and we have
\begin{equation}
\int_{\Omega} \left[
 \left(\theta + \frac{1}{2}\right)\phi^{n+1} - 2\theta\phi^n
      + \left(\theta-\frac{1}{2}\right)\phi^{n-1}
\right]\mathscr{H}^{n+\theta}
=m\Delta t\int_{\Omega}\mathscr{H}^{n+\theta}\nabla^2\mathscr{H}^{n+\theta}.
\label{equ:int_1}
\end{equation}
Multiplying $\Delta t\left.\frac{\partial\phi}{\partial t} \right|^{n+\theta}$
to equation \eqref{equ:alg_disc_2} and integrating over $\Omega$ leads to
\begin{equation}
\begin{split}
&
\int_{\Omega}\mathscr{H}^{n+\theta}\left[
  \left(\theta + \frac{1}{2}\right)\phi^{n+1} - 2\theta\phi^n
      + \left(\theta-\frac{1}{2}\right)\phi^{n-1}
\right] \\
& = -\lambda\int_{\Omega}\nabla^2\phi^{n+\theta} \left[
  \left(\theta + \frac{1}{2}\right)\phi^{n+1} - 2\theta\phi^n
      + \left(\theta-\frac{1}{2}\right)\phi^{n-1}
\right] \\
& \quad
+ S\int_{\Omega} \left(\phi^{n+1}-2\phi^n+\phi^{n-1} \right) \left[
  \left(\theta + \frac{1}{2}\right)\phi^{n+1} - 2\theta\phi^n
      + \left(\theta-\frac{1}{2}\right)\phi^{n-1}
\right] \\
& \quad
+ \frac{r^{n+\theta}}{\sqrt{E[\bar{\phi}^{n+\theta}]}}\int_{\Omega}h(\bar{\phi}^{n+\theta})\left[
  \left(\theta + \frac{1}{2}\right)\phi^{n+1} - 2\theta\phi^n
      + \left(\theta-\frac{1}{2}\right)\phi^{n-1}
\right].
\end{split}
\label{equ:int_2}
\end{equation}
Multiplying $2r^{n+\theta}\Delta t$ to equation \eqref{equ:alg_disc_3}
leads to
\begin{multline}
  2r^{n+\theta}\left[
    \left(\theta + \frac{1}{2}\right)r^{n+1} - 2\theta r^n
      + \left(\theta-\frac{1}{2}\right)r^{n-1}
      \right] \\
  =\frac{r^{n+\theta}}{\sqrt{E[\bar{\phi}^{n+\theta}]}}\int_{\Omega}h(\bar{\phi}^{n+\theta})
  \left[
    \left(\theta + \frac{1}{2}\right)\phi^{n+1} - 2\theta\phi^n
      + \left(\theta-\frac{1}{2}\right)\phi^{n-1}
      \right].
  \label{equ:int_3}
\end{multline}

Summing up equations \eqref{equ:int_1} and \eqref{equ:int_3},
and then subtracting equation \eqref{equ:int_2},
results in
\begin{equation}
  \begin{split}
    &
  2r^{n+\theta}\left[
    \left(\theta + \frac{1}{2}\right)r^{n+1} - 2\theta r^n
      + \left(\theta-\frac{1}{2}\right)r^{n-1}
      \right] \\
  &=
  -m\Delta t\int_{\Omega}\nabla\mathscr{H}^{n+\theta}\cdot\nabla\mathscr{H}^{n+\theta}
  -\lambda\int_{\Omega}\nabla\phi^{n+\theta}\cdot\left[
    \left(\theta + \frac{1}{2}\right)\nabla\phi^{n+1} - 2\theta\nabla\phi^n
      + \left(\theta-\frac{1}{2}\right)\nabla\phi^{n-1}
      \right] \\
  & \quad
  -S\int_{\Omega} \left(\phi^{n+1}-2\phi^n+\phi^{n-1} \right) \left[
  \left(\theta + \frac{1}{2}\right)\phi^{n+1} - 2\theta\phi^n
      + \left(\theta-\frac{1}{2}\right)\phi^{n-1}
      \right]
  \end{split}
  \label{equ:dist_eng_pre_1}
\end{equation}
where we have performed integration by part on
the right-hand-side (RHS) of equation \eqref{equ:int_1} and
the first term on the RHS of equation \eqref{equ:int_2}, and
used equations \eqref{equ:alg_disc_2}, \eqref{equ:alg_disc_4}
and \eqref{equ:alg_disc_5}.

Use the relations \eqref{equ:relation_2} and \eqref{equ:relation_3}
to transform the corresponding terms in \eqref{equ:dist_eng_pre_1},
and then collect related terms, and one would arrive at
the discrete energy balance equation \eqref{equ:discrete_energy_balance}.

\section*{Appendix B. Proof of Theorem \ref{thm:thm-3}}

Substituting $\psi_2^{n+1}$ from \eqref{equ:phi_2_equ} into
equations \eqref{equ:psi_2_equ} and \eqref{equ:psi_2_bc}
leads to
\begin{subequations}
\begin{equation}
  \nabla^2\left(\nabla^2\phi_2^{n+1}\right)
  -\frac{S}{\lambda\omega_0}\nabla^2\phi_2^{n+1}
  + \frac{\gamma_0}{\lambda\omega_0m\Delta t}\phi_2^{n+1}
  =\frac{1}{2\lambda}\nabla^2b^n,
  \label{equ:B-1}
\end{equation}
\begin{equation}
  \mathbf{n}\cdot\nabla\left(\nabla^2\phi_2^{n+1} \right)
  =\frac{1}{2\lambda}\mathbf{n}\cdot\nabla b^n,
  \quad \text{on}\ \partial\Omega,
  \label{equ:B-2}
\end{equation}
\end{subequations}
where we have used equation \eqref{equ:phi_2_bc}
and the relation
$
\alpha\left(\alpha + \frac{S}{\lambda\omega_0} \right)
= -\frac{\gamma_0}{\lambda\omega_0m\Delta t}
$.
The system of equations \eqref{equ:B-1}, \eqref{equ:B-2} and
\eqref{equ:phi_2_bc} is equivalent to the system
consisting of equations \eqref{equ:psi_2_equ}--\eqref{equ:phi_2_bc}.
By integrating equation \eqref{equ:B-1} over the domain $\Omega$,
we conclude that $\phi_2^{n+1}$ has the property
\begin{equation}
  \int_{\Omega} \phi_2^{n+1} = 0,
  \label{equ:phi_2_int}
\end{equation}
where we have used the divergence theorem and
the equations \eqref{equ:B-2} and \eqref{equ:phi_2_bc}.

Define function $\xi(\mathbf{x})$ by
\begin{equation}
  \left\{
  \begin{split}
    &
    \nabla^2\xi = \phi_2^{n+1},  \\
    &
    \mathbf{n}\cdot\nabla \xi = 0, \quad \text{on} \ \partial\Omega, \\
    &
    \int_{\Omega} \xi = \frac{\omega_0m\Delta t}{2\gamma_0}\int_{\Omega} b^n.
  \end{split}
  \right.
  \label{equ:def_xi}
\end{equation}
Let
\begin{equation}
  \Phi = \nabla^2\phi_2^{n+1} - \frac{S}{\lambda\omega_0}\phi_2^{n+1}
  + \frac{\gamma_0}{\lambda\omega_0m\Delta t}\xi
  -\frac{1}{2\lambda}b^n.
  \label{equ:def_Phi}
\end{equation}
Then equations \eqref{equ:B-1}, \eqref{equ:B-2} and \eqref{equ:phi_2_int}
are transformed into
\begin{subequations}
  \begin{align}
    &
    \nabla^2\Phi = 0, \\
    &
    \mathbf{n}\cdot\nabla\Phi = 0, \quad \text{on}\ \partial\Omega, \\
    &
  \int_{\Omega}\Phi = 0,
  \end{align}
\end{subequations}
where we have used \eqref{equ:def_xi} and \eqref{equ:phi_2_bc}.
So we conclude that $\Phi= 0$ and
\begin{equation}
  \nabla^2\phi_2^{n+1} - \frac{S}{\lambda\omega_0}\phi_2^{n+1}
  + \frac{\gamma_0}{\lambda\omega_0m\Delta t}\xi
  =\frac{1}{2\lambda}b^n.
  \label{equ:B-4}
\end{equation}
Taking the $L^2$ inner product between this equation
and $\phi_2^{n+1}$ and integrating by part, we get
\begin{equation}
  \frac{1}{2\lambda}\int_{\Omega} b^n\phi_2^{n+1}
  = -\int_{\Omega}\left|\nabla\phi_2^{n+1}\right|^2
  -\frac{S}{\lambda\omega_0}\int_{\Omega}\left|\phi_2^{n+1} \right|^2
  -\frac{\gamma_0}{\lambda\omega_0m\Delta t}\int_{\Omega}\left|\nabla\xi \right|^2
  \leqslant 0,
\end{equation}
where we have used the divergence theorem and equation \eqref{equ:def_xi}.

\section*{Acknowledgement}
This work was partially supported by
NSF (DMS-1318820, DMS-1522537).

\bibliographystyle{plain}
\bibliography{nphase,obc,mypub,nse,sem,contact_line,interface,multiphase}

\end{document}